\documentclass[11pt]{article}

\usepackage{geometry,graphicx,amsmath,amssymb,amsfonts,mathdots,mathrsfs,caption,subcaption, setspace, wrapfig}
\usepackage[usenames,dvipsnames,svgnames,table]{xcolor}
\usepackage{graphicx}
\usepackage{jheppub}
\usepackage{epsfig}
\usepackage{epstopdf}
\usepackage{longtable}
\usepackage{braket}
\usepackage{wasysym}
\usepackage{comment}
\usepackage{amsmath}
\usepackage{slashed}

\usepackage{multicol}
\usepackage{multirow}
\usepackage{xspace}
\usepackage{xcolor}
\usepackage{comment}

\graphicspath{{../}{./Figures/}}
\newcommand{\met}{\ensuremath{\not\!\!E_T}\xspace}

\begin{document}


\title{Extending LHC Coverage to Light Pseudoscalar Mediators and Coy Dark Sectors}
\author[]{Jonathan Kozaczuk,}
\author[]{Travis A. W. Martin}
\affiliation[]{TRIUMF, 4004 Wesbrook Mall, Vancouver, Canada V6T 2A3}

\emailAdd{jkozaczuk@triumf.ca}
\emailAdd{tmartin@triumf.ca}

\abstract{Many dark matter models involving weakly interacting massive particles (WIMPs) feature new, relatively light pseudoscalars that mediate dark matter pair annihilation into Standard Model fermions.  In particular, simple models of this type can explain the gamma ray excess originating in the Galactic Center as observed by the Fermi Large Area Telescope.  In many cases the pseudoscalar's branching ratio into WIMPs is suppressed, making these states challenging to detect at colliders through standard dark matter searches.  Here, we study the prospects for observing these light mediator states at the LHC without exploiting missing energy techniques. While existing searches effectively probe pseudoscalars with masses between  5 -- 14 GeV and above 90 GeV, the LHC reach can be extended to cover much of the interesting parameter space in the intermediate 20 -- 80 GeV mass range in which the mediator can have appreciable Yukawa-like couplings to Standard Model fermions but would have escaped detection by LEP and other experiments. Models explaining the Galactic Center excess via a light pseudoscalar mediator can give rise to a promising signal in this regime through the associated production of the mediator with bottom quarks while satisfying all other existing constraints. We perform an analysis of the backgrounds and trigger efficiencies, detailing the cuts that can be used to extract the signal.  A significant portion of the otherwise unconstrained parameter space of these models can be conclusively tested at the 13 TeV LHC with 100 fb$^{-1}$, and we encourage the ATLAS and CMS collaborations to extend their existing searches to this mass range.}

\maketitle

\section{Introduction}

Light, weakly-interacting massive particles (WIMPs) are a particularly compelling class of particle dark matter (DM) candidates.  The case for WIMPs with masses close to the electroweak scale has been strengthened by recent observations of an excess in gamma rays originating from the Galactic Center (GC) by the Fermi Large Area Telescope \cite{Goodenough:2009gk, Hooper:2010mq, Boyarsky:2010dr, Hooper:2011ti, Abazajian:2012pn, Hooper:2013rwa, Gordon:2013vta, Abazajian:2014fta, GCrecent, Zhou:2014lva}. This signal has garnered much recent attention, since its morphology closely resembles that expected from dark matter pair annihilation into bottom quarks \cite{GCrecent, Calore:2014xka}, though other final states can also provide a good fit when systematics are properly taken into account \cite{Agrawal:2014oha}. Moreover, the signal suggests a WIMP annihilation rate close to that required in the early universe for a thermal relic to saturate the observed dark matter density \cite{GCrecent}, and the excess is difficult to explain in terms of astrophysical backgrounds alone \cite{GCrecent, Cholis:2014lta}.  This has led many to believe that the Fermi GC signal may represent the first (indirect) observation of dark matter to date.  

A common and well-motivated class of models that can explain the observed excess features dark matter annihilating through a light pseudoscalar with Yukawa-like couplings to Standard Model fermions \cite{coy, Ipek:2014gua, Cheung:2014lqa, Guo:2014gra}.  For example, these states appear generically in two Higgs doublet models and their extensions \cite{Gunion:1989we}, as well as pseudo Nambu Goldstone bosons associated with the spontaneous breaking of a new global symmetry \cite{Batell:2009jf, Mimasu:2014nea, Dolan:2014ska}. Their couplings to Standard Model fermions can arise at tree- or loop-level (see e.g. Ref.~\cite{Shuve} for an example with heavy vector-like fermions).  Since they couple to the visible sector, such pseudoscalars can constitute a portal to the dark sector, mediating the annihilation of dark matter (DM) into SM final states \cite{Kozaczuk:2013spa,coy,Ghorbani:2014qpa,  Buckley:2014fba, Harris:2014hga, Arina:2014yna, Hektor:2014kga}. 

Understanding how dark matter interacts with the visible sector is a crucial part of the current dark matter program.  Direct detection experiments \cite{Agnese:2013jaa, Akerib:2013tjd, Agnese:2014aze, Angloher:2014myn} and the observation of a Standard Model-like 125 GeV Higgs with a small invisible decay width \cite{Aad:2014iia, Chatrchyan:2014tja} have severely constrained $Z$- and Higgs boson-mediated scenarios \cite{Dolan:2014ska}.  As a result, much recent work has been devoted to studying various possibilities for new mediator particles coupling weakly to the Standard Model degrees of freedom.  Of these possibilities, pseudoscalars stand apart for several reasons.  For one, they do not predict sizable spin-independent direct detection signals, in contrast with scalar and vector mediators. Furthermore, current collider constraints on new pseudoscalar particles are generally weaker than those on new scalar and vector degrees of freedom \cite{Berlin:2014tja, Hooper:2014fda}. 

If the GC excess is indeed a signal of dark matter annihilation, and if the annihilation is mediated by a new pseudoscalar particle, it is both important and timely to consider how one might probe such scenarios at colliders.  Much progress has already been made on this front. Based on the topology and kinematics of the dominant dark matter annihilation channel, scenarios explaining the GC excess with pseudoscalar mediators can be grouped into roughly three types, each with distinct prospects for collider discovery:
\begin{enumerate}

\item Models which rely on dark matter annihilating into on-shell mediators \cite{Boehm:2014bia, Abdullah:2014lla, Martin:2014sxa,Berlin:2014pya, Ko:2014gha, Cerdeno:2015ega}.  In this case, the annihilation rate into SM fermions factorizes and the coupling of the pseudoscalar mediator to SM degrees of freedom can be very small.  Prospects for direct collider searches are often dim in this case, but there may be other handles on these models provided by direct detection, as well as fixed target and other precision experiments \cite{Boehm:2014bia, Abdullah:2014lla, Martin:2014sxa,Berlin:2014pya}.

\item Scenarios featuring a pseudoscalar mediator with a significant invisible branching fraction \cite{Shuve, Agrawal:2014una, Huang:2014cla, Alves:2014yha, Cahill-Rowley:2014ora, Harris:2014hga, Buckley:2014fba}. This results in distinctive missing energy signatures at the LHC which can be effectively probed by $b\bar{b}+$MET, mono-jet, and other existing and planned LHC searches, as studied in detail in e.g. Refs.~\cite{Alves:2014yha, Shuve, Buckley:2014fba, Harris:2014hga}.

\item Scenarios in which the pseudoscalar mediator is expected to have a small branching fraction into dark matter particles \cite{coy, Ipek:2014gua, Cheung:2014lqa, Arina:2014yna, Guo:2014gra, Hektor:2014kga}.  This can occur when the coupling between the dark matter and the mediator is small relative to the coupling of the mediator to Standard Model degrees of freedom, or when on-shell decays of the mediator into WIMP pairs is not kinematically allowed. Such scenarios can be more difficult to probe directly at the LHC than case 2, since they lack a distinctive missing energy signature \cite{coy}. In concrete models of this type, rare Higgs decays can be constraining, however the resulting limits can be straightforwardly avoided in many instances, as can limits from LEP, the Tevatron, and $B$-physics experiments (see e.g. Refs.~\cite{Ipek:2014gua, Cheung:2014lqa}).  While a signal would arise in indirect detection experiments, it has been shown that the dark matter and mediator in this case might avoid detection elsewhere \cite{coy}.  This rather grim scenario is appropriately known as ``Coy dark matter".

\end{enumerate}

In this study we will focus our attention on case 3 above, as it is a generic yet largely unconstrained possibility, as we discuss below.  We will restrict our attention to light mediators, with masses below 90 GeV, as pseudoscalars with larger masses are already probed by existing LHC Higgs searches. Furthermore, light pseudoscalars are very attractive from the standpoint of the Galactic Center excess, since they can provide an efficient resonant annihilation channel for the light dark matter masses suggested by the signal and, in some cases, allow for a $p$-wave annihilation channel into pairs of mediators to drive down the relic abundance without violating constraints from dwarf spheroidal observations \cite{Dolan:2014ska}.  In this situation, on-shell decays of the pseudoscalar to pairs of dark matter particles are suppressed and WIMP production at the LHC through the mediator will be negligible. Our strategy will be to extend LHC coverage to such scenarios by probing the light pseudoscalar directly through its interactions with the Standard Model degrees of freedom.  The discovery of such a new particle would constitute a great step forward in our understanding of the dark sector and open up many possibilities for further study, including more dedicated experiments to probe its coupling to dark matter directly.  

As we discuss below, the GC excess can suggest an appreciable mediator coupling to down-type fermions.  Consequently, we focus on the associated production of the mediator with a $b$-jet or $b\bar{b}$ pair.  We will assume that the mediator couples to Standard Model fermions with strength proportional to their mass, as in models with minimal flavor violation (MFV). We find that, for a significant range of mediator masses and couplings consistent with the GC excess, a promising signal is predicted in the 1--2 $b+a$ production modes, with $a\rightarrow \tau^+\tau^-$. We also explore the possibility of $a\rightarrow \mu^+ \mu^-$ decays, which is more promising for low masses and likely features lower systematic uncertainties.  Existing searches for pseudoscalars motivated by the Minimal Supersymmetric Standard Model (MSSM) and Next-to-MSSM (NMSSM) currently probe mediator masses down to 90 GeV and in the low-mass region between 5--14 GeV. However, we find that coverage can be extended to pseudoscalars in the intermediate mass range (between 20--80 GeV), which are promising for explaining the GC excess and would have evaded detection by LEP. We encourage both ATLAS and CMS to expand their analysis to include this region. In this study, we detail the cuts and kinematic variables that can be used to reduce the large backgrounds and show the extent to which the parameter space in these models can be conclusively tested at the 13 TeV LHC with 100 fb$^{-1}$ of integrated luminosity. We demonstrate this using a simplified model and show the application of our results to the otherwise unconstrained parameter space of the NMSSM that is consistent with the excess (the NMSSM can be mapped directly onto our model).

It is important to emphasize that, although we will focus on pseudoscalars mediating dark matter annihilation consistent with the GC signal, our study can be applied much more generally to any model featuring light mediators with significant coupling to isospin-down Standard Model fermions.  Since we assume that the invisible branching fraction of the pseudoscalar is small, our analysis of the predicted collider signatures does not depend on the pseudoscalar's coupling to dark matter, nor on the nature of the dark matter itself.

This study is organized as follows: in Sec.~\ref{sec:model}, we discuss the simplified model used for our analysis, its relationship to the GC excess, and the existing constraints on light pseudoscalars. The following section (Sec.~\ref{sec:production}), details the collider signatures of the new mediator, as well as the backgrounds and trigger efficiencies relevant for our analysis. Our results for the LHC discovery potential of light psuedoscalar mediators are presented and discussed in Sec.~\ref{sec:results}, with further details of the analysis contained in Appendices \ref{sec:appa}, \ref{sec:appb}, and  \ref{sec:appc}.  We then apply these results to the NMSSM in Sec.~\ref{sec:NMSSM}, showing that the searches we propose here can cover much of the parameter space consistent with the excess and that is currently unconstrained by other experimental searches.  Finally, we summarize and conclude in Sec.~\ref{sec:summary}.

\section{A Simplified Model}\label{sec:model}

For our analysis, we follow Ref.~\cite{coy} and consider a light pseudoscalar that couples to Dirac fermion dark matter, $\chi$, and to Standard Model fermions, with effective Lagrangian
\begin{equation}\label{eq:Lag}
\mathcal{L}_{\rm int}\supset -i \frac{g_{\rm DM}}{\sqrt{2}}a\bar{\chi}\gamma^5\chi-i\sum_{i=u,c,t}\frac{g_u y_i}{\sqrt{2}}a\bar{f}_i\gamma^5f_i - i\sum_{\substack{i=d,s,b,\\ e,\mu,\tau}}\frac{g_d y_i}{\sqrt{2}}a\bar{f}_i\gamma^5f_i,
\end{equation}
where $y_i=m_i/v$ are the SM Yukawa couplings, with $v=174$ GeV.  We have assumed that the pseudoscalar $a$ couples to the SM fermions with strength proportional to their masses. The pseudoscalar couplings to up- and down-type fermions are further assumed to depend on the overall scaling factors, $g_{u,d}$, which we take to be the same for all down- or up-type fermions \footnote{These assumptions need not be the case to explain the GC excess, and our results can be applied beyond this set-up by appropriately rescaling the pseudoscalar production cross-sections and branching ratios.}. These factors appear e.g. in Two Higgs Doublet models (2HDMs) and their extensions; in a Type II 2HDM, $g_d=1/g_u=\tan\beta$, where $\beta$ is the ratio between the two $SU(2)$ Higgs vacuum expectation values.  With the addition of a singlet that mixes with the $SU(2)$ doublets, the effective couplings become $g_u=\cot \beta \cos\theta$ and $g_d=\tan\beta \cos\theta$, where $\theta$ is the mixing angle between the $SU(2)$ and singlet pseudoscalars. 

Note that Ref.~\cite{coy} considered the case in which $g_d=g_u=1$. This situation is very difficult to probe at colliders. Explaining the Fermi GC signal with $g_u=g_d=1$ can require rather large values of $g_{\rm DM}$, unless the annihilation is quite close to the $s$-channel resonance.  Often in ultra-violet (UV) complete models, a sizable value for $g_{\rm DM}$ occurs together with low mass WIMPs in parametric regions featuring a large invisible branching fraction of the Standard Model-like Higgs \cite{Kozaczuk:2013spa}, which is not observed. On the other hand, for pseudoscalar--WIMP couplings that are not too large, the Galactic Center excess suggests enhanced couplings to down-type fermions, as we show below. This situation is much more promising from the standpoint of LHC searches and, in some cases, is not probed by existing searches.

\subsection{Explaining the Excess}
Given the Lagrangian in Eq.~\ref{eq:Lag}, the zero-temperature $s$-channel annihilation rate for dark matter through a pseudoscalar into SM fermion pair $f_i \bar{f}_i$ is
\begin{equation}
\langle \sigma v\rangle _i = \frac{N_{C,i}}{8\pi}\frac{g_{\rm DM}^2 g_{i}^2 y_i^2 m_{\chi}^2}{(m_a^2-4m_{\chi}^2)^2+m_a^2\Gamma_a^2}\sqrt{1-\frac{m_{i}^2}{m_{\chi}^2}},
\end{equation}
where $m_i$, $N_{C,i}$ are the mass and color factor of the decay states, $g_i$ is either $g_{u,d}$ depending on the fermion, and $\Gamma_a$ is the total width of the mediator.  Throughout this study we will assume that the dominant DM annihilation channel is $\chi \bar{\chi} \rightarrow b \bar{b}$. This mode has received the most attention in explaining the GC excess, and although a recent analysis has pointed out that other channels can also explain the signal \cite{Agrawal:2014oha}, annihilation into a  $b\bar{b}$ pair still provides a very good fit to the data. 

There have been several recent developments in determining which annihilation channels, WIMP masses, and annihilation rates best fit the Fermi data. For the $b\bar{b}$ channel, most previous analyses had suggested that $m_{\chi}$ should fall roughly in the range 35 GeV $\lesssim m_{\chi}\lesssim$ 50 GeV with annihilation rate $\langle \sigma v \rangle \simeq 2$--$6\times10^{-26}$ cm$^3/$s  \cite{coy,GCrecent} (the required annihilation rate for self-conjugate dark matter would be reduced by a factor of two relative to these values). However, there are large systematic uncertainties associated with the propagation of gamma rays in the Galactic Center that must be taken into account. The impact of these systematics was first studied in Ref.~\cite{Cholis:2014lta}, and subsequently by Ref.~\cite{Agrawal:2014oha}, which performed a detailed analysis incorporating several different models for the diffuse gamma ray background supplied by the Fermi collaboration.  The end result is that the range of WIMP masses and annihilation modes statistically consistent with the excess increased significantly once these systematic uncertainties were taken into account. In particular, the range for WIMP masses annihilating primarily into $b\bar{b}$ was extended to \cite{Agrawal:2014oha} 
\begin{equation}
35 \hspace{0.1cm}{\rm GeV} \lesssim m_{\chi} \lesssim 165  \hspace{0.1cm}{\rm GeV}, \hspace{0.5cm} \chi \bar{\chi}\rightarrow b\bar{b}.
\end{equation}
for specific values of the annihilation rate.

\begin{figure*}[!t]
\centering
\mbox{\includegraphics[width=0.45\textwidth,clip]{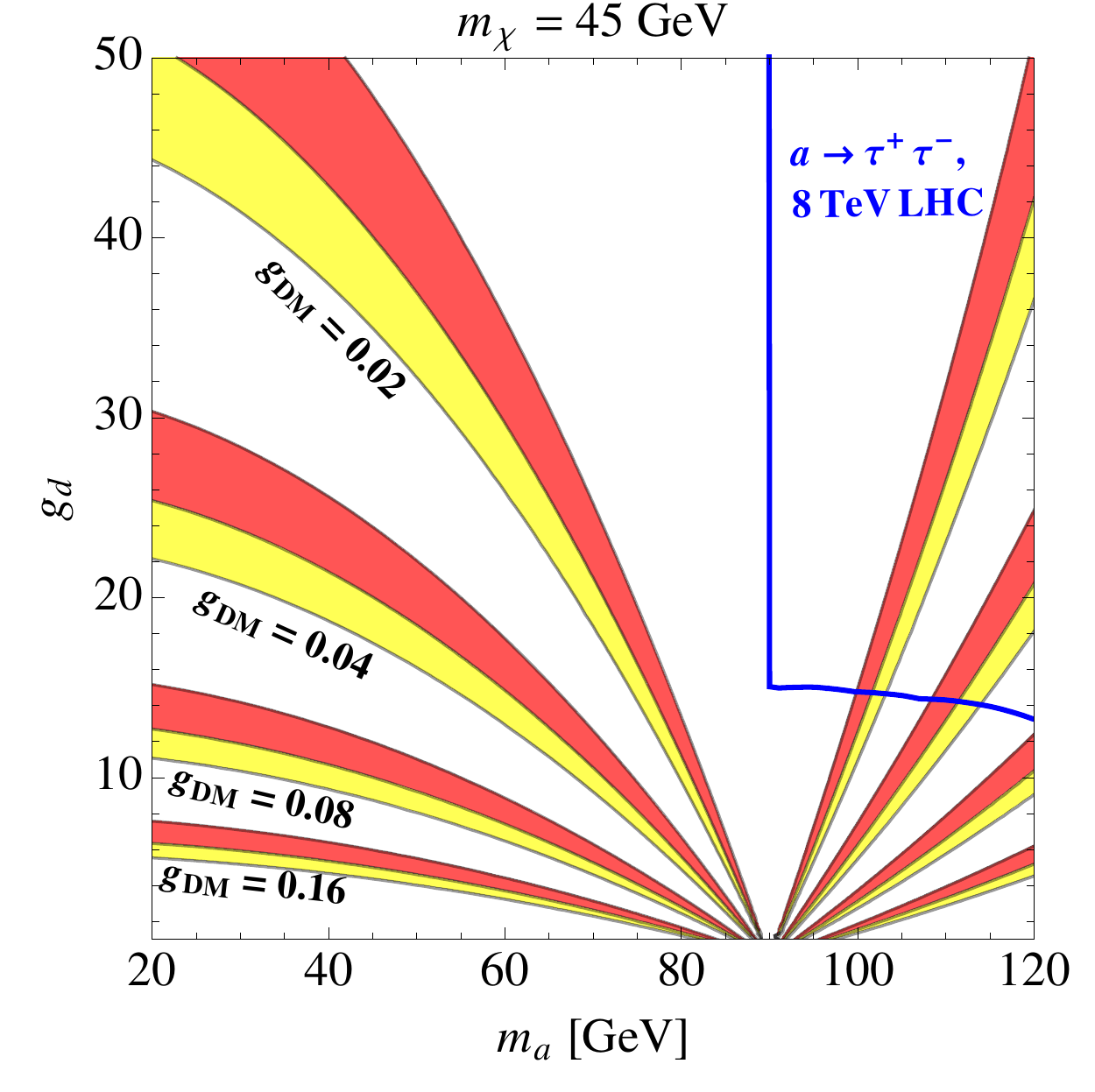}\hspace*{1cm}\includegraphics[width=0.45\textwidth,clip]{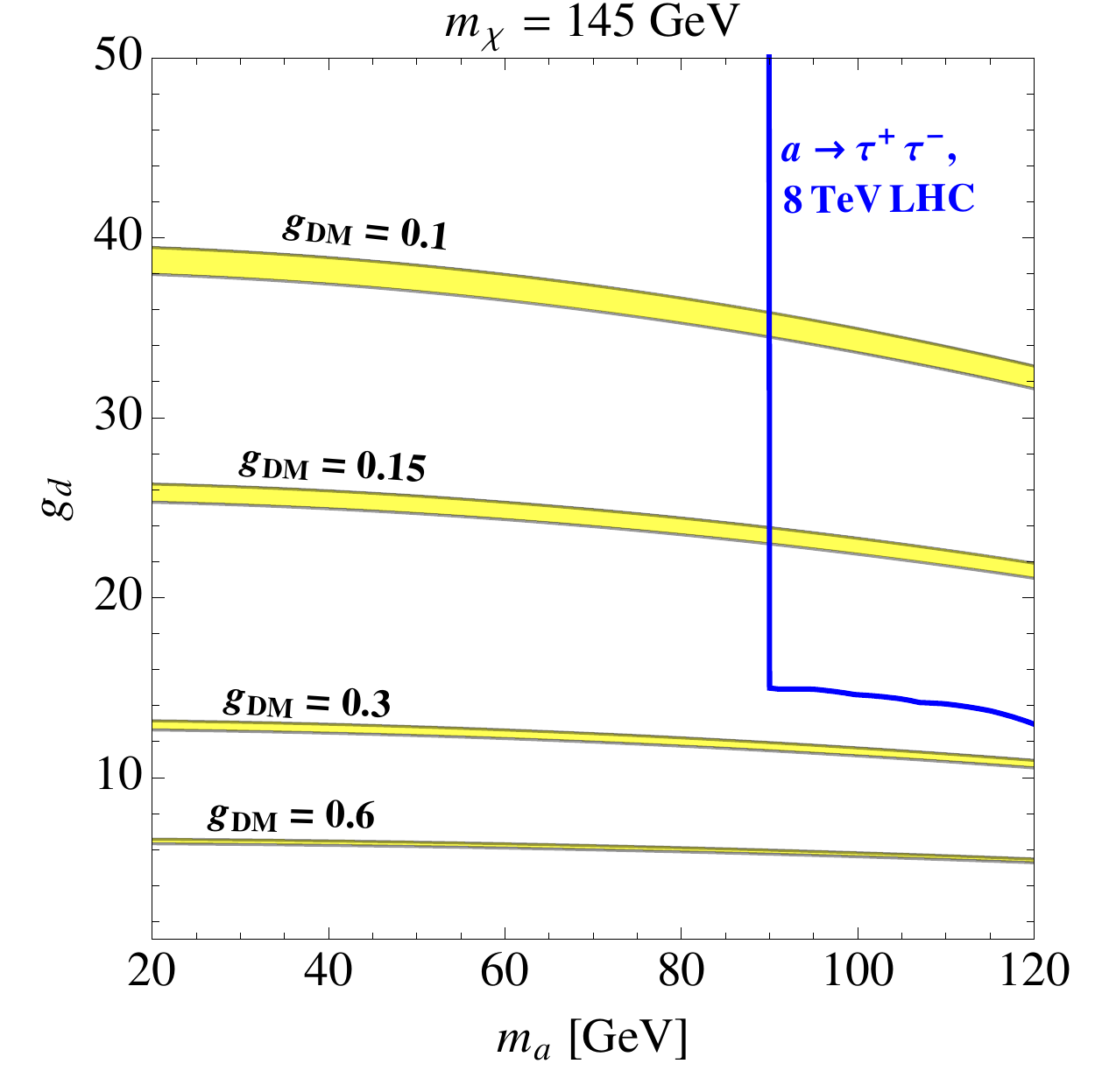}}
\caption{\label{fig:DM_CS}\small Value of the mediator coupling to down-type SM fermions (relative to that of a Standard Model-like Higgs boson of the same mass) required to explain the GC excess as a function of $m_a$, for several values of $g_{DM}$ and masses $m_\chi =45$, 145 GeV, assuming that $\chi$ saturates the relic abundance.  The shaded regions are compatible with the signal, with the red (upper) regions in each band excluded by the recent dwarf spheroidal constraints from Fermi \cite{dSph}, and the yellow (lower) regions with an annihilation rate compatible with both the excess and the constraints. The upper bound on $g_d$ from existing LHC searches for $a\rightarrow \tau^+\tau^-$ is shown in blue.}
\end{figure*}

Across this mass window, the signal from the Galactic Center suggests a clear range of values for the coupling constants of the mediator to SM states for a given $g_{\rm DM}$ in this setup.  Since the annihilation cross-section and pseudoscalar width are dominated by down-type interactions (for $BR(a\rightarrow \chi\bar{\chi})\ll1$), the only significant parametric dependence is on $g_{\rm DM}$, $g_d$, $m_{\chi}$ and $m_a$.  The down-type scale factor $g_d$ required to explain the GC excess for $m_{\chi}=45$ GeV and $m_{\chi}=145$ GeV (close to the best fit mass for the Fermi model (d) from Ref.~\cite{Agrawal:2014oha}) is shown by the bands on the left and right hand side of Fig.~\ref{fig:DM_CS}, respectively, as a function of $m_a$ for various values of $g_{\rm DM}$.  The range of annihilation rates allowed in the low mass case (LHS) is taken from Ref.~\cite{Cholis:2014lta}, while the allowed values in the high mass case (RHS) are taken from Ref.~\cite{Agrawal:2014oha}. In both cases the local dark matter density is assumed to be $\rho_{\astrosun}=0.4$ GeV/cm$^3$. The preferred regions depend on $\mathcal{J}\equiv \bar{J}/\bar{J}_0$, the ratio of the angularly--averaged integral over the line-of-sight dark matter density $\rho_{\rm DM}(r)$, given by
\begin{equation}
\bar{J}(\psi)=\frac{1}{\Delta \Omega}\int_{\Delta \Omega} d\Omega \int_{\rm l.o.s.}ds \hspace{0.1cm} \rho_{\rm DM}(r)^2,
\end{equation}
to the canonical value $\bar{J}_0$. For the low mass ($m_{\chi}=45$ GeV) case we take $\mathcal{J}=1$, while for $m_{\chi}=145$ GeV we take $\mathcal{J}=0.3$, which is within the systematic uncertainties discussed in Ref.~\cite{Agrawal:2014oha}. The latter choice allows for an annihilation rate close to the canonical thermal freeze-out value ($\langle \sigma v \rangle \simeq 4.4\times 10^{-26}$ cm$^3/$s for Dirac fermion dark matter \cite{Steigman:2012nb}) and consistent with the Fermi signal while evading the dwarf spheroidal constraints, discussed below.

For reasonable choices for $g_{\rm DM}$, the value of $g_d$ must be quite large to account for the GC excess, unless the masses are tuned to fall very close to the resonance. In addition, reducing the $\chi$ abundance has the effect of increasing the preferred value of $g_d$ for a given $g_{\rm DM}$. The regions of parameter space with large $g_d$, in many cases preferred by the signal, predict a significant mediator production cross section at the LHC in association with bottom quarks. Also, the pseudoscalar's invisible branching fraction is small across the entire parameter space, except for low $g_d$ and large $g_{\rm DM}$. For $m_a<2m_{\chi}$ an on-shell pseudoscalar cannot decay to a pair of WIMPs, while for $m_a>2m_{\chi}$ we find that $BR(a\rightarrow \chi \bar{\chi})>0.1$ only for $g_d\lesssim 4$ for $g_{\rm DM}=0.1$ in the $m_{\chi}=45$ GeV case, since everywhere else $g_d$ ($g_{\rm DM}$) is too large (small) for this decay to contribute appreciably to the total width.

It is important to note that the Fermi collaboration recently released updated limits on the dark matter annihilation rate from observations of dwarf spheroidal (dSph) galaxies \cite{dSph}. The resulting constraints\footnote{There are also potential constraints on dark matter interpretations of the excess coming from observations of the cosmic ray antiproton spectrum \cite{Cirelli:2014lwa}. These constraints depend sensitively on the propagation of the charged cosmic rays through the galaxy, which is difficult to model and results in large uncertainties on the predicted flux \cite{Hooper:2014ysa}. With conservative choices for the propagation model, the signal can be shown to be consistent with the current limits \cite{Cirelli:2014lwa}.  Similarly, radio observations of the Galactic Center region can also be consistent with a dark matter interpretation of the excess \cite{Cholis:2014fja}.} are in mild tension with a dark matter explanation of the excess, however there is still a large amount of parameter space capable of explaining the GC excess that survives this constraint. This is shown in Fig.~\ref{fig:DM_CS}, in which the red bands show the impact of the dwarf spheroidal limits (points in these bands could potentially explain the excess but are excluded at 95\% C.L.). Meanwhile the yellow bands show points consistent with both the GC excess and dSph limits. Note that in the high mass case all points consistent with the excess are compatible with the dSph constraints for our particular choice of $\mathcal{J}$.

One concern may be that, since the recent dSph constraints disfavor larger annihilation rates, some points with light WIMP masses consistent with the GC excess and dSph limits will tend to produce too large a relic abundance. The dark matter relic density is set by the annihilation rate at finite temperature, which can differ from that at $T=0$. In particular, for $s$-channel annihilation through a pseudoscalar with $m_a<2 m_{\chi}$, the annihilation rate at $T=0$ is greater than that at freeze-out ($T_{f.o.}\sim m_{\chi}/20$). The upper limit on the annihilation rate, set by the Fermi dSph results, is below the required annihilation rate at freeze-out for $m_{\chi}\lesssim 100$ GeV, naively disfavoring this region. However, there are several well-known and straightforward exceptions to this reasoning \cite{Griest:1990kh}. For example, $p$-wave processes with contributions to the total annihilation rate scaling as $v_{DM}^2$ (with $v_{DM}$ the relative dark matter velocity) will become important at freeze-out, increasing the annihilation rate at $T_{f.o.}$ but not altering the $T=0$ prediction.  An example of such a process generically expected along with light mediators is $\chi\bar{\chi} \rightarrow a a$ (this is another virtue of the light pseudoscalar scenario).   Other scenarios allowing for an enhanced annihilation rate at $T_{f.o.}$ relative to that at late times include those with additional co-annihilation channels or featuring $m_a>2m_{\chi}$ so that $\langle \sigma v\rangle_{T=0}<\langle \sigma v\rangle_{T=T_{f.o.}}$. Thus, although in some cases the dSph limits may result in requiring some additional tuning or model-building to achieve the correct DM relic abundance, dark matter explanations of the excess, particularly those involving $s$-channel annihilation through a relatively light pseudoscalar, are alive and well.  Note that this tension largely disappears above $m_{\chi}\sim 100$ GeV, since the dSph upper bound is above the canonical WIMP cross-section in this region (although one should verify that contributions to the annihilation rate at freeze-out from the other states in the theory do not over-dilute the relic density).  

In summary, dark matter annihilating through a relatively light pseudoscalar can explain the Galactic Center excess and be compatible with the recent dwarf spheroidal limits from Fermi. In all most discussed and shown above we expect $BR(a\rightarrow \chi\bar{\chi})\ll1$, either because $m_a<2m_{\chi}$, $g_{DM}\ll 1$, or both.  This implies a low likelihood of observing the pseudoscalar through missing energy signals at the LHC. In the following subsection, we describe some of the other existing constraints on the parameter space and highlight the need for direct coverage of these scenarios at the LHC.  

\subsection{Existing Constraints}

Our goal will be to ascertain to what extent LHC searches can cover the parameter space shown in Fig.~\ref{fig:DM_CS} that is not currently probed by LHC searches \cite{Aad:2014vgg, Khachatryan:2014wca, Chatrchyan:2012am, Chatrchyan:2013qga, Aad:2014ioa} . To our knowledge, there are currently no \emph{direct} constraints on the parameter space of our simplified model with 15 GeV $\lesssim m_a \lesssim 90$ GeV. By this, we mean that there do not exist constraints depending only on the pseudoscalar's coupling to SM fermions in this mass range.  There are indeed several other \emph{indirect} constraints, but these are inherently dependent on other degrees of freedom in the UV complete theory and can be straightforwardly avoided in many cases.  We will present explicit examples of points evading all of the searches discussed below but still predicting an observable LHC signal in the NMSSM in Sec.~\ref{sec:NMSSM}.

Pseudoscalar mediators with GeV-scale masses predict highly suppressed direct dark matter detection cross-sections. At tree-level, the pseudoscalar only interacts spin-dependently with nuclei. Using the expressions and results found in Refs.\cite{coy, Freytsis:2010ne}, we find that the spin-dependent scattering cross-section for dark matter off of nuclei via the pseudoscalar is far below the reach of current and planned experiments ($\sigma_{\rm SD}\lesssim 10^{-48}$ cm$^2$) across the parameter space we consider.  This is thanks to the $1/m_a^4$ suppression in $\sigma_{SD}$ in this regime. Also, while spin-independent scattering can occur via one loop diagrams, this contribution is also much too small to be observed. The difficulty in observing dark matter interacting with the visible sector primarily through a pseudoscalar in direct detection experiments in indeed one of the main reasons such models are understood to be {\em coy}.

Light pseudoscalars can also be constrained by flavor observables\footnote{Other precision tests, such as the pseudoscalar contributions to the muon $g-2$, do not impact the parameter space we consider \cite{Domingo:2008bb}, although a light pseudoscalar with very large $g_d$ can, in some cases, help reconcile the observed $(g-2)_{\mu}$ with the SM prediction \cite{Hektor:2014kga}. Also, precision electroweak measurements are typically unconstraining, since pseudoscalar contributions to the gauge boson self-energies first appear at two loops.}. Loop diagrams involving the pseudoscalar can generate effective flavor-changing vertices \cite{Hiller:2004ii, Dolan:2014ska}. The limits are severe for pseudoscalars lighter than the $B$ and $\Upsilon$ meson scale simply because the mediator can be produced on-shell in decays. For $m_a\gtrsim 10$ GeV, the constraints are very significantly relaxed, with the most stringent arising from LHCb \cite{Aaij:2013aka} and CMS \cite{Chatrchyan:2013bka} measurements of $BR(B_s\rightarrow \mu^+\mu^-)$. For $m_a \gg m_B$, the limits are approximately 
\begin{equation}
\label{eq:Bsmumu}
g_d \lesssim \frac{3 m_a}{10 \hspace{0.1cm} {\rm GeV}}
\end{equation}
considering the pseudoscalar contribution alone \cite{Dolan:2014ska}. This constraint would naively appear to directly constrain some of the parameter space shown in Fig.~\ref{fig:DM_CS}, however, the new contributions to $B_s\rightarrow \mu^+\mu^-$ are strongly model-dependent \cite{Altmannshofer:2012ks}. For example, in supersymmetric UV completions of our model, such as the NMSSM, there are several new contributions which enter with opposite sign to that from the $a$-induced vertex. Thus, cancellations can occur over large portions of the parameter space allowing for light pseudoscalars with large couplings to SM fermions (i.e. above the naive upper bound of Eq.~\ref{eq:Bsmumu}) \cite{Cheung:2014lqa}, once again highlighting the need for direct probes of this parameter space.

For light mediators with $2m_a<m_{h}$ ($h$ is the 125 GeV SM-like Higgs), exotic Higgs decays to pseudoscalar pairs can affect the Higgs width and signal rates \cite{Curtin:2013fra, Curtin:2014pda}, which are constrained by both ATLAS \cite{ATLAS_exotic} and CMS \cite{CMS_exotic}.  Evidence for $h\rightarrow aa$ decays was also searched for at LEP \cite{Schael:2010aw} and the Tevatron \cite{Abazov:2009yi}. Such decay modes can also be very effectively probed at the High Luminosity LHC  \cite{Curtin:2014pda}. Indeed, this has long been recognized as an important potential discovery channel of NMSSM pseudoscalars at colliders \cite{Dobrescu:2000jt, Dermisek:2005ar, Dermisek:2007yt, Carena:2007jk, Belyaev:2010ka}.  However, these constraints depend on the $haa$ coupling which, in some cases, can be made appropriately small in realistic models \cite{Kozaczuk:2013spa, Cheung:2014lqa}, especially those in which the pseudoscalar coupling to Standard Model fermions does not arise through mixing with the SM-like Higgs \cite{Shuve}. Alternatively, simply taking $m_a>m_h/2$ avoids these constraints altogether.  

Another indirect constraint arises from LEP searches for $e^+e^-\rightarrow ha$ production \cite{Schael:2006cr}.  While these results prohibit MSSM-like pseudoscalars lighter than 90 GeV for all values of $\tan\beta$, these bounds depend on the $Zha$ coupling, which is model-dependent, and can again be straightforwardly avoided \cite{Cheung:2014lqa}.  For example, in a Type II 2HDM with an additional singlet (2HDM+S), the $Zh_ia$ coupling scales as
\begin{equation}
g_{Zh_ia}\sim \cos\theta \left(S_{i1}\sin\beta-S_{i2}\cos\beta\right)
\end{equation}
where $S_{i1}$ and $S_{i2}$ are the corresponding entries in the matrix diagonalizing the $3\times 3$ CP-even mass matrix with the Higgs bosons ordered in mass (see e.g. Eq.~2.22 in Ref.~\cite{Ellwanger:2009dp}). Contrasting to $g_d\sim \cos \theta \tan\beta$, we see that the simple limit $\cos\theta \ll 1$, $\tan\beta\gg1$ can result in an appreciable $g_d$ with a significantly suppressed $g_{Zha}$.  

Finally, existing MSSM Higgs boson searches at the Tevatron \cite{Abazov:2008zz, Aaltonen:2009vf, Abazov:2010ci, Abazov:2011jh, Aaltonen:2011nh} and the LHC \cite{Aad:2014vgg, Khachatryan:2014wca, Chatrchyan:2013qga} constrain $g_d$ for $m_a>90$ GeV, but in an effort to avoid the large backgrounds encountered for lighter masses, and because LEP had already ruled out MSSM-like pseudoscalars with masses below 90 GeV, the published limits do not extend below the $Z$ mass.  There are also searches for light ($m_a\lesssim 15$ GeV) pseudoscalars at CMS \cite{Chatrchyan:2012am}, motivated by certain limits of the NMSSM. However, the 15 GeV $\lesssim m_a\lesssim $90 GeV mass range remains currently untested \footnote{There are also searches for $a\rightarrow \gamma \gamma$ that probe masses down to $m_a=65$ GeV \cite{Aad:2014ioa}, however, as we will see below, the production cross-section in the diphoton mode is very small in the parameter space we consider and thus significantly below the existing limits.}.

Although the collider limits on a light pseudoscalar can be avoided, one might also be concerned about the consistency of this scenario once the model is UV-completed. Our Lagrangian is not invariant under $SU(2)_L\times U(1)_Y$, and so, given a particular UV completion, one should also check that constraints on the other states can be satisfied while demanding a light pseudoscalar.  In 2HD+S models, for example, most constraints on the rest of the Higgs sector can be satisfied by simply taking the charged Higgs mass to be moderately heavy (a few hundred GeV) with an appropriate choice of $\tan\beta$ \cite{Aad:2014vgg, Khachatryan:2014wca}. Such requirements are consistent with light pseudoscalars and sizable $g_d$, as shown in e.g. Ref.~\cite{Cheung:2014lqa} and in Sec.~\ref{sec:NMSSM} for the NMSSM. 

Perhaps surprisingly, there is a significant gap in coverage for light pseudoscalars with appreciable couplings to SM fermions, as arise in models explaining the GC excess or otherwise. This situation has room for improvement.  In the remaining portion of this paper, we will investigate to what extent searches similar to those already existing for heavy MSSM Higgs bosons and for light NMSSM pseudoscalars can directly probe the parameter space motivated by the Galactic Center excess.  This task requires a careful treatment of the backgrounds below the $Z$ mass. As we will show below, the backgrounds can be substantially reduced by using a suitably chosen sequence of kinematic cuts.

\section{Light Mediators at the LHC}\label{sec:production}
\subsection{Production and Signals}
\begin{figure}[!t] 

	\includegraphics[width=0.47\textwidth]{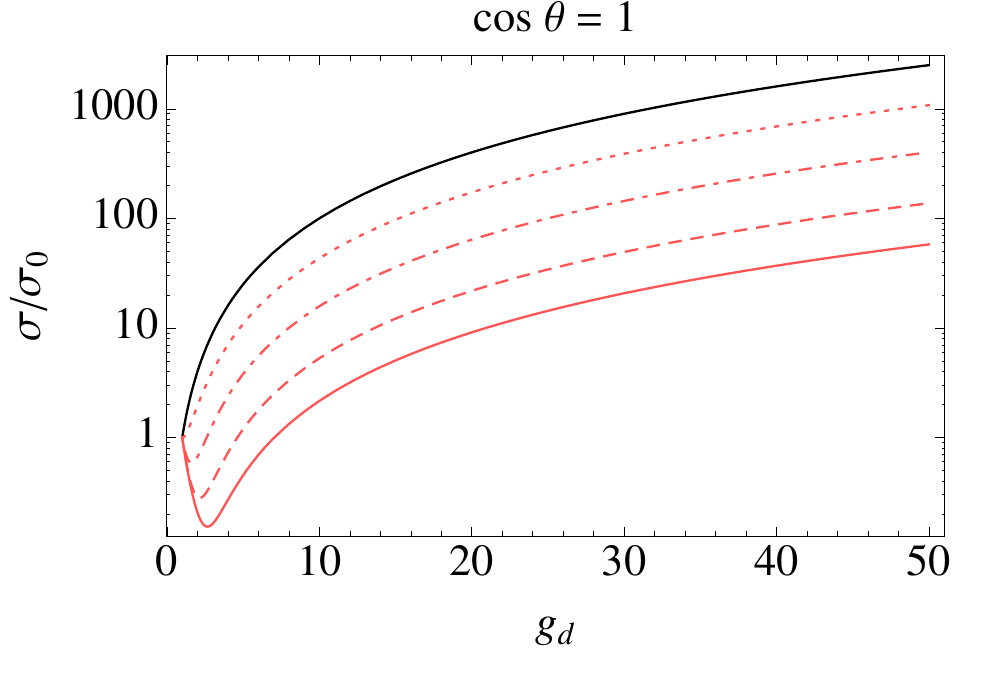} \includegraphics[width=0.47\textwidth]{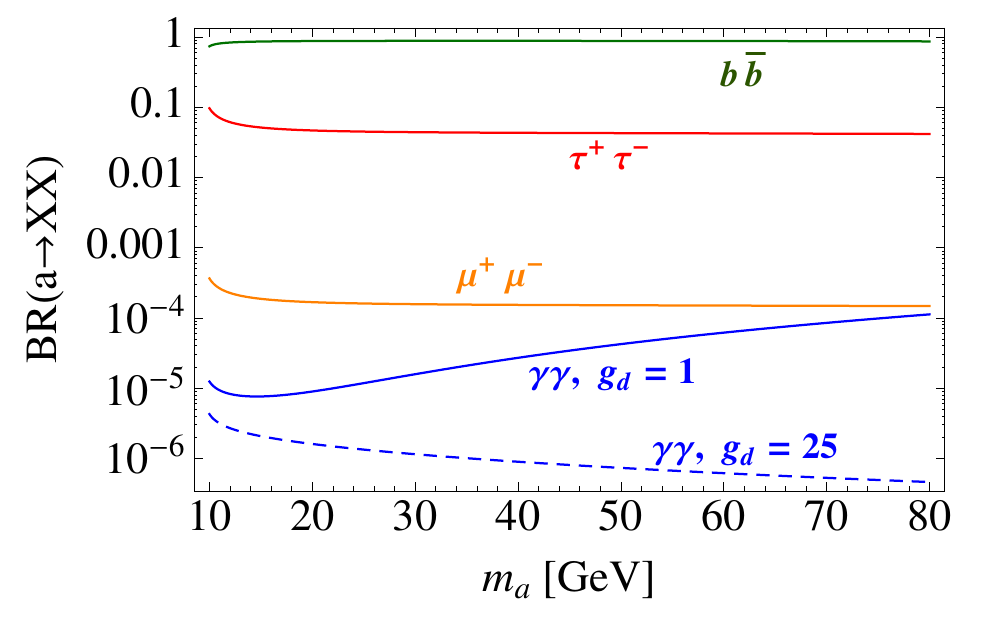}

\caption{Left: The enhancement of the inclusive $b\bar{b}a$ and gluon fusion production cross-section relative to those with $g_d=g_u^{-1}=1$ as a function of $g_d$. The dotted, dash-dotted, dashed, and solid red lines correspond to the enhancement in ggF production for $m_a=20$, 40, 60, and 80 GeV respectively.  The corresponding enhancement for $b\bar{b}$ associated production is shown by the solid black curve (the enhancement is independent of $m_a$). Right: Branching fraction of the pseudoscalar into various final states (assuming $BR(a\rightarrow \chi \bar{\chi})$ is negligible).  Note that the branching ratios into fermions are nearly independent of $g_d$ (since the total width is set primarily by $a\rightarrow b\bar{b}$, $\tau^+\tau^-$ decays), while the $a\rightarrow \gamma\gamma$ partial width is substantially suppressed for $g_d>1$.} \label{fig:CS}
\end{figure}

Heavy neutral Higgs bosons in two Higgs doublet models are being searched for via a variety of experimental signatures, including gluon fusion (ggF) production, or production in association with top or bottom quarks~\cite{Aad:2014vgg, Khachatryan:2014wca}. These canonical Higgs-type searches become much more difficult below the $Z$ threshold, where the backgrounds increase dramatically. Fortunately, as we have shown in Sec.~\ref{sec:model} above, light pseudoscalar mediators consistent with the Galactic Center excess can have enhanced couplings to down-type Standard Model fermions relative to those expected for a Standard Model-like Higgs boson of the same mass. This results in an enhanced production cross-section in modes involving $b$ quarks, and (potentially) in the gluon fusion channel relative to the Standard Model-like case. This situation is depicted on the left hand side of Fig.~\ref{fig:CS}, which shows as an example the enhancement of both the inclusive $b\bar{b}a$ (black) and gluon fusion (red) production cross-sections with $g_d=g_u^{-1}$ (i.e. $\cos\theta=1$), relative to those with $g_d=g_u=1$ ($\sigma_0$), as a function of $g_d$\footnote{This relationship between $g_u$ and $g_d$ holds in the MSSM, but not the NMSSM. In the latter case, $g_u = g_d^{-1} \cos^2\theta$, which results in values of $g_u$ that suppress the top quark contribution to the ggF loop process even further than we have considered here.}. The enhancement of the $b\bar{b}a$ cross-section is independent of $m_a$, as it only depends on $g_d$ for a given $m_a$, while the differently styled red curves correspond to $\sigma_{\rm ggF}/\sigma_{{\rm ggF},0}$ for different values of the pseudoscalar mass.  The enhancement is substantially larger in the $b\bar{b}a$ mode across the parameter space, which suggests focusing on production processes involving $b$ quarks rather than the gluon fusion process. 

We consider the branching ratio of the pseudoscalar into various final states, assuming $BR(a\rightarrow \chi\bar{\chi})$ is negligible, on the right hand side of Fig.~\ref{fig:CS}. The pseudoscalar's branching fraction into photons is small and is further suppressed for $g_d>1$ which, when combined with the increased backgrounds for $m_a<m_Z$, suggests that diphoton searches will likely be unable to probe the low-mass pseudoscalar mediators we are interested in. On the other hand, while the favored decay is into a $b\bar{b}$ pair, searches for such resonances would contend with large, pure QCD backgrounds to exploit this mode. Thus, to avoid large backgrounds while maintaining a reasonable signal, and to maximize the enhancement of the production cross-section, we propose a search for the pseudoscalar in second and third generation dilepton ($\tau^+\tau^-$ and $\mu^+\mu^-$) pair production in association with one or two $b$-jets.  Of course this strategy requires that the pseudoscalar couples to leptons, which is typical in extended two Higgs doublet models, but need not be the case \cite{Dolan:2014ska}.

Similar searches have been considered by both ATLAS \cite{Aad:2014vgg} and CMS \cite{Khachatryan:2014wca}, but are focused on higher mass resonances motivated by two Higgs doublet models and the MSSM, where the mass region of interest is greater than about 90~GeV~\cite{Schael:2006cr} due to LEP searches and precision constraints on heavy Higgs bosons. Also, previous theoretical studies in the context of the NMSSM have investigated the potential for the LHC to probe light pseudoscalars with somewhat similar searches \cite{Almarashi:2012ri, Almarashi:2010jm, Almarashi:2011hj, Almarashi:2011bf, Bomark:2014gya, King:2014xwa}. However, these investigations did not incorporate trigger and detector effects, and did not analyze the effects of cuts on the signal and backgrounds in detail, which is a major component of this work and crucial for obtaining an observable signal. While  Ref.~\cite{Bomark:2014gya}  arrives at largely negative conclusions regarding $b\bar{b}a$ production (at least in the NMSSM with partial universality), our analysis suggests a much more positive picture once appropriate cuts are implemented. 

It is worthwhile to point out that the CMS search in Ref. \cite{Chatrchyan:2012am} finds sensitivity down to $g_d \sim 3$ for masses up to $m_a\sim 14$~GeV in the gluon fusion mode with $a\rightarrow \mu^+\mu^-$. One might be inclined to conclude that this search channel could simply be extrapolated to larger masses in the scenarios of interest. However, this is unlikely to be the case. Fig. \ref{fig:CS} shows that the gluon fusion production cross-section is actually suppressed for $1<g_d\lesssim10$ as compared to its value with $g_d=1$, given our assumptions about the couplings.  The suppression increases with $m_a$ and is due to the decreased top quark loop contribution that is otherwise dominant for heavier masses.  In addition, due to the kinematic beta factor $\sqrt{1-(\frac{2m_i}{m_a})^2}$, the $b\bar{b}$ branching ratio is suppressed for smaller values of $m_a$, resulting in an increase in the $\mu^+\mu^-$ branching fraction.  For example, $BR(a\rightarrow \mu^+\mu^-)$ is enhanced by almost a factor of 2 at $m_a=10$~GeV versus $m_a>20$~GeV. Thus, for the scenarios we consider, production modes involving down-type fermions at tree-level would appear more promising than those relying on gluon fusion production and decays to muons, although different assumptions about the coupling structure could alter this conclusion. For a related analysis of the potential LHC reach in the 0$b$ mode in $Z'$ models, see e.g. Ref.~\cite{Hoenig:2014dsa}.

In the remainder of this section, we discuss the challenges and strategies for examining low mass pseudoscalars with enhanced couplings to down-type fermions, $g_d>1$.

We implemented our simplified model in \texttt{FeynRules 2.0}~\cite{Alloul:2013bka}, and generated both our signals and backgrounds at leading order (LO) using \texttt{MadGraph5+aMC@NLO}~\cite{Alwall:2014hca}. We then used \texttt{Pythia 6.4}~\cite{Sjostrand:2006za} to decay the $\tau$ leptons and hadronize the $b$-jets, and incorporated initial and final state radiation, with an appropriate scale used for the MLM matching of hard element and radiated jets. Detector simulation for trigger and tagging was performed using \texttt{Delphes 3.0}~\cite{deFavereau:2013fsa}. Trigger effects were implemented as step-function cuts at the analysis stage, though some minimum kinematic requirements were enforced at the generation phase.

\begin{figure}[!t]
   \begin{center}
	\includegraphics[width=0.7\textwidth]{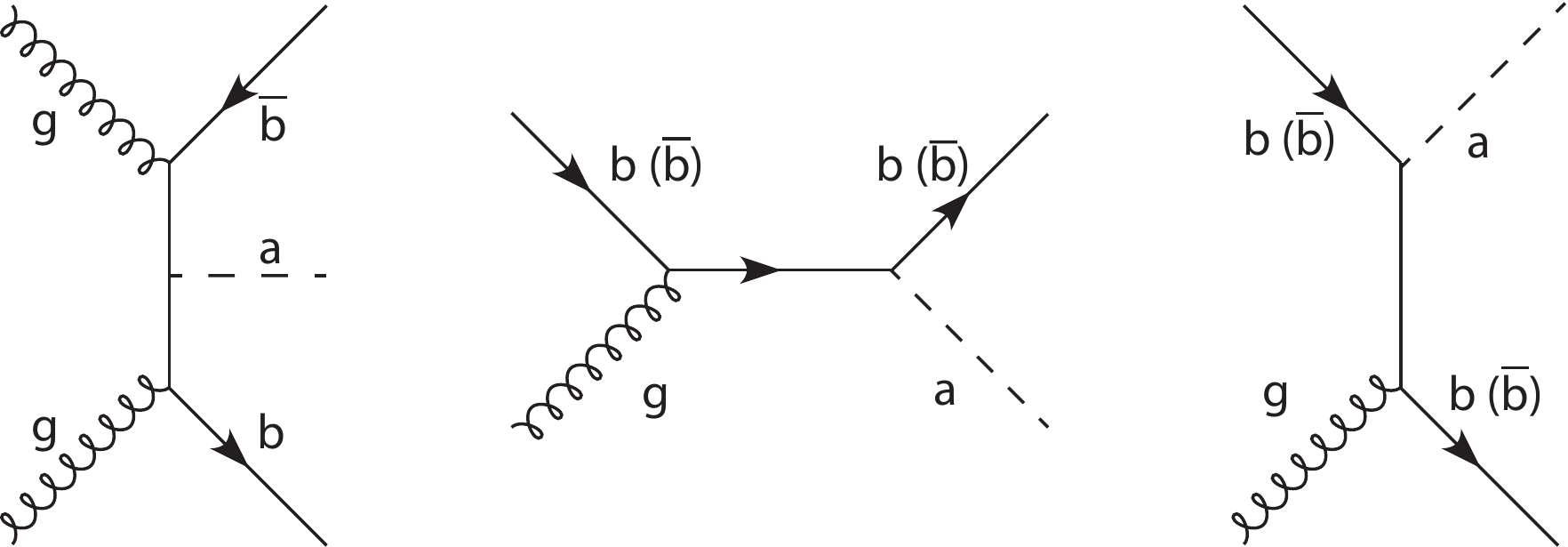}
   \end{center}
\caption{Some of the diagrams contributing to the production of the pseudoscalar, $a$, at the LHC. The two rightmost diagrams arise in the 5FS.} \label{fig:production}
\end{figure}

Diagrams for some of the primary production modes for the signal are shown in Figure \ref{fig:production}. To avoid the appearance of potentially large logarithms arising from the phase space integration over collinear final state quarks, the semi-inclusive $b(\bar{b})a$ events were generated with $b$ quarks included in the parton distribution functions (pdfs) of the proton. This is known as the ``five flavor scheme" which effectively re-sums the large logarithms \cite{Barnett:1987jw, Olness:1987ep, Dicus:1988cx}. Exclusive $b\bar{b}a$ events were generated without the inclusion of the $b$ pdfs since the resulting contributions are doubly pdf-suppressed and subleading when compared to the gluon induced processes. To avoid double counting between the two-body, $b(\bar{b})a$, production and the three-body, $b\bar{b}a$, production mode where one of the $b$-jets is collinear with the proton beam, the three-body production mode was generated with a minimum $p_T^b>5$~GeV.

There are several technical difficulties associated with accurately calculating the two-body $b(\bar{b})a$ production cross-section at hadron colliders, which have received much attention in the literature \cite{Dicus:1998hs, Campbell:2002zm,  scale_dep, Dawson:2003kb,  Campbell:2004pu, Dawson:2004sh, Dawson:2005vi, Wiesemann:2014ioa}. In particular, the leading order production cross-sections are known to exhibit a substantial dependence on the renormalization and factorization scales, $\mu_r$ and $\mu_f$, respectively \cite{scale_dep, Dawson:2005vi}. For our signal generation, we consider dynamic scales defined by
\begin{equation} \label{eq:scale}
\mu_r=\mu_f= \frac{f}{4}\sum_i\sqrt{m_i^2+p_{T_i}^2}
\end{equation}
where $f$ is an overall scaling factor, and $i$ refers to the produced $b$'s and $a$. This is in keeping with previous analyses in the context of Standard Model-like Higgs production \cite{ scale_dep, Maltoni:2003pn, Harlander:2003ai,  Maltoni:2012pa, Wiesemann:2014ioa}. We considered the impact of the scale dependence by varying the overall scaling factor in the range $[1/2,2]$, which resulted in a 2-20\% change in the production cross section, with larger effects occurring for smaller values of $m_a$. This is consistent with the range typically found in the literature \cite{Campbell:2002zm, scale_dep, Campbell:2004pu}.

To further validate the results of our leading order calculation, we have compared our LO result for the dominant ($g b(\bar{b})\rightarrow b(\bar{b}) a$) production mode to the next-to-leading order (NLO) result calculated in the five flavor scheme implemented in the program \texttt{MCFM} \cite{MCFM} for several choices for $\mu_{f,r}$ (we neglect the difference between scalar and pseudoscalar production which are small \cite{Wiesemann:2014ioa}). We find that our LO results exhibit reasonable agreement with the NLO result, falling within a factor of 1--2 across the parameter space we consider. Additionally, there are theoretical uncertainties related to the specific choice of parton distribution functions, which have been shown to be of order $\sim 5\%$ for low masses \cite{Dawson:2005vi}, as well as some residual renormalization scheme dependence (\texttt{MadGraph} uses an on-shell scheme, while e.g. \texttt{MCFM} uses $\overline{\rm MS}$). To account for these effects, Appendix~\ref{sec:appc} takes a conservative approach and explores the effect of a factor of 2 over-estimation in our signal and, separately, a factor of 2 under-estimation in the backgrounds. Our overall conclusions are not significantly affected by this re-scaling, and so we believe them to be quite robust.

For an experimental search, we consider three possible leptonic tagging channels: SR1 requires one electron and one muon; SR2 requires one lepton ($e$ or $\mu$) and one hadronic $\tau$; SR3 requires two muons. SR1 is motivated by excellent trigger response, while SR2 is motivated by the larger branching ratios and SR3 is motivated by a resonance search methodology in the di-muon invariant mass spectrum that allows for the use of data-driven backgrounds. In all three signal regions, we also require 1-2 $b$-jet tags, and no light jets, where light jets are defined as $p_T > 40$~GeV. The signals are therefore inclusive for light jets with $p_T < 40$~GeV, such as those that are commonly generated from ISR effects. These tagging requirements significantly suppress fake backgrounds arising from vector boson production in association with light jets.

We assume the default CMS tagging efficiencies that are implemented in Delphes 3.0, which are as follows. For tagging, electrons are required to have $p_T>10$~GeV and $|\eta|<2.5$. Within the inner region of the detector, $|\eta|<1.5$, we assume a tagging efficiency of $\epsilon_e=0.95$, while for the outer region but with $|\eta|<2.5$ we assume $\epsilon_e=0.85$. The rate at which jets fake electrons is taken to be $\not{\!\epsilon}_e^j = 0.0001$ and uniform over the whole detector. For muons, we require that candidates have $p_T>10$~GeV and $|\eta|<2.4$. Since our analysis involves only low $p_T$ muons, we take a fixed tagging efficiency of $\epsilon_\mu=0.95$, which is appropriate for $p_T^\mu<1000$~GeV. For the tagging of hadronic taus, we require $|\eta|<2.5$ and take a fixed tagging efficiency of $\epsilon_\tau=0.4$ with a fake rate for mistagging a light jet as a hadronic tau of $\not{\!\epsilon}_\tau^j = 0.001$. The tagging of $b$-jets occurs only where $p_T^b>15$~GeV and $|\eta|<2.5$, with an efficiency of $\epsilon_b=0.5\tanh(0.03p_T - 0.4)$ within the inner detector, $|\eta|<1.2$, and $\epsilon_b=0.4\tanh(0.03p_T - 0.4)$ up to the boundary of $|\eta|<2.5$. Light jets are taken to fake $b$-jets at a rate of $\not{\!\epsilon}_b^j = 0.001$, while $c$-jets faking $b$-jets follow a formula similar to the $b$-tagging efficiency but with coefficients of 0.2 and 0.1 for the two regions, respectively.

\subsection{Trigger Effects}
Since the signal typically produces very soft jets and leptons, trigger effects are very important to consider. To account for the effect that trigger has on our results, we have implemented a variety of triggers as a step-function cut based on what we believe are reasonable off-line triggers for CMS\footnote{Dilepton triggers are motivated from discussions with James Hirschauer of the CMS collaboration. Final trigger details for LHC13/14 are not currently available.}. The following primary triggers are potentially relevant to our study:
\begin{itemize}
\item $1e$: single electron with $p_T > 35$~GeV;
\item $1\mu$: single muon with $p_T > 25$~GeV;
\item $2\mu$: di-muon leading with $p_T > 17$~GeV, subleading $p_T > 10$~GeV;
\item $e\tau_h$: electron $+$ hadronic tau with $p_T^\tau > 45$~GeV, $p_T^e > 19$~GeV;
\item $\mu\tau_h$: muon $+$ hadronic tau with $p_T^\tau > 40$~GeV, $p_T^e > 15$~GeV;
\item $e\mu$: leading electron $+$ muon with $p_T^e > 23$~GeV, $p_T^\mu > 10$~GeV;
\item $\mu e$: electron $+$ leading muon with $p_T^e > 12$~GeV, $p_T^\mu > 23$~GeV;
\end{itemize}
We also include other triggers, such as those involving photons, jets, $\tau_h$ plus MET, and $b$-jets, but these provide a negligible effect on the signal events (i.e. $<0.3\%$ of signal events pass all the non-primary triggers combined)  and so are not included in the above list. The non-primary triggers pass a significant portion of the backgrounds, however, which necessitates their inclusion, but this indicates that these events have distinctive signatures that can be eliminated from the analysis by kinematic cuts.

Due to the low mass of the pseudoscalar in our search, a significant number of the production events will not pass the trigger. Since we are not privy to the details of the final triggers, we consider the effect of varying the muon $p_T$ thresholds for the triggers that include a primary muon. These triggers have the greatest likelihood for discretionary variation in a dedicated experimental search, and are the most important due to having the lowest inclusive cross sections and thus $p_T$ thresholds. We analyzed the cross section of signal events that pass each of the primary trigger cuts ($\sigma_{SRx}^{T_y}$) as a fraction of the cross section of generated events ($\sigma_{SRx}^{gen} = \sigma_{gen}\times BR(\tau^+\tau^- \rightarrow SRx)\times\epsilon_{SRx}$) with the same tagging signature in each signal region, independently:
\begin{equation}
\label{eq:trig}
R_{SRx}^{\rm T_y} = \frac{\sigma_{SRx}^{T_y}}{\sigma_{SRx}^{gen}}
\end{equation}
where SRx refers to the signal region and $T_y$ refers to the specific trigger. This ratio can be considered as a sort of trigger efficiency. Of note, we found that the $e+\tau_h$ and $\mu+\tau_h$ triggers did not pass any of a preliminary 200k generated events, likely due to the hard cut on the $p_T$ of the $\tau_h$ and the low mass of the pseudoscalar. Since the hadronic tau has a large fake background from mistagged light jets, we do not anticipate that the trigger threshold for hadronic tau $p_T$ will be improved enough to make these triggers worthwhile to consider. While the fake rate of jets for electrons is smaller than for hadronic taus, we believe it is unlikely that any significant improvement in the electron trigger thresholds will be implemented as there would still be a larger increase in the inclusive cross section than for similar changes in the muon trigger thresholds.

\begin{figure}[!h]
   	\begin{center}
	\hspace{-.5cm}
	\includegraphics[width=0.5\textwidth]{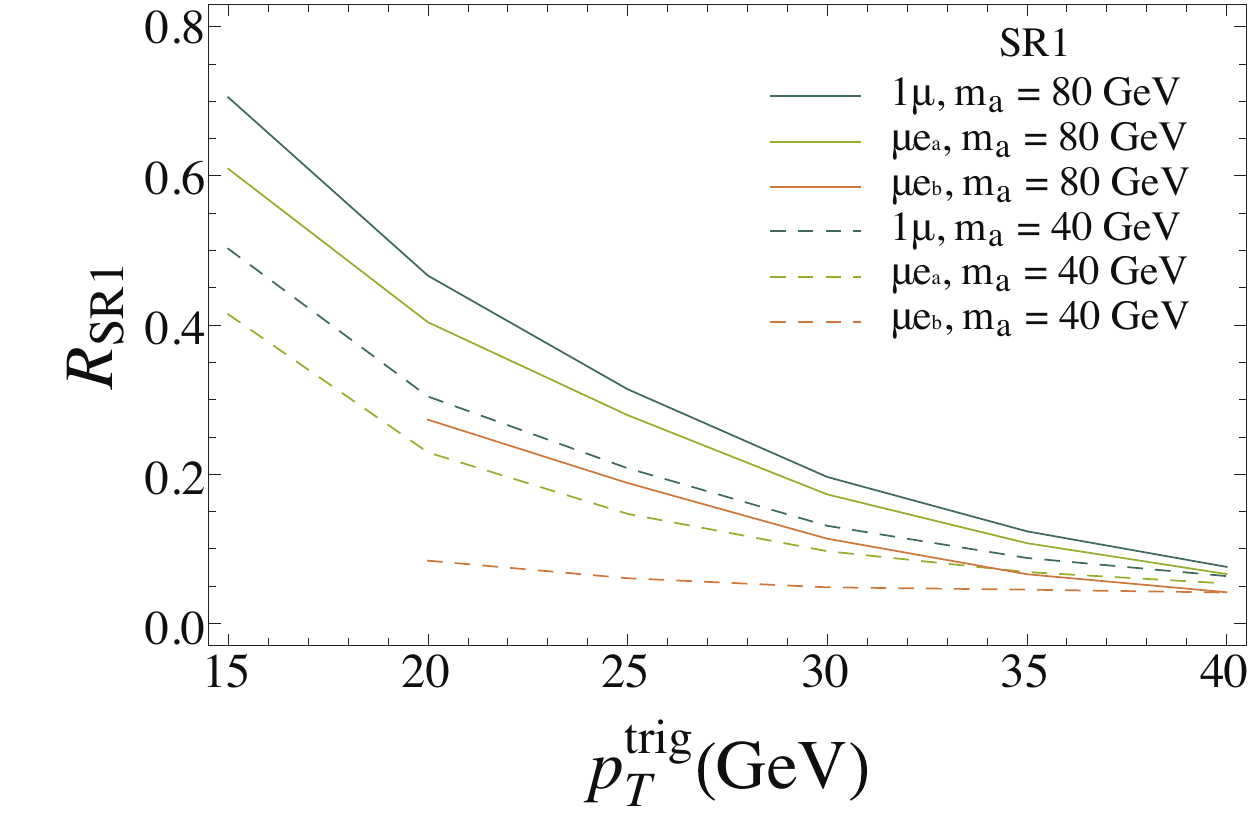}
	\end{center}
	\hspace{-0.5cm}
	\includegraphics[width=0.5\textwidth]{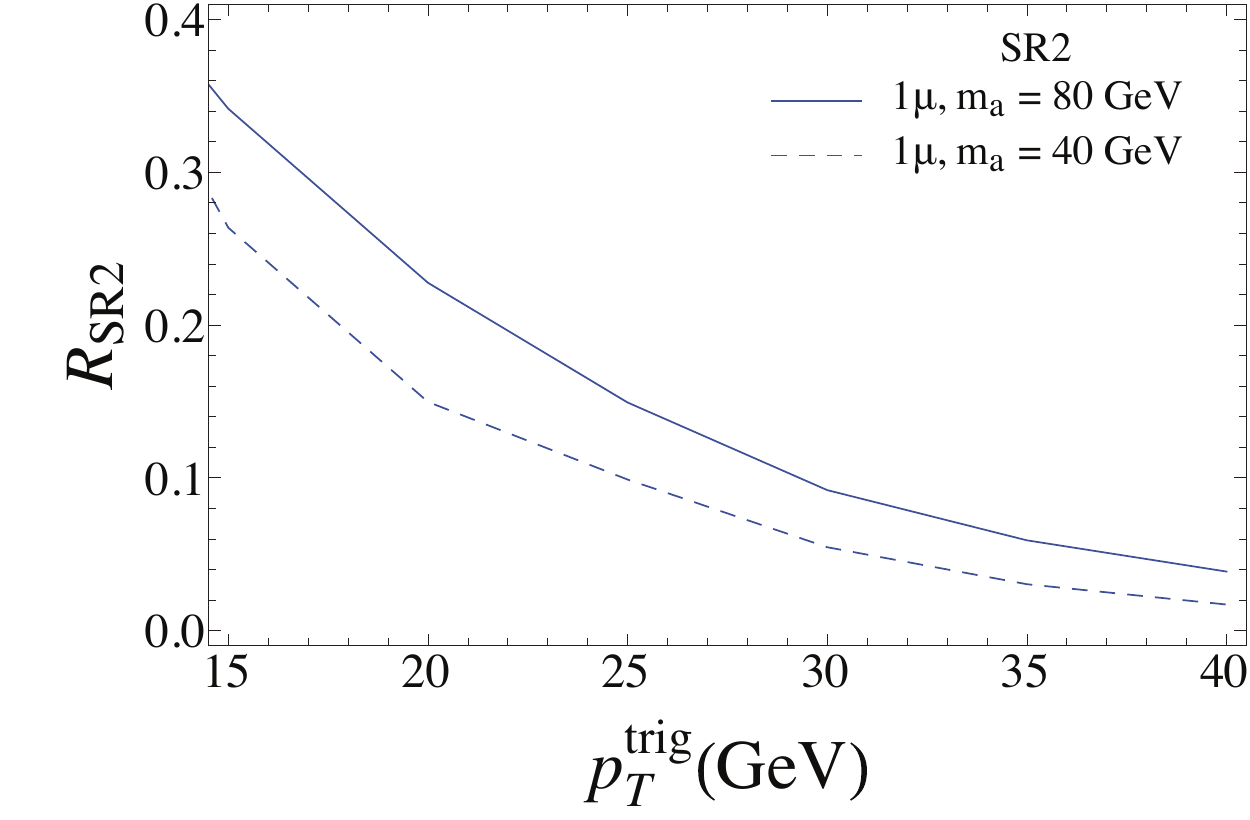}
	\includegraphics[width=0.5\textwidth]{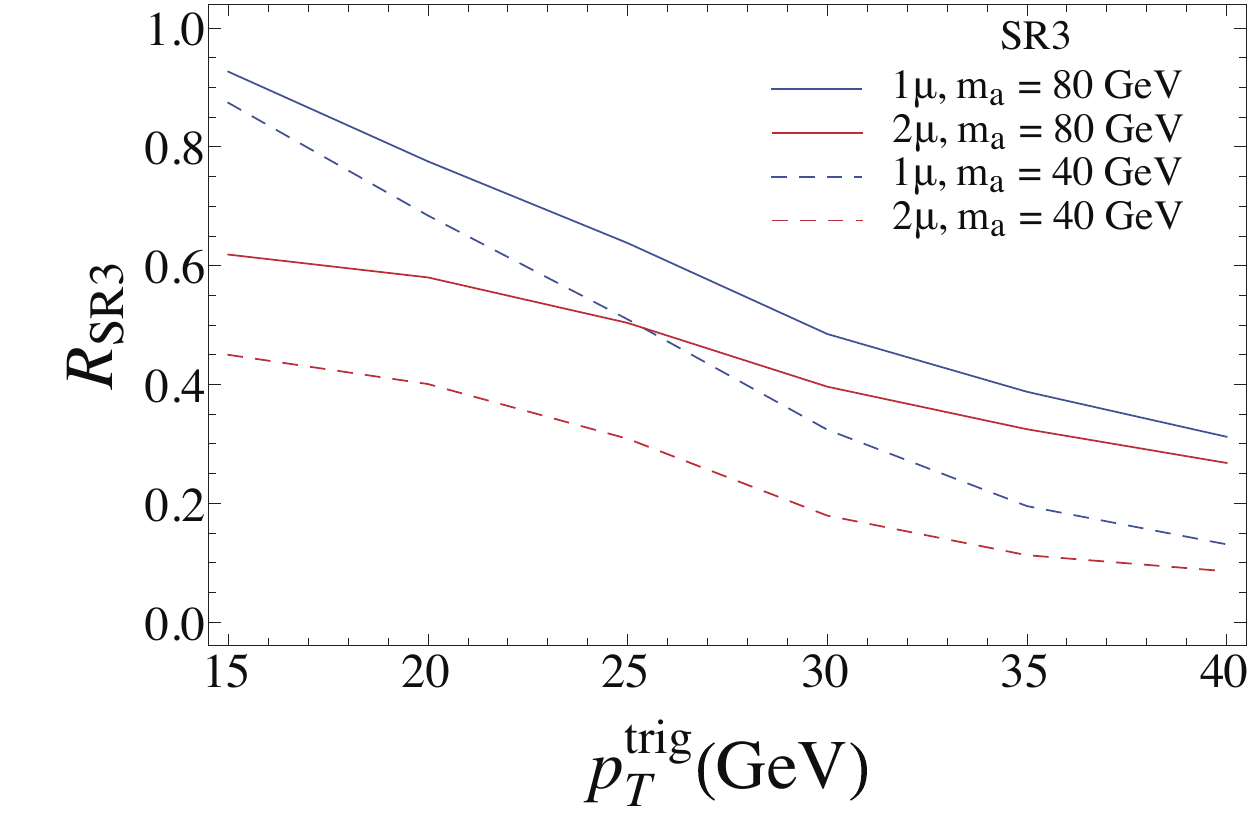}
\caption{Trigger ratios for each signal region, normalized to the produced and tagged cross section, based on varying the leading muon $p_T$. The $\mu e_a$ trigger is based on a subleading electron $p_T=12$ GeV, while $\mu e_b$ is based on a subleading electron $p_T=17$ GeV. The $2\mu$ trigger for SR3 is based on a subleading muon $p_T=15$ GeV rather than the $p_T=10$~GeV discussed in the text, as the trigger response for the lower subleading $p_T$ is very similar to the single muon rate due to the minimum $p_T$ settings in the event generation stage and tagging thresholds.} \label{fig:trigger}
\end{figure}

The summary of the trigger efficiency ratios in Eq. \ref{eq:trig} for the default implemented triggers is shown in Table \ref{tab:trigger}, while an analysis of the effect of varying the threshold for the muon $p_T$ in the $1\mu$, $2\mu$ and $\mu e$ triggers for each of the three signal regions is shown in Figure \ref{fig:trigger}. A na{\" i}ve interpretation of this figure suggests that the single muon trigger includes a larger fraction of the signal than $\mu e$ or $2\mu$ triggers, but it is important to note that the single muon inclusive cross section at the LHC is significantly larger than the muon+electron or dimuon inclusive cross sections, and thus will typically have a higher $p_T$ threshold than the other triggers and a lower trigger efficiency, as shown in Table \ref{tab:trigger}.

\begin{table}
\begin{center}

\begin{tabular}{ | c | c | c | c | c | c | c | c |}
\hline
SRx	& 	$m_a$ (GeV) & $1e$ & $1\mu$ & $2\mu$ & $e+\mu$ & $\mu+e$ & all \\ \hline
\multirow{4}{*}{SR1}	&	20	&	0.03	&	0.12	&	0.00	&	0.16	&	0.13	&	0.30\\\cline{2-8} 
				&	40	&	0.05	&	0.21	&	0.00	&	0.22	&	0.15	&	0.39\\\cline{2-8} 
				&	60	&	0.06	&	0.21	&	0.00	&	0.25	&	0.23	&	0.44\\\cline{2-8} 
				&	80	&	0.14	&	0.31	&	0.00	&	0.36	&	0.33	&	0.59\\\hline\hline
\multirow{4}{*}{SR2}	&	20	&	0.01	&	0.06	&	0.00	&	0.00	&	0.00	&	0.07\\\cline{2-8} 
				&	40	&	0.03	&	0.10	&	0.00	&	0.00	&	0.00	&	0.13\\\cline{2-8} 
				&	60	&	0.05	&	0.12	&	0.00	&	0.00	&	0.00	&	0.17\\\cline{2-8} 
				&	80	&	0.06	&	0.15	&	0.00	&	0.00	&	0.00	&	0.21\\\hline\hline 
\multirow{4}{*}{SR3}	&	20	&	0.00	&	0.41	&	0.75	&	0.00	&	0.00	&	0.75\\\cline{2-8} 
				&	40	&	0.00	&	0.51	&	0.78	&	0.00	&	0.00	&	0.78\\\cline{2-8} 
				&	60	&	0.00	&	0.55	&	0.86	&	0.00	&	0.00	&	0.86\\\cline{2-8} 
				&	80	&	0.00	&	0.64	&	0.86	&	0.00	&	0.00	&	0.86\\\hline 

\end{tabular}
\caption{The ratio of cross section that passes the trigger cut to the generated cross section for 200k generated events. Kinematic dependent tagging efficiencies are already incorporated into the cross sections. All leptons ($e$, $\mu$, $\tau$) are generated with a minimum $p_T>10$~GeV, but $\tau$ decays to leptons can result in a $p_T^{e,\mu} < 10$~GeV. The columns in this table are not necessarily independent, as it is possible for an event to simultaneously pass multiple triggers.
}
\label{tab:trigger}
\end{center}
\end{table}

\subsection{Backgrounds and Their Reduction}
Since the QCD backgrounds at the LHC are significant, the fake rate of jets as electrons, hadronic $\tau$-jets, and $b$-jets are important to take into account. Additionally, backgrounds with similar kinematics to the signal we examine produce soft leptons that may not be identified as easily or may fall outside of the central region of the detectors where tagging is possible. Thus, backgrounds producing more than two leptons, where one is not tagged, may contribute to the signal regions. To account for these effects, we include backgrounds that produce between one and three leptons ($e$, $\mu$, $\tau$), and 0-2 $b$-jets, in association with 1-3 light jets (with $n_b+n_j  \leq 3$), since our signal is inclusive to low $p_T$ light jets. The following background processes are generated:
\begin{itemize}
\item $p p \rightarrow \gamma^*/Z + b\bar{b} + (0,1)j$, $\gamma^*/Z \rightarrow \ell^+\ell^-$;
\item $p p \rightarrow \gamma^*/Z + b(\bar{b}) + (0,1,2)j$, $\gamma^*/Z \rightarrow \ell^+\ell^-$;
\item $p p \rightarrow \gamma^*/Z + (0,1,2,3)j$, $\gamma^*/Z \rightarrow \ell^+\ell^-$;
\item $p p \rightarrow W^\pm + b\bar{b} + (0,1)j$, $W^\pm \rightarrow \ell^\pm \nu_\ell (\bar{\nu}_\ell)$;
\item $p p \rightarrow W^\pm + b(\bar{b}) + (0,1,2)j$, $W^\pm \rightarrow \ell^\pm \nu_\ell (\bar{\nu}_\ell)$;
\item $p p \rightarrow W^\pm + (0,1,2,3)j$, $W^\pm \rightarrow \ell^\pm \nu_\ell (\bar{\nu}_\ell)$;
\item $p p \rightarrow W^+W^- + b\bar{b} + (0,1)j$, $W^+ \rightarrow \ell^+ \nu_\ell$, $W^- \rightarrow \ell^{\prime -} \bar{\nu}_{\ell^\prime}$;
\item $p p \rightarrow W^+W^- + b(\bar{b}) + (0,1,2)j$, $W^+ \rightarrow \ell^+ \nu_\ell$, $W^- \rightarrow \ell^{\prime -} \bar{\nu}_{\ell^\prime}$;
\item $p p \rightarrow W^+W^- + (0,1,2,3)j$, $W^+ \rightarrow \ell^+ \nu_\ell$, $W^- \rightarrow \ell^{\prime -} \bar{\nu}_{\ell^\prime}$;
\item $p p \rightarrow ZW^\pm + b\bar{b} + (0,1)j$, $W^\pm \rightarrow \ell^\pm \nu_\ell (\bar{\nu}_\ell)$, $Z \rightarrow \ell^{\prime +}\ell^{\prime -}$;
\item $p p \rightarrow ZW^\pm + b(\bar{b}) + (0,1,2)j$, $W^\pm \rightarrow \ell^\pm \nu_\ell (\bar{\nu}_\ell)$, $Z \rightarrow \ell^{\prime +}\ell^{\prime -}$;
\item $p p \rightarrow ZW^\pm + (0,1,2,3)j$, $W^\pm \rightarrow \ell^\pm \nu_\ell (\bar{\nu}_\ell)$, $Z \rightarrow \ell^{\prime +}\ell^{\prime -}$;
\end{itemize}
where $\ell = (e, \mu, \tau)$ and $j$ are light jets ($u$, $d$, $s$, $c$, $g$) that can come from associated production. Each entry in the list above is produced with the number of quoted jets, and MLM matching and merging is incorporated to avoid double counting of the light jet production with the initial state radiation (MLM matching with $XQCUT = 15$ and $QCUT = 20$). The two largest contributions to our backgrounds are the inclusive $Z$ production modes (included in the first three entries) and $t\bar{t}$ (included in the seventh entry), but these are effectively reduced by kinematic cuts. The kinematic distributions of the signal and backgrounds are included in Appendix \ref{sec:appa}.

Based on the kinematic distributions we examined, we have identified a number of possible cuts that improve the signal significance. These cuts are focused on reducing the $t\bar{t}$ and $Z+nj$ backgrounds. The $t\bar{t}$ and other backgrounds with $W^+W^-$ lepton production can be reduced with cuts that involve the $\met$ measurement, including a direct $\met$ cut, as well as the transverse mass $m_T=\sqrt{2p_T^{{\rm2nd}\hspace{.08cm}\ell}\met(1-\cos\theta)}$. Backgrounds with a $Z$ resonance can be reduced by a cut on the dilepton invariant mass, $m_{\ell\ell}$. In addition, a large fraction of the backgrounds producing both leptons and jets have a large total $p_T$. Thus, we consider cuts on the scalar sum $H_T = \displaystyle\sum_\ell p_T^\ell + \sum_b p_T^b$, and $\not{\!\!H}_T^\ell = \displaystyle\sum_\ell p_T^\ell + \met$.

In the case of SR3, dilepton invariant mass cuts are implemented in a fixed range. While the branching ratio to dimuons is small ($<0.1\%$), the $a\rightarrow \tau^+\tau^- \rightarrow \mu^+\mu^-+\met$ branching ratio is similarly small, and the invariant mass peak of the direct decay is reconstructible with low smearing. Thus, it may be possible to observe the pseudoscalar with a resonance search methodology. For SR3, we consider only events within a 2-3~GeV invariant mass bin centred at the mass of the pseudoscalar. In contrast, the analysis for SR1 and SR2 are based on a cut-and-count methodology, since the dilepton peak is significant smeared out due to the loss of information from the neutrinos originating from the $\tau$ decays. For these signal regions, we do not employ a narrow invariant mass window and instead employ $m_{\ell\ell}$ cuts to exclude backgrounds only.

The cuts for SR1 and SR2 are considered separately in each of two distinct scenarios: hard cuts are better for high luminosity searches and have a greater overall reach, while soft cuts are better for low luminosity searches. Kinematic threshold values for the considered cuts were chosen by maximizing $\sigma_{sig}*L/\sqrt{\sigma_{sig}*L+\sigma_{bkg}*L+\epsilon_{sys}^2\sigma_{bkg}^2*L^2}$, for a systematic uncertainty of $\epsilon_{sys} = 0.2$ and luminosity of $L=100$/fb, while maintaining $\sigma_{sig}^{cut}/\sigma_{sig}^{tot}\sim0.5(0.8)$ for hard (soft) cuts for $m_a>40$~GeV. The dimuon signal region, SR3, is analyzed assuming only a single cut scenario, as background events with $m_{\ell\ell} \sim m_a$ generally have similar acceptance rates to the signal.

The final cut values for each signal region are:
\begin{itemize}
\item SR1 hard: leading $p_T^\ell < 30$~GeV, $12 < m_{\ell\ell} < 35$~GeV, $H_T < 90$~GeV, $\not{\!\!H}_T^\ell <80$~GeV.
\item SR1 soft: leading $p_T^\ell < 40$~GeV, $12 < m_{\ell\ell} < 45$~GeV, $H_T < 140$~GeV, $m_T<40$~GeV, $\not{\!\!H}_T^\ell <120$~GeV.
\item SR2 hard: leading $p_T^\ell < 40$~GeV, $12 < m_{\ell\ell} < 45$~GeV, $H_T < 130$~GeV, $\not{\!\!H}_T^\ell <100$~GeV.
\item SR2 soft: $12 < m_{\ell\ell} < 60$~GeV, $H_T < 190$~GeV, $m_T < 45$~GeV, $\not{\!\!H}_T^\ell <140$~GeV.
\item SR3: leading $p_T^\ell < 50$~GeV, $H_T < 120$~GeV, $\not{\!\!H}_T^\ell <120$~GeV.
\end{itemize}
The expected search reach using these cuts is given in the next section. Further details about the acceptance rates for each cut are provided in Appendix \ref{sec:appb}. Alternative approaches for determining the cut regions, such as those incorporating repeated algorithmic refinements of the phase space, would optimize cuts for a single mass value and be unable to account for the full range of parameters we explore. Maximizing the acceptance rate for $m_a = 40$~GeV would result in a poorer reach in $g_d$ values for $m_a = 80$~GeV, for example. We feel our approach is more appropriate for a general search strategy.

\begin{figure}[!h]
   \begin{center}
\includegraphics[width=0.45\textwidth]{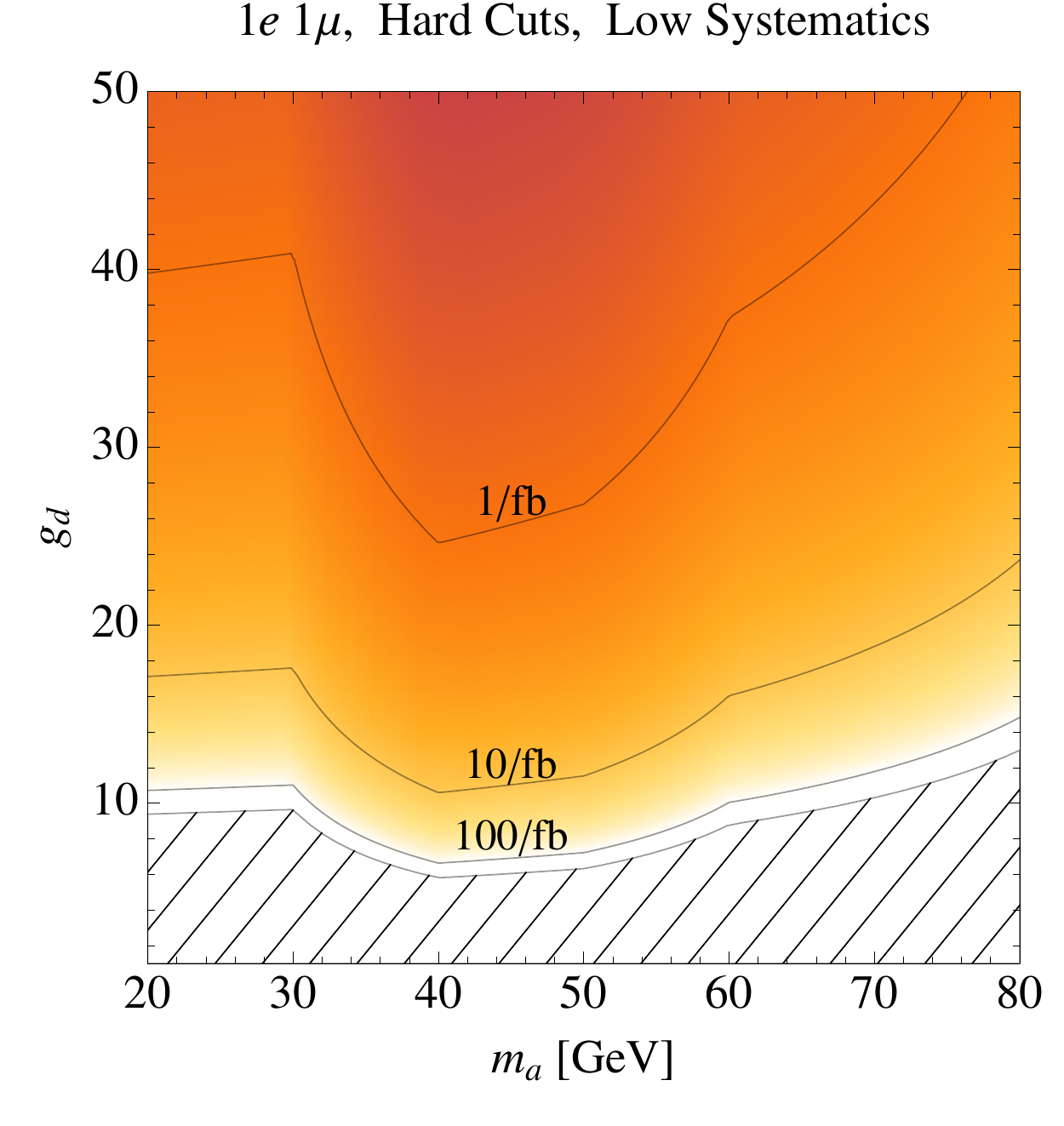}
\includegraphics[width=0.45\textwidth]{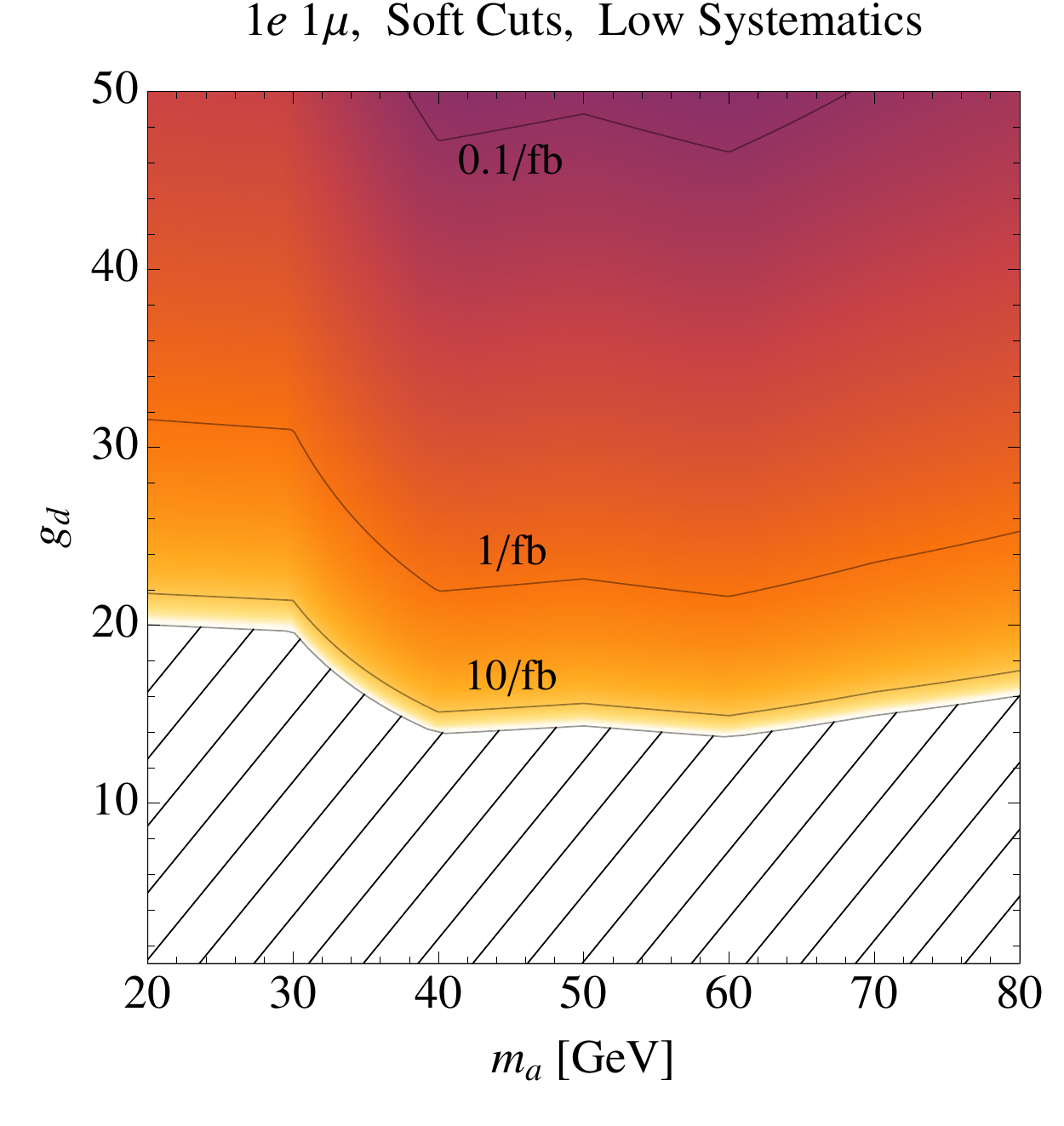}\\
\includegraphics[width=0.45\textwidth]{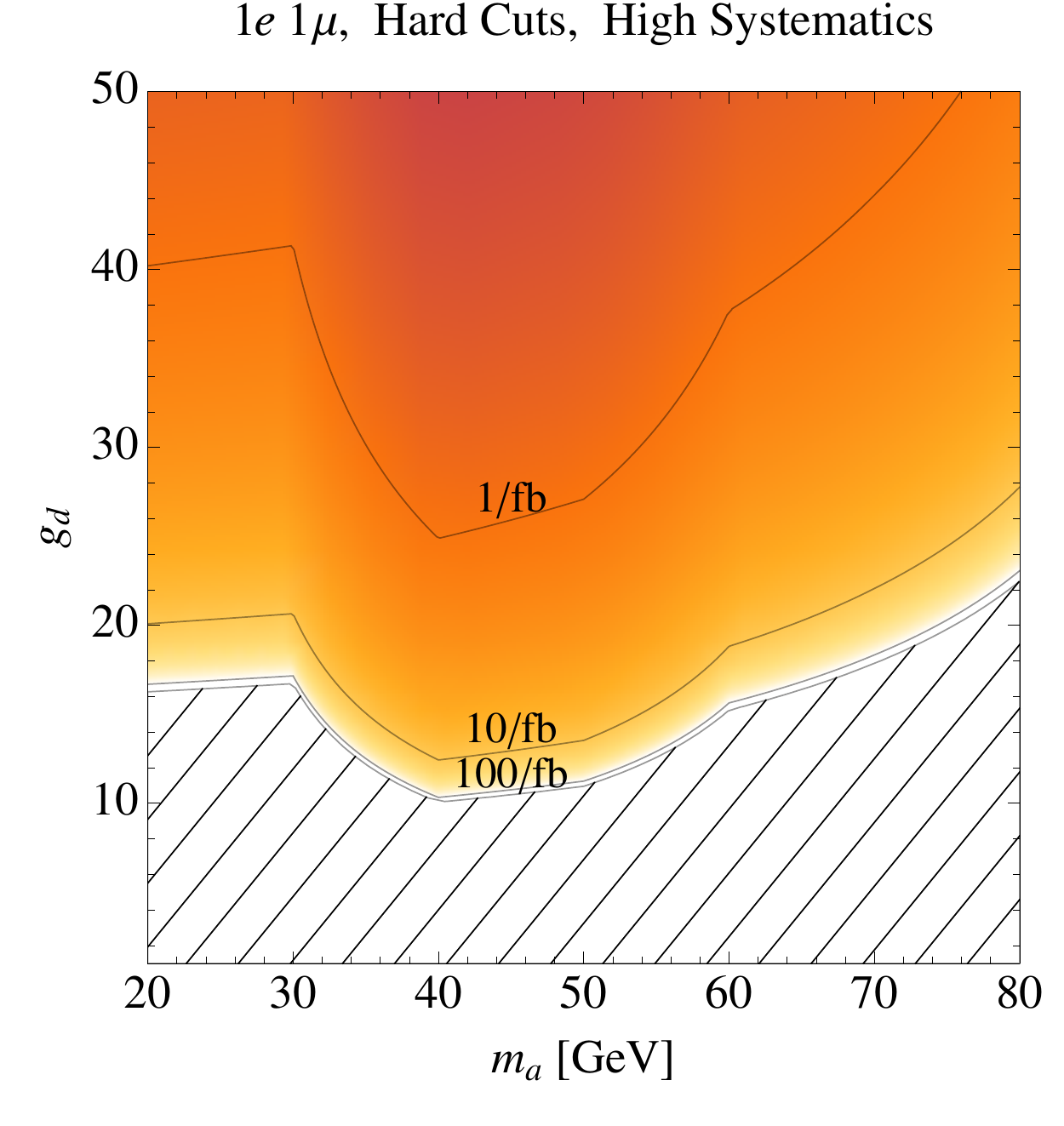}
\includegraphics[width=0.45\textwidth]{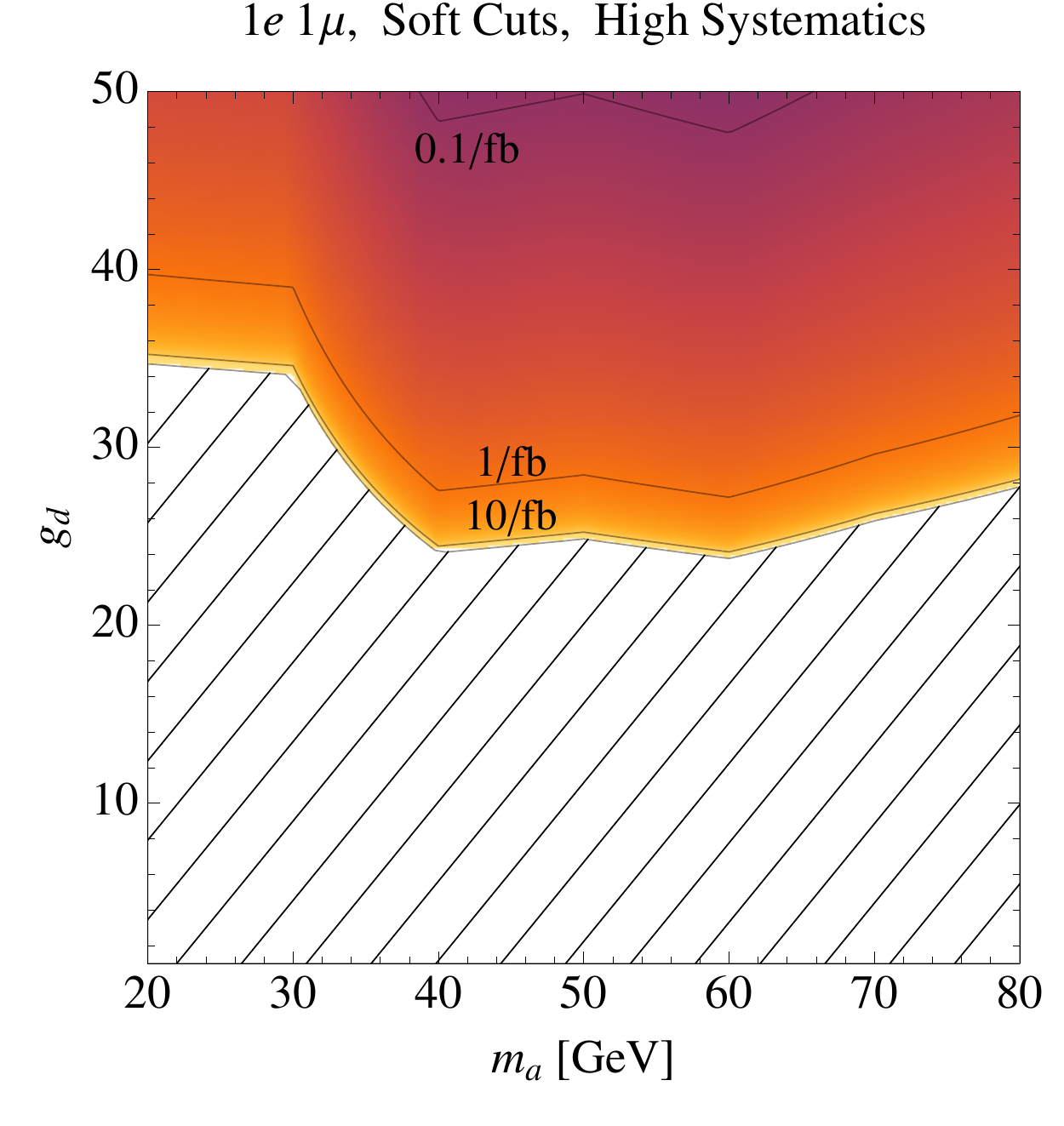}
   \end{center}
\caption{Discovery potential for the SR1 signal region with hard (left) and soft (right) cuts for $\epsilon_{sys}=0.1 (0.3)$ (top (bottom)). Hatched region is the region where no discovery is possible, regardless of luminosity, due to systematic uncertainties. The shading and labeled contours correspond to constant values of $\log(L\times \mathrm{fb})$ needed to achieve $k=3$.} \label{fig:1e1udisc}
\end{figure}

\begin{figure}[!h]
   \begin{center}
\includegraphics[width=0.45\textwidth]{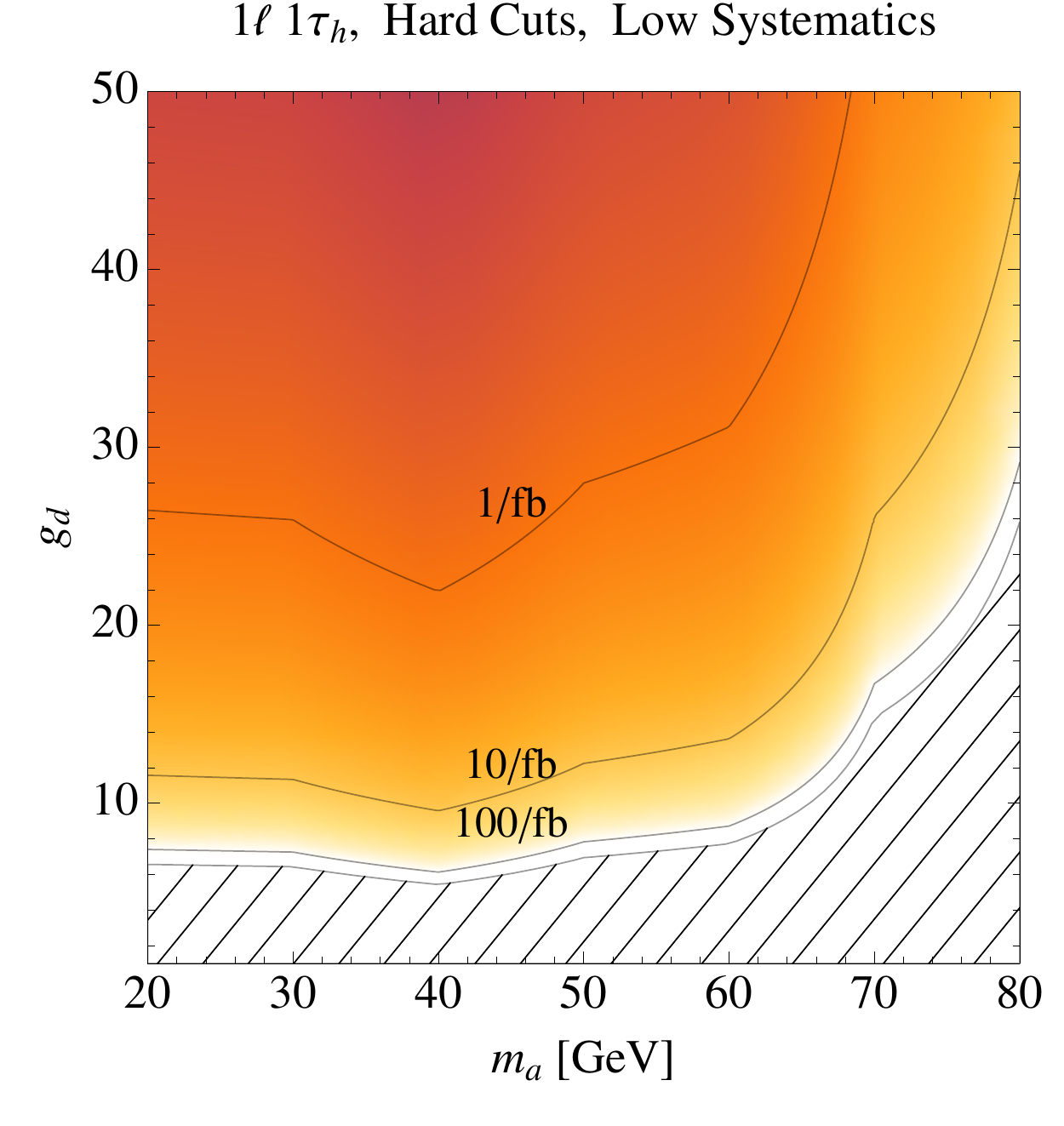}
\includegraphics[width=0.45\textwidth]{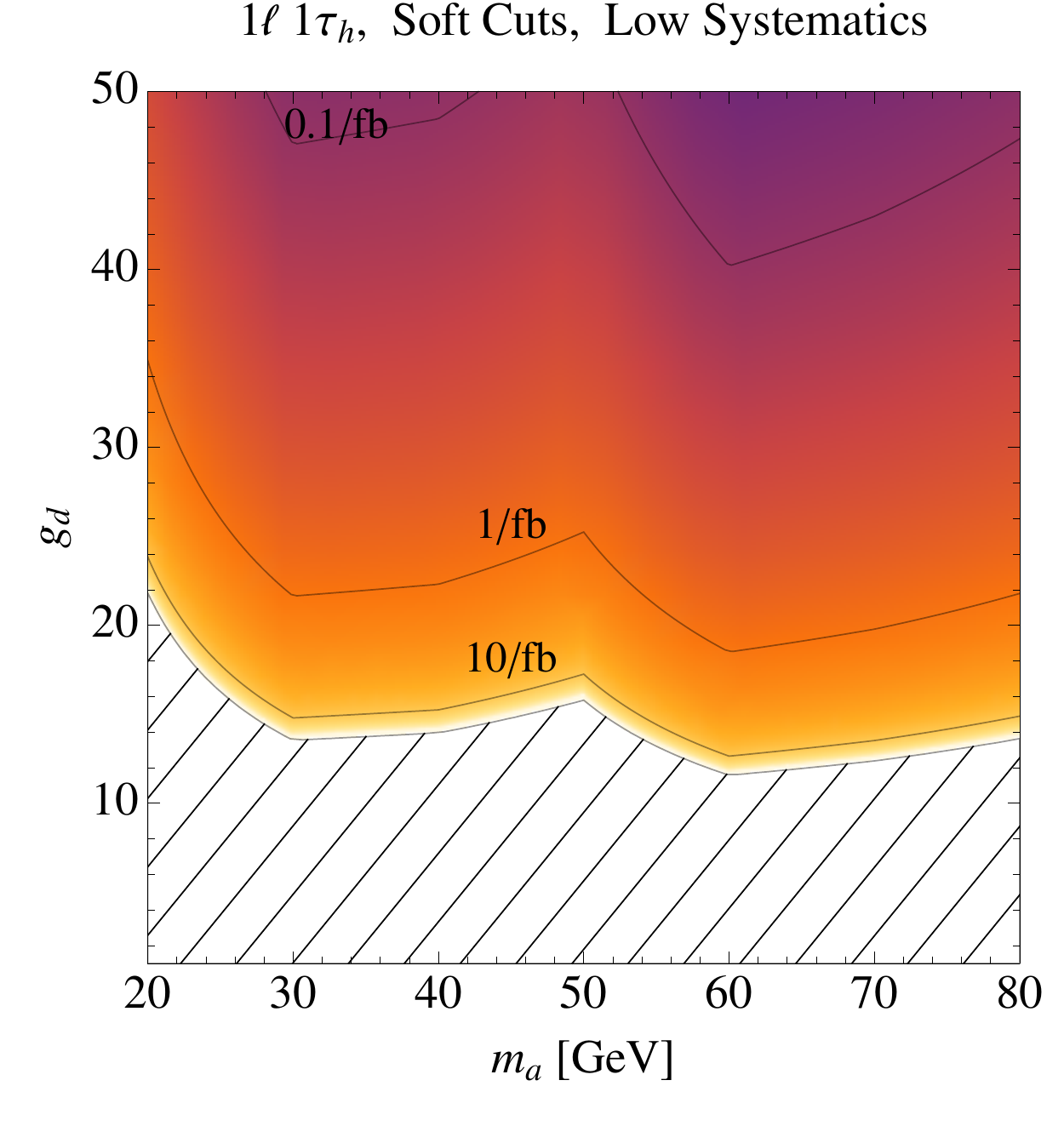}\\
\includegraphics[width=0.45\textwidth]{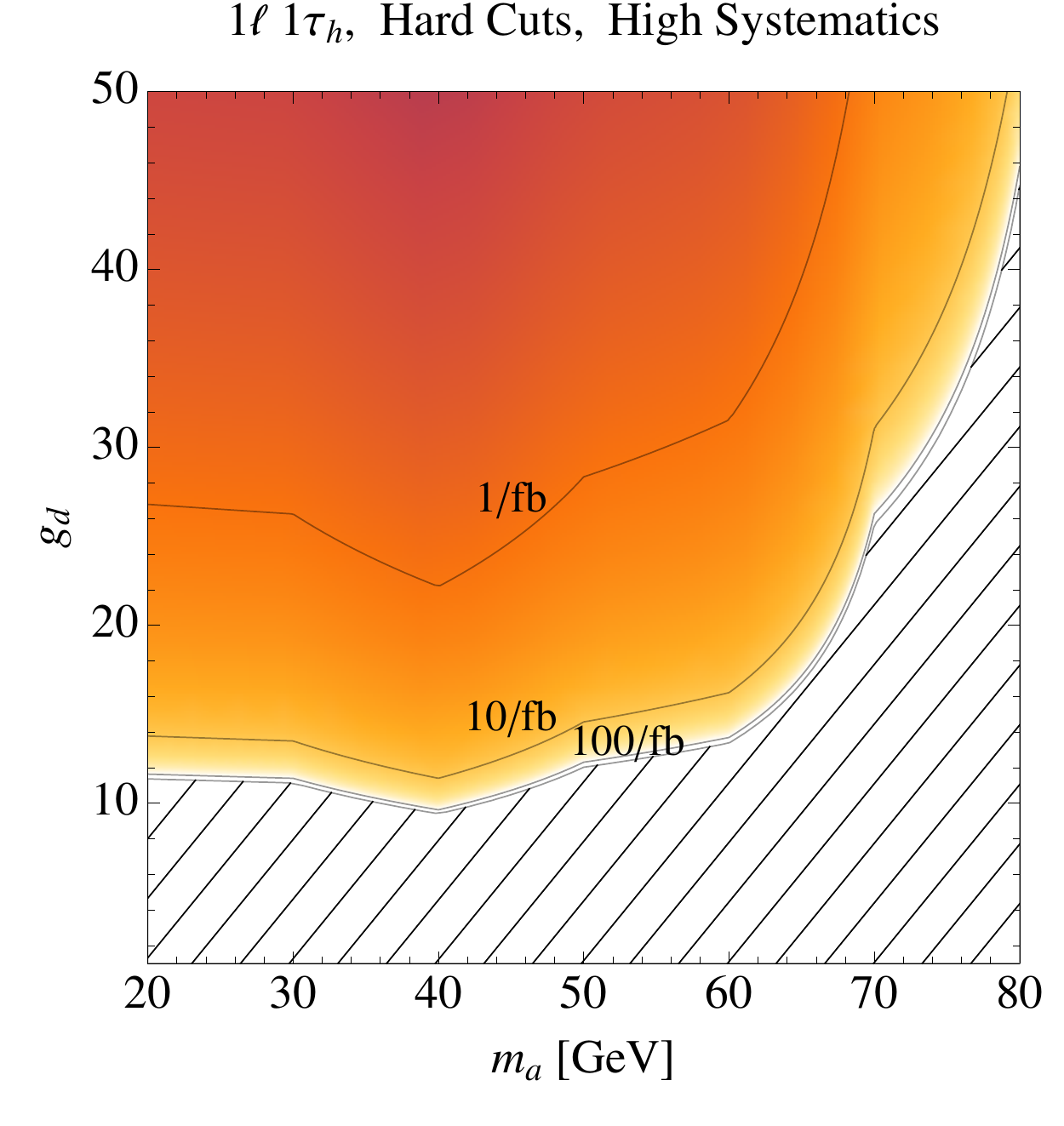}
\includegraphics[width=0.45\textwidth]{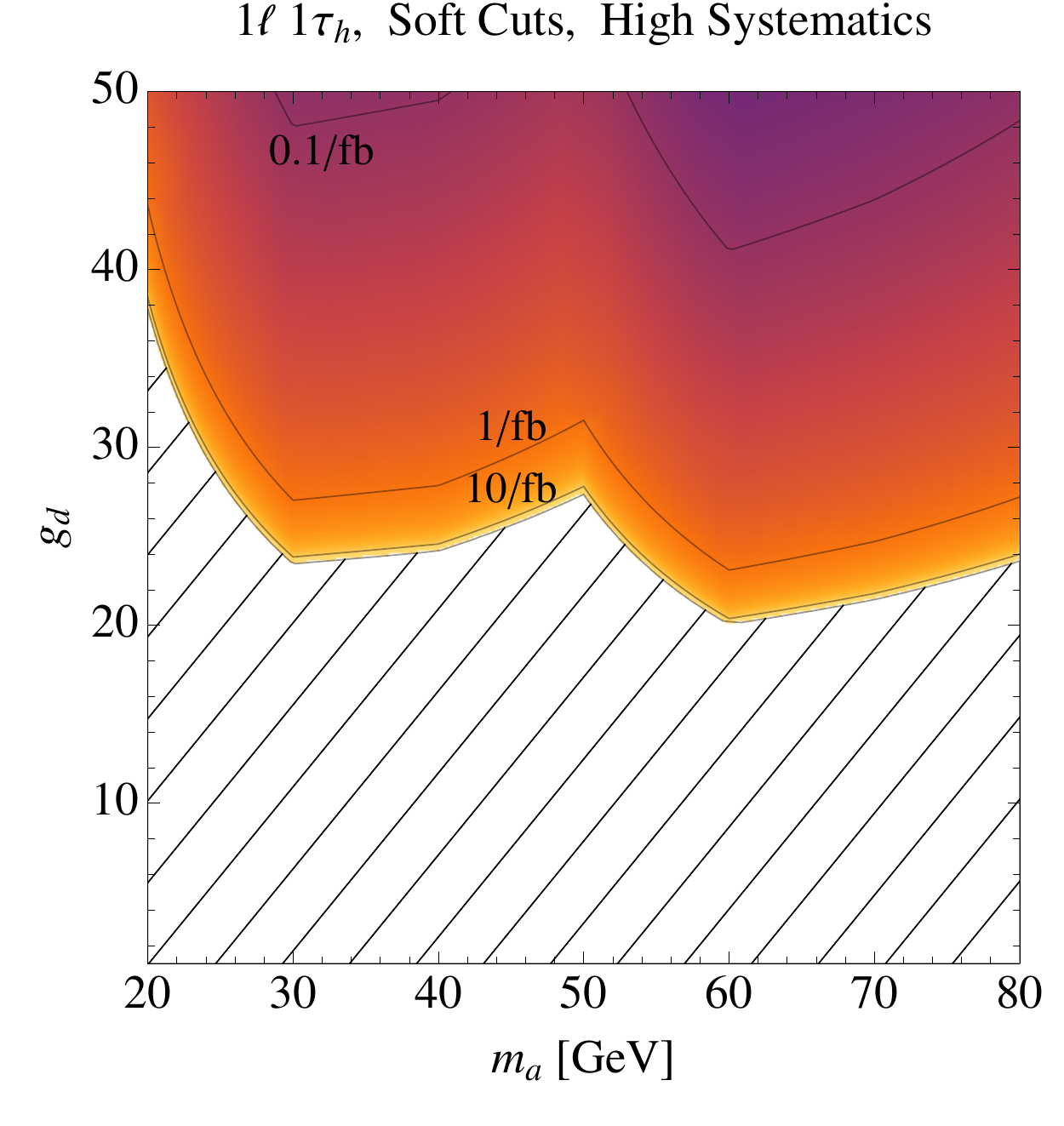}
   \end{center}
\caption{Discovery potential for the SR2 signal region with hard (left) and soft (right) cuts for $\epsilon_{sys}=0.1 (0.3)$ (top (bottom)). Hatched region is the region where no discovery is possible, regardless of luminosity, due to systematic uncertainties. The shading and labeled contours correspond to constant values of $\log(L\times \mathrm{fb})$ needed to achieve $k=3$.} \label{fig:1l1tdisc}
\end{figure}

\begin{figure}[!h]
   \begin{center}
\includegraphics[width=0.45\textwidth]{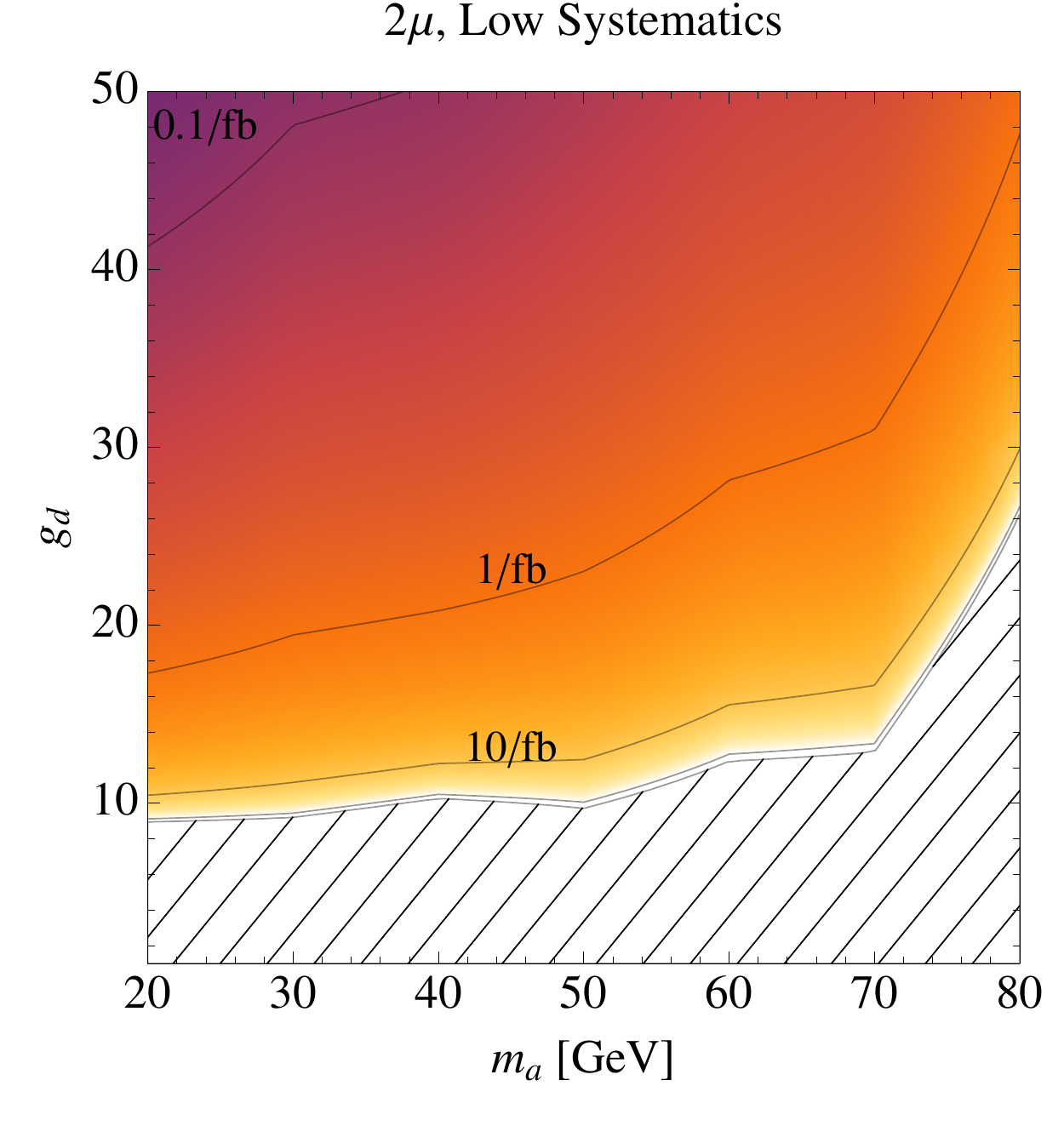}
\includegraphics[width=0.45\textwidth]{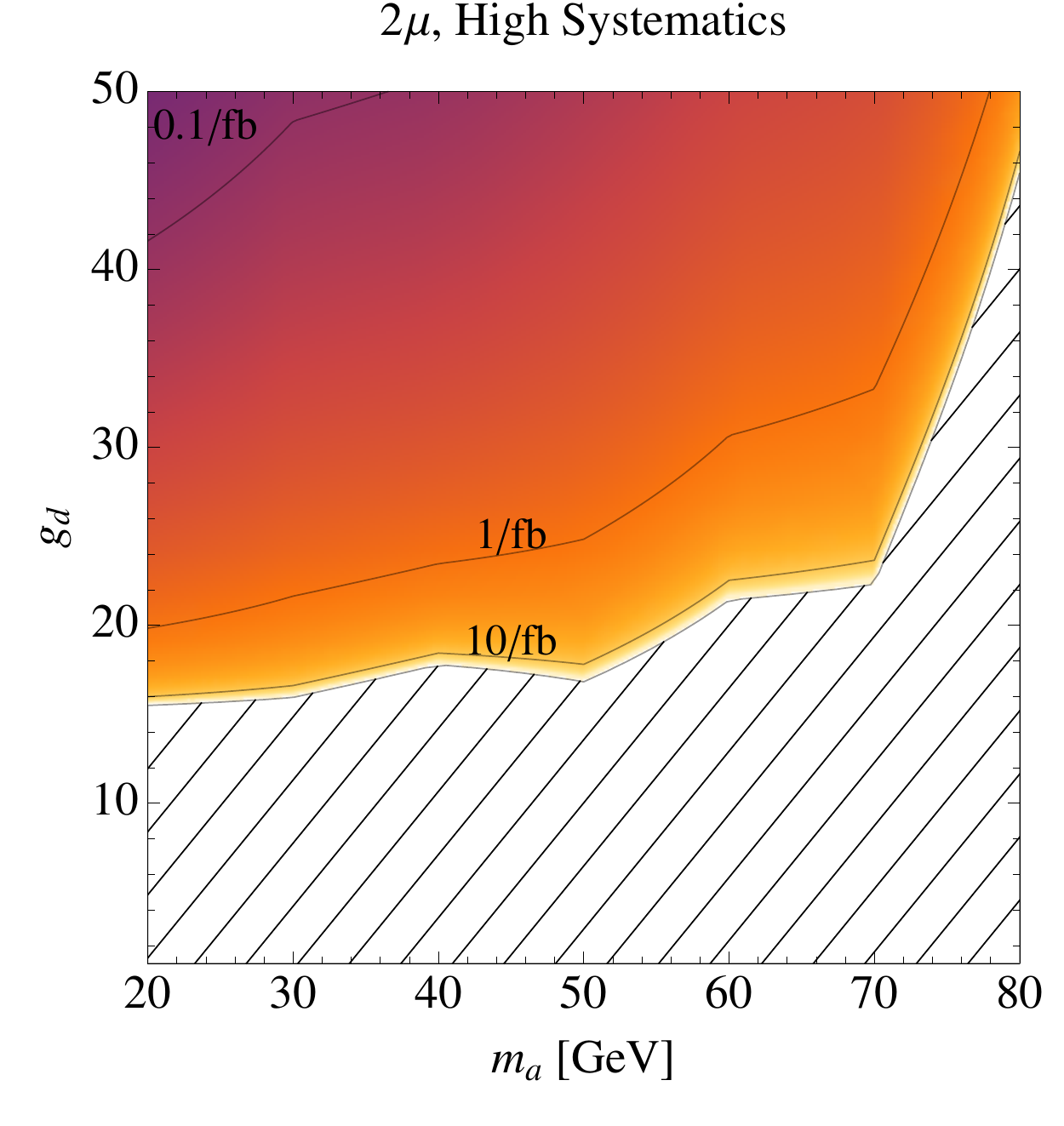}
   \end{center}
\caption{Discovery potential for the SR3 signal region for $\epsilon_{sys}=0.1 (0.3)$ (left (right)). Hatched region is the region where no discovery is possible, regardless of luminosity, due to systematic uncertainties. The shading and labeled contours correspond to constant values of $\log(L\times \mathrm{fb})$ needed to achieve $k=3$.} \label{fig:1u1udisc}
\end{figure}

\section{Results}\label{sec:results}

We can now investigate the extent to which the light pseudoscalar parameter space consistent with the Fermi signal can be probed by the searches we propose.

Due to the low pseudoscalar mass region of interest in this study, as well as the cut-and-count search method for SR1 and SR2, systematic uncertainties are a particularly challenging aspect of performing this search. To estimate the effect of systematic uncertainties, we consider two scenarios in addition to our two cut (hard/soft) scenarios -- low systematics, with $\epsilon_{sys} = 10\%$, and high systematics, with $\epsilon_{sys}=30\%$. Our analysis of the discovery potential is based on a signal significance, given by
\begin{equation}
k=\frac{N_s}{\sqrt{N_s+N_b+\epsilon_{sys}^2N_b^2}},
\end{equation}
where $N_s = \sigma_s*L$ and $N_b = \sigma_b*L$ are the number of signal and background events, respectively, after cuts for a given integrated luminosity, $L$. Contours of constant luminosity are plotted in Figures \ref{fig:1e1udisc}, \ref{fig:1l1tdisc} and \ref{fig:1u1udisc}. For small enough values of $g_d$, systematic uncertainties dominate the signal, and we expect that greater luminosity will be insufficient to illuminate any signal. Note that we have also verified that each signal data set considered has at least 5 events after cuts.

The soft cut scenarios of SR1 and SR2 are optimized for early searches with low luminosity, but suffer from a larger systematics-dominated region, since the total backgrounds are much larger. Thus, their ability to exclude the parameter space ends at approximately $L=10/$fb integrated luminosity. Alternatively, hard cuts scenarios have a better reach with exclusions from $L=100/$fb, though larger luminosity will be unlikely to push this boundary any further.

As discussed, the expected sensitivites for each case are affected by three primary components: production, trigger and cuts. Production rates decrease with increasing mass, $m_a$, reducing the overall cross section and number of events at the LHC. In contrast to this, trigger response improves for heavier pseudoscalars, but has a significant effect on the lighter pseudoscalar scenario. However, the pseudoscalar is produced in association with $b$ quarks, which results in a boost to the $a$ that allows a large enough fraction of events to pass trigger and thereby make the search viable. Lastly, eliminating backgrounds resulting from the $Z$ peak results in a choice of cut thresholds that has a larger impact on events from heavier pseudoscalar masses, especially for the hard cut scenarios. These issues combined result in the typical shape observed in Figures \ref{fig:1e1udisc} and \ref{fig:1l1tdisc}, with reduced exclusion reach for both the lowest and highest mass scenarios.

The dimuon search uses a different approach, incorporating a pseudo-resonance search methodology. While we do not fit a line-shape over the background and compare the signal, we employ a narrow invariant mass window with a sliding center that effectively estimates the result from such an approach. In practice, an approach that fits a line to the continuum background will reduce systematic uncertainties that are associated with the cut-and-count methodology, which requires simulations to estimate the backgrounds. As a result, we suspect that the low systematics scenario in Figure \ref{fig:1u1udisc} is potentially a more realistic case, in contrast with the other signal regions, where low systematics may be overly optimistic.

As a result of the relatively large width of the SM $Z$, combined with detector smearing effects, a dimuon resonance at 80~GeV will contend with increased backgrounds from the $Z$ peak (which is why we do not consider heavier masses). If we assume similar systematic uncertainties for each signal type, then the most promising reach for the high $m_a$ region is in the $1e1\mu$ signal regions, while the $1\ell1\tau$ signal regions are more promising for the low $m_a$ regime. Note that the reach in the dimuon signal region is not as promising as the others for any part of the parameter space \emph{under the assumption of similar systematics}. As mentioned, however, systematic uncertainties in the dimuon search will likely be smaller than in the other modes, and so all signal regions combine to form a complimentary and robust search strategy.

Comparing Figures~\ref{fig:DM_CS} and \ref{fig:1e1udisc}--\ref{fig:1u1udisc}, we see that the searches we propose will cover a significant portion of the otherwise unconstrained parameter space consistent with the Galactic Center excess in scenarios with light pseudoscalar mediators, even with rather low integrated luminosity.  This region is both theoretically and phenomenologically well-motivated, and we encourage both ATLAS and CMS to consider searches along the lines of those presented here.

\section{Application to the NMSSM}\label{sec:NMSSM}
To illustrate the usefulness of our results in a UV-complete model, we can consider how our searches impact the $\mathbb{Z}_3$-symmetric NMSSM parameter space consistent with the excess.  To set our conventions, we take the superpotential to be
\begin{equation}
W=W_{\rm MSSM}|_{\mu=0}+\lambda \widehat{S}\widehat{H}_u\cdot \widehat{H}_d + \frac{\kappa}{3}\widehat{S}^3
\end{equation}
with soft supersymmetry breaking terms given by
\begin{equation}
\Delta V_{soft}=m^2_{H_u} \left|H_u\right|^2+m_{H_d}^2\left|H_d\right|^2+m_{S}^2\left|S\right|^2+\lambda A_{\lambda}H_u\cdot H_d S + \frac{1}{3}\kappa A_{\kappa}S^3.
\end{equation}
Hatted quantities are chiral superfields. The lightest pseudoscalar mass eigenstate can be written in terms of the $SU(2)$ and singlet pseudoscalar ($A$ and $a_s$, respectively) as
\begin{equation}
a=A \cos\theta + a_s \sin\theta
\end{equation}
with effective couplings 
\begin{equation}
g_u=\cos\theta \cot\beta, \hspace{.2cm} g_d = \cos\theta \tan \beta.
\end{equation}
For sizable $\tan\beta$ and $\cos \theta$ not too small, $g_d$ will be larger than 1. Our conventions follow those found in Refs.~\cite{Ellwanger:2009dp, Kozaczuk:2013spa}, to which we refer the Reader for further details regarding the spectrum.

There have been two scenarios proposed in the $\mathbb{Z}_3$-invariant NMSSM to explain the GC excess involving neutralino annihilation into SM particles through a light singlet-like pseudoscalar \cite{Cheung:2014lqa, Guo:2014gra} (see also Ref.~\cite{Cahill-Rowley:2014ora} for an analysis of the general NMSSM, which in some cases may also be probed by the searches we present).  The first involves a mixed singlino/Higgsino-like neutralino, which, to achieve a Standard Model- (and not singlet-) like 125 GeV Higgs, requires the lightest pseduoscalar to be a nearly pure singlet \cite{Cheung:2014lqa} (i.e. $\cos\theta \ll 1$).  Since the pseudoscalar couplings to SM fermions are suppressed, to explain the GC excess this scenario requires $m_a\approx 2 m_{\chi}$ to within about a GeV precision, as well as additional $Z$-mediated contributions to the annihilation rate in the early universe to drive down the relic abundance.  This would seem quite finely tuned, requiring a fortunate conspiracy of parameters to achieve.  Instead, we focus on the second possibility, namely that the neutralino is bino/Higgsino-like. In this case, the singlet component of the 125 GeV Higgs is naturally small, and so the lightest pseudoscalar can feature a more significant amount of mixing between the singlet and $SU(2)$ states.  As a result, the requirement that the neutralino annihilation is on resonance is relaxed, allowing one to consider a much larger range of masses not precisely tuned to $m_a\approx 2m_{\chi}$ \cite{Cheung:2014lqa}.

It is worth mentioning that analyses of the NMSSM subsequent to Ref.~\cite{Cheung:2014lqa} have found somewhat different results, favoring the singlino/Higgsino scenario \cite{Guo:2014gra, Cao:2014efa}.  However, taking the systematics into account in fitting the Fermi signal \cite{Cholis:2014lta, Agrawal:2014oha}, we find that the bino/Higgsino scenario is fully compatible with both the GC signal and the Fermi dwarf spheroidal limits. Another reason the bino/Higgsino scenario may have been disfavored in Ref.~\cite{Cao:2014efa} is that the large pseudoscalar couplings to the SM fermions in the bino/Higgsino scenario are constrained by rare meson decays, in particular $B_s\rightarrow \mu^+\mu^-$.  As pointed out in Ref.~\cite{Cheung:2014lqa}, these constraints can be avoided rather straightforwardly by taking advantage of mild cancellations between the various SUSY contributions to $BR(B_s\rightarrow \mu^+\mu^-)$.  Such points can be difficult to sample in a large scan of the parameter space, as employed in Refs.~\cite{Guo:2014gra, Cao:2014efa}.  However, we have verified that the bino/Higgsino scenario is still in fact viable when taking these constraints into account, as claimed in Ref.~\cite{Cheung:2014lqa}.  

\begin{figure*}[!t]
\centering
\mbox{\hspace*{-1cm}\includegraphics[width=0.55\textwidth,clip]{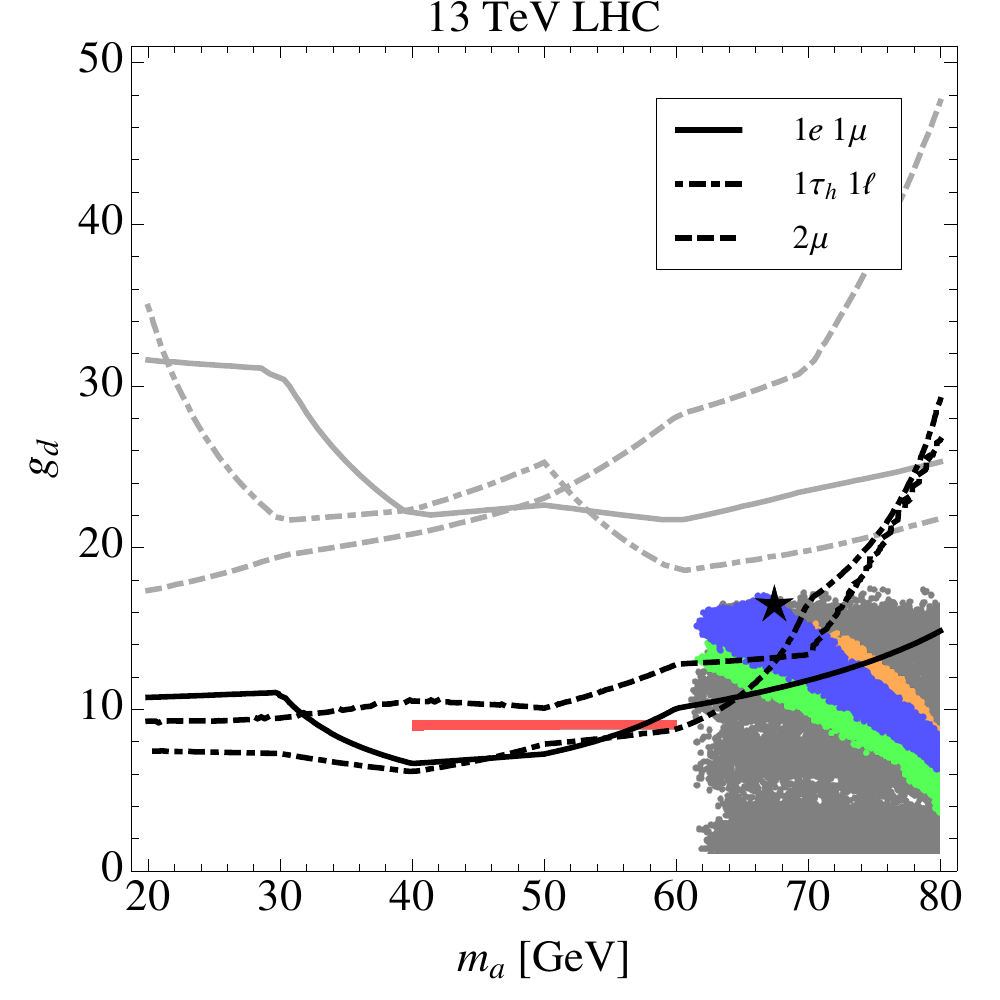}}
\caption{\label{fig:NMSSM_results}\small Application of our results to the $\mathbb{Z}_3$-symmetric NMSSM.  The black (gray) contours correspond to the reach at 100 fb$^{-1}$ (1 fb$^{-1}$) for the hard (soft) cut scenarios and low systematics in the various search channels. The gray points are the result of a Markov Chain Monte Carlo scan of the parameter space (described in the text) consistent with all existing phenomenological constraints with no requirements on the LSP relic abundance or annihilation rate with parameters as in Eq.~\ref{eq:vals} and $m_A=550$ GeV. The green, blue, and orange points correspond to points capable of explaining the Fermi signal and consistent with the recent dwarf spheroidal constraints for $m_A=500$, 550, and 600 GeV, respectively. The red band is an example of the NMSSM parameter space found to be consistent with the excess in Ref.~\cite{Cheung:2014lqa}.  The sample point of Table~\ref{tab:bm} below is indicated with a star. Note that it may be possible to choose parameters minimizing the $h a a$ coupling to fill in the $m_a<m_h/2$, $g_d>1$ region, which we did not attempt in our scan.}
\end{figure*}

The bino/Higgsino explanation for the GC excess maps directly onto our simplified model (only that the WIMP is a Majorana, instead of Dirac, fermion).  To illustrate the effect of our searches on the viable bino/Higgsino parameter space of the NMSSM, we performed a Markov Chain Monte Carlo scan of the parameter space using \texttt{NMSSMTools 4.4.0} \cite{NMSSMTools}, interfaced with \texttt{micrOmegas 3.1} \cite{MicrOmegas}. Motivated by the parameter space presented in Ref.~\cite{Cheung:2014lqa}, we fixed 
\begin{equation} \label{eq:vals}
\begin{aligned}
&\lambda = 0.05, \hspace{0.2cm} \mu= 615 \hspace{0.1cm}{\rm GeV}, \hspace{0.2cm} m_A=550 \hspace{0.1cm}{\rm GeV}, \\
&M_1=45 \hspace{0.1cm}{\rm GeV}, M_2=1 \hspace{0.1cm}{\rm TeV}, M_3=2 \hspace{0.1cm}{\rm TeV},\\
&M_{Q_3}=M_{U_3}=7.5\hspace{0.1cm}{\rm TeV},\hspace{0.2cm} A_t = \sqrt{6} M_{Q_3},\hspace{0.2 cm} M_{D_{1,2,3}}=5.5 \hspace{0.1cm}{\rm TeV}
\end{aligned}
\end{equation}
with all other soft masses and triscalar couplings at 1 TeV, while varying $\tan\beta$, $\kappa$, and $A_{\kappa}$. We required all points to satisfy all existing constraints discussed earlier and implemented in \texttt{NMSSMTools}. The results of the scans are shown, along with our results for the LHC reach across the parameter space, in Fig.~\ref{fig:NMSSM_results}.  The gray points were generated without requiring the lightest supersymmetric particle (LSP) to explain the Galactic Center excess or satisfy constraints on its relic abundance. The green, blue, and orange points correspond to $m_A=500$, 550, 600 GeV and feature a bino-like LSP with a relic abundance compatible with WMAP and Planck measurements (including a 2$\sigma$ theoretical uncertainty) \cite{Kozaczuk:2013spa}
\begin{equation}
0.091\leq \Omega h^2\leq 0.138
\end{equation}
and compatible with both the Galactic Center excess and the dwarf constraints,
\begin{equation}
1.0 < \frac{\langle\sigma v \rangle}{1\times 10^{-26} \hspace{.1cm}{\rm cm}^3/{\rm s}} < 1.5,
\end{equation}
for self-conjugae dark matter.
Points satisfying these constraints typically have small, but non-negligible, $p$-wave suppressed contributions at freeze-out, such as those involving the $Z$ (but still consistent with limits on the invisible $Z$ width).  This slightly reduces the relic abundance relative to the value suggested by  $\chi\chi \rightarrow a\rightarrow b \bar{b}$ annihilation alone and allows these points to circumvent the dSph limits.  Note that we did not attempt to minimize the $h a a$ coupling, and so no points were found with $2m_a<m_h$ and $g_d>1$. However, it might be possible to reach this parametric regime \cite{Kozaczuk:2013spa} as suggested in Ref.~\cite{Cheung:2014lqa}, whose results we show along with ours in Fig.~\ref{fig:NMSSM_results} by the red band. These values were taken from Fig.~6 of Ref.~\cite{Cheung:2014lqa} for $m_{\chi}=35$ GeV, while our scan was performed assuming $m_{\chi}\approx M_1=45$ GeV.  Table~\ref{tab:bm} provides the detailed spectrum information for an example parameter space point consistent with the GC excess and which would be probed by $a\rightarrow \tau^+\tau^-$, $\mu^+\mu^-$ at the 13 TeV LHC. This point is marked by the black star in Fig.~\ref{fig:NMSSM_results}. Note also that our scan did not find points with $g_d>18$. Larger values of $g_d$ are typically excluded by LHC limits on the heavy MSSM-like pseudoscalar for the values of $\tan\beta$ sampled. In theories that do not rely on mixing with the SM-like Higgs, these constraints, as well as those from $h\rightarrow a a$ decays, are often significantly relaxed or absent.

 \begin{table}[!tc]
\centering
 \begin{tabular}{| c   c c  c  c  c  c   c  |}
\hline
 \hspace{.2cm} $\lambda$  \hspace{.2cm} &  \hspace{.2cm} $\kappa$  \hspace{.2cm} &  \hspace{.2cm} $A_{\kappa}$  \hspace{.2cm} &  \hspace{.2cm} $\tan\beta$   \hspace{.2cm} &  \hspace{.2cm} $m_A$ \hspace{.2cm} & \hspace{.2cm} $\mu$  \hspace{.2cm} &  $M_1$ & $M_2$    \\

0.05  & 0.52 &-8.5 & 21.8 & 550 & 615 & 45 & 1000   \\

\hline
\hline

$ m_h$ & $m_a$ &  $m_{\chi}$ &  $g_d$ & $\Omega h^2$ & $\langle \sigma v \rangle$ [cm$^3/$s] &  $\sigma_{\rm SI}$ [cm$^2$] & $\sigma_{\rm SD}$ [cm$^2$] \\

125.8 & 67.5 & 44.1 & 16.4 & 0.137 & $1.48\times 10^{-26}$& $4.3\times 10^{-46}$ & $3.9\times 10^{-44}$ \\
\hline
\end{tabular}
\caption{\label{tab:bm} Example parameter space point in the NMSSM capable of explaining the GC excess and consistent with the Fermi dwarf spheroidal limits. All dimensionful parameters are in GeV unless otherwise stated. The remaining parameters are set to the values shown in Eq.~\ref{eq:vals}. This point would likely be probed by the searches we propose at the 13 TeV LHC with 100 fb$^{-1}$ of integrated luminosity. }
\end{table}

The contours in Fig.~\ref{fig:NMSSM_results} show the sensitivity of our proposed searches to the NMSSM parameter space consistent with the Galactic Center excess at both 1 fb$^{-1}$ and 100 fb $^{-1}$. A significant portion of the favored region with sizable $g_d$ would be probed by the 13 TeV LHC at these luminosities.  Even more reach would be expected at the 14 TeV LHC. Our searches are complementary to $h\rightarrow a a$ observations as well as existing LHC searches for MSSM Higgs bosons and would access regions of the parameter space not currently probed by other experiments, providing a potential window into a dark sector difficult to access otherwise.

\section{Summary and Conclusions \label{sec:summary}}

Many dark matter models feature WIMPs that can be very difficult to observe at colliders. Scenarios of this type can be consistent with the Galactic Center excess observed by the Fermi Large Area Telescope. Exploring these ``coy dark sectors" at the LHC suggests a shift away from missing transverse energy signals and towards direct signatures of the particle(s) mediating the interaction of the dark matter with the Standard Model. 

Models involving pseudoscalar mediators and consistent with the GC excess can be of the coy variety.  A good fit to the Fermi signal can be provided by relatively light WIMPs annihilating through a pseudoscalar into $b$ quarks. In many realistic scenarios this suggests substantial couplings of the mediator to down-type Standard Model fermions. The signal favors WIMP masses in excess of $\sim 35$ GeV, while current collider bounds often imply pseudoscalar masses below 90 GeV (provided they satisfy constraints from LEP). An interesting and currently untested explanation of the GC signal thus involves a pseudoscalar with mass below about 90 GeV with sizable couplings to down-type fermions and small branching fraction into WIMPs.  The latter is generically small in this scenario since the on-shell decay of the mediator into dark matter is often kinematically forbidden and because the pseudoscalar's coupling to WIMPs is relatively small.  Our study has attempted to extend LHC coverage to this scenario by taking advantage of the mediator's enhanced couplings to Standard Model fermions (relative to those of a SM-like Higgs boson of the same mass) and studying the production and decays of the pseudoscalar involving down-type final states.

To this end, we explored signals that include one to two $b$-jets and with either $\tau$ or $\mu$ lepton pairs in the final state. We employed a simplified model, in which we assumed that the couplings of the pseudoscalar to Standard Model fermions were proportional to their mass, modulo common scaling factors for down- and up-type fermions.  While this need not be the case, this situation is common in UV completions involving Type II 2 Higgs doublet models, as in supersymmetry.  Our results can be applied to models with different coupling structures by a straightforward re-scaling of the production cross-section and branching ratios.

Due to the rather low pseudoscalar masses we consider, trigger is an important factor in the search reach. We thus performed an analysis of the trigger response of the signal, and explored cuts that were effective in improving the signal significance. Our search strategy comprises a signal excess analysis for the $1e1\mu+1-2b$ and $1\ell1\tau+1-2b$ modes, including low luminosity (soft cuts) and high luminosity (hard cuts) scenarios, and a dilepton resonance search in the $\mu^+\mu^-+1-2b$ signal. Since signal excess searches suffer from large systematics from comparisons to simulated instead of data driven backgrounds, we also analyzed the impact of systematic uncertainties on the LHC reach in all three signal modes.

In the most optimistic scenarios, we find that the LHC should be able to explore values of the reduced pseudoscalar coupling to down-type fermions as low as $g_d\sim8$ for 100/fb of integrated luminosity at $\sqrt{s}=$13 TeV. Even in more pessimistic scenarios with higher systematics, we find that the LHC should be able to explore down to $g_d \sim 10$ for some values of $m_a$. This reach, however, is highly dependent on the trigger settings, and so we strongly recommend that the experimental collaborations attempt to account for this type of signal when finalizing their trigger thresholds for leptons, particularly those triggers for muons. The parameter space in the NMSSM  not covered by $h\rightarrow a a$ searches, with $m_a\sim60-80$~GeV, should be explorable to some extent, and further optimization of the search strategy could focus on this narrow region of masses. More generally, the searches we propose are highly complementary to those already existing at the LHC or elsewhere, highlighting their importance in the interest of fully covering the parameter space in question.

In summary, light pseudoscalars with significant couplings to Standard Model fermions are well-motivated mediators for dark matter annihilation and arise in many models, including those explaining the Fermi Galactic Center excess.  In many cases, these new particles would have evaded previous searches but should be testable at the LHC.  Significant regions of the parameter space can be explored even with low luminosity, and so this signal presents a possibility for ongoing examination throughout the full LHC program.

\section*{Acknowledgements}
We would like to thank Dugan O'Neil, James Hirschauer, Dan Hooper, and John Ng for helpful discussions. We are also very grateful to David E. Morrissey for insightful conversations throughout this project and for providing feedback on this manuscript.  This work was supported by the National Science and Engineering Research Council of Canada~(NSERC).

\appendix

\section{Appendix: Kinematic Distributions\label{sec:appa}}

The kinematic distributions of signals and backgrounds are included in this section. For the signals, 1.5 million events are generated for each of the three-body processes, $\sigma(p p \rightarrow b \bar{b} a)BR(a \rightarrow \ell^+ \ell^-)$, and two-body processes, $\sigma(p p \rightarrow b(\bar{b}) a)BR(a \rightarrow \ell^+ \ell^-)$. For the backgrounds, each process discussed in Section \ref{sec:production} was generated with at least 1.5 million events, with some generated at higher multiplicity in order to achieve sufficient statistics to be confident on the distributions. All distributions are generated from events remaining after applying trigger level cuts and then signal region tagging. Backgrounds are plotted additively, such that each successively larger background is added to the previous backgrounds. All distributions combine the generated events from multiple processes, such that the bin value for a single event is dependent on the specific process that resulted in the event generated. For this reason, two successive bins containing a single generated event each may have different weights.

\begin{figure}[!h]
   \begin{center}
\includegraphics[width=0.31\textwidth]{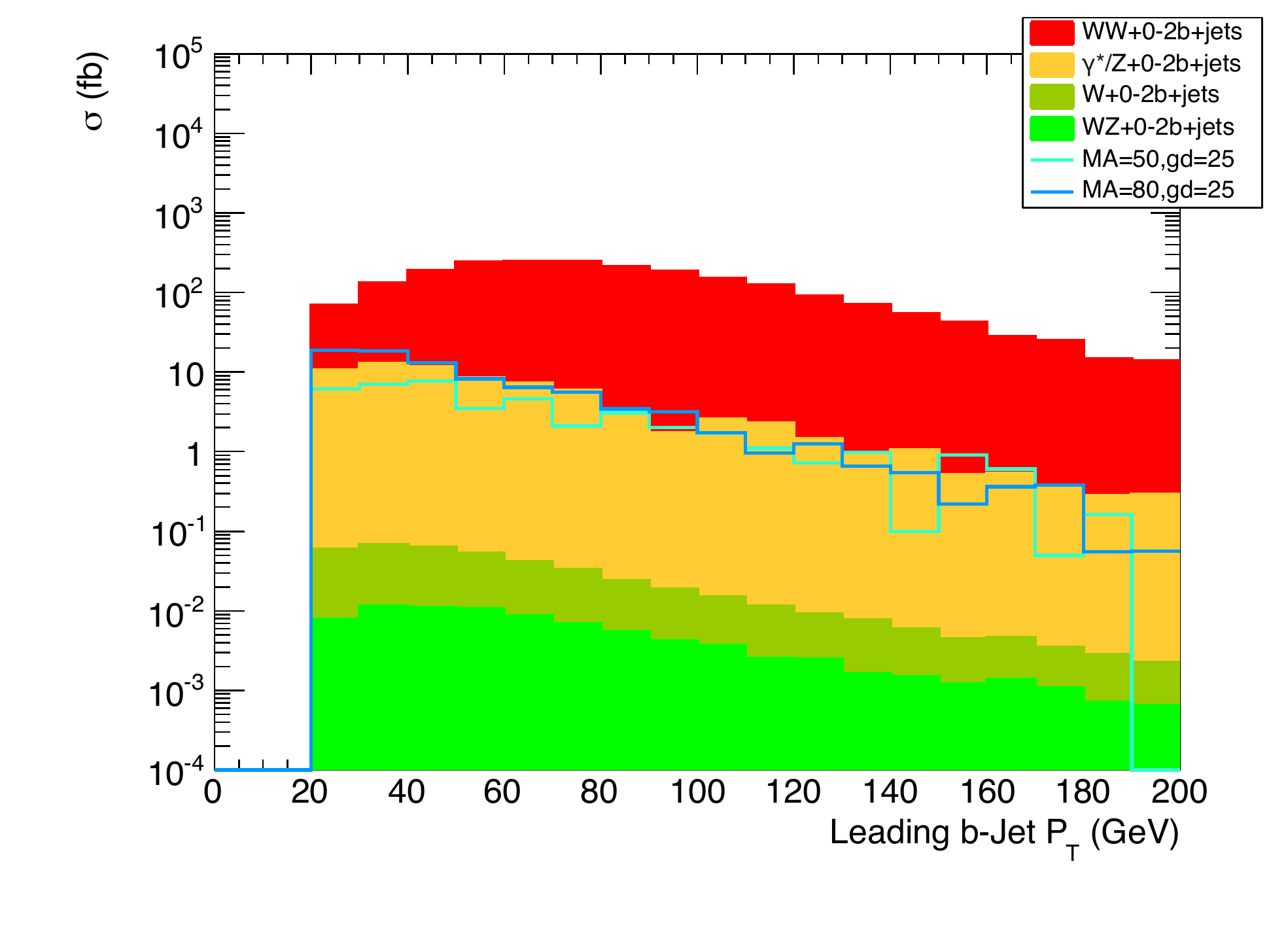}
\includegraphics[width=0.31\textwidth]{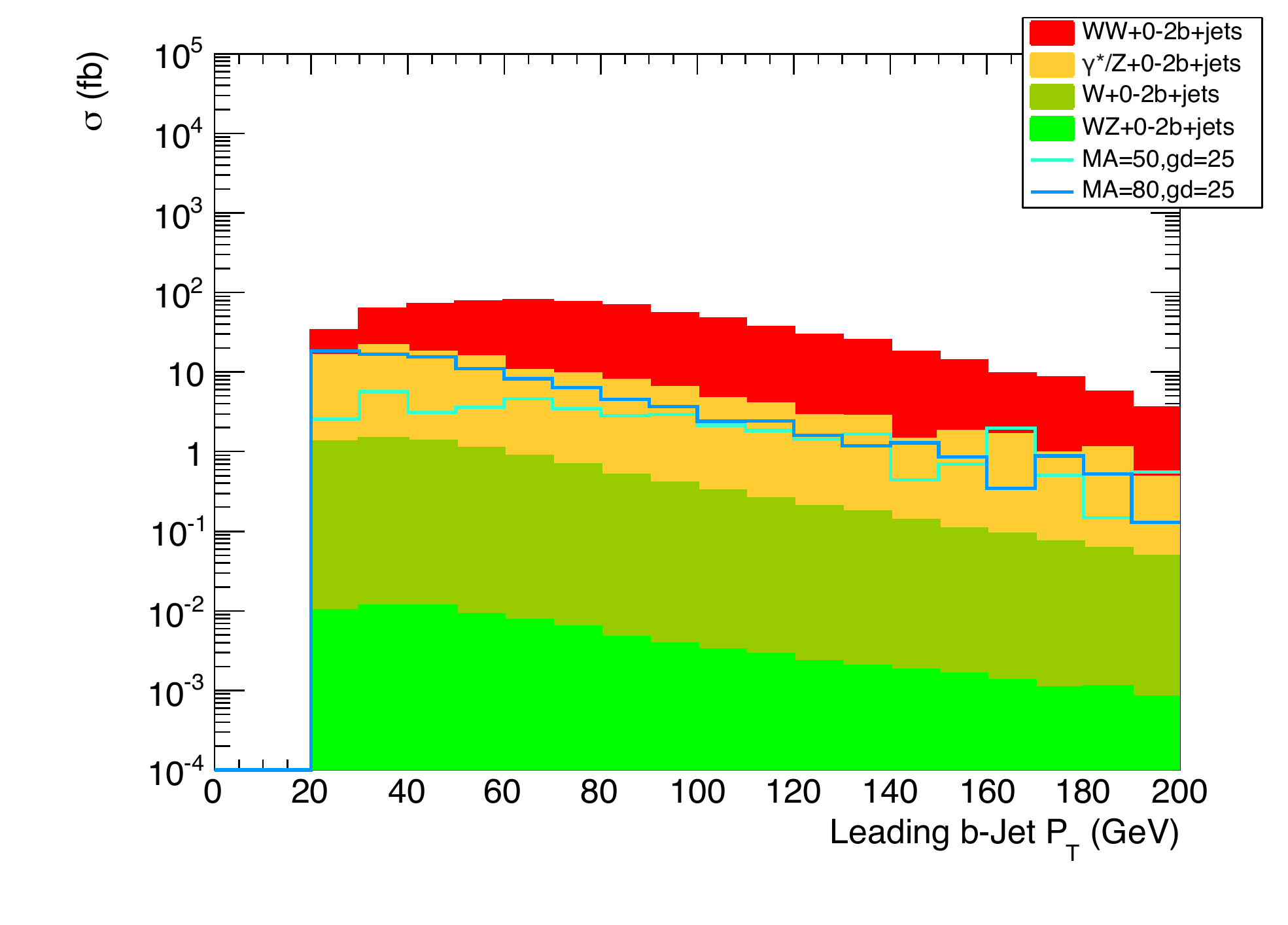}
\includegraphics[width=0.31\textwidth]{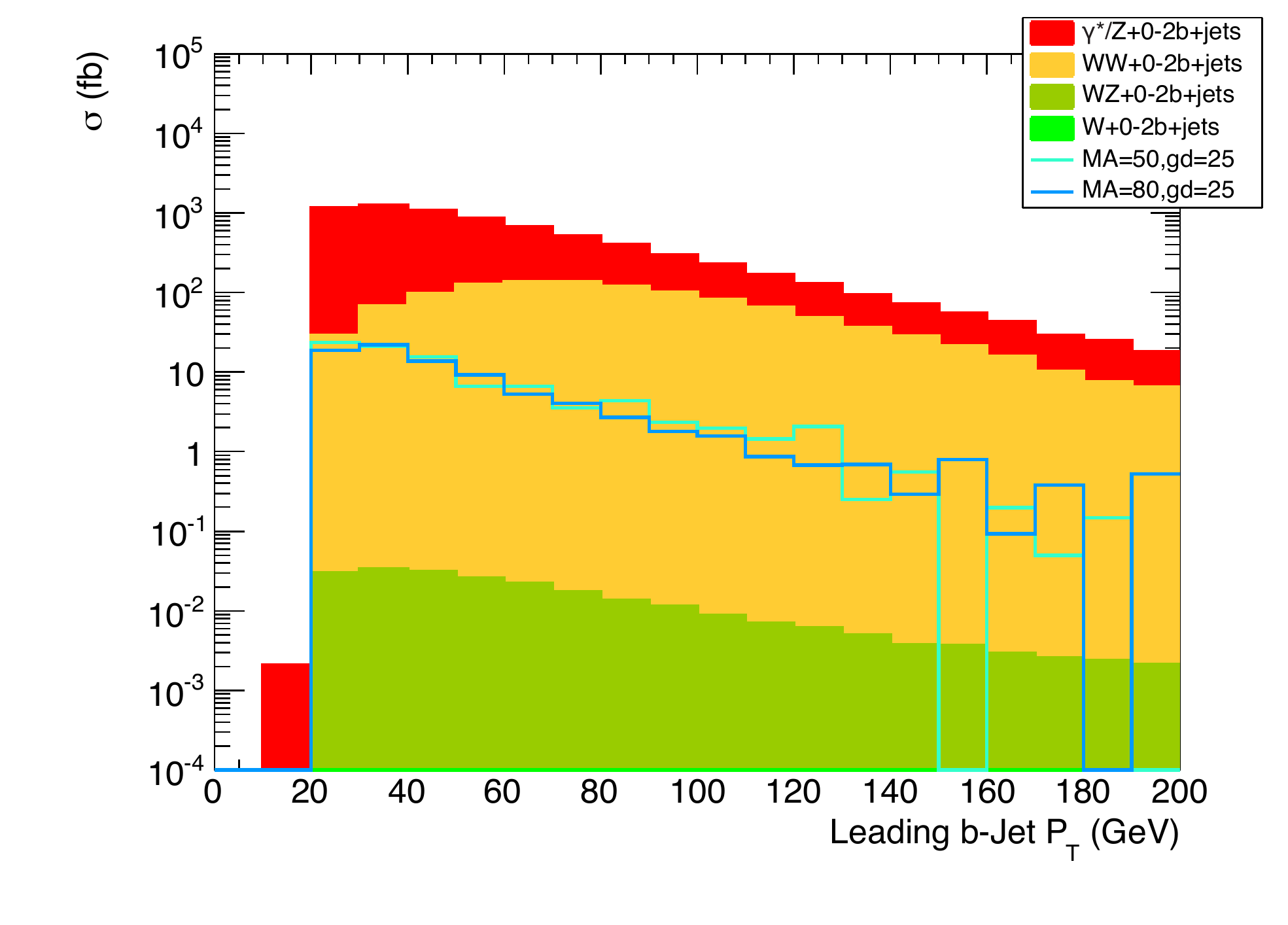}
   \end{center}
\caption{Leading $b$-jet $p_T$ distribution. Trigger cuts and tagging are applied, but no other kinematic cuts are applied. From left to right, figures are for SR1 ($1e1\mu$), SR2 ($1\ell1\tau_h$) and SR3 ($2\mu$). } \label{fig:bptmax}
\end{figure}

\begin{figure}[!h]
   \begin{center}
\includegraphics[width=0.31\textwidth]{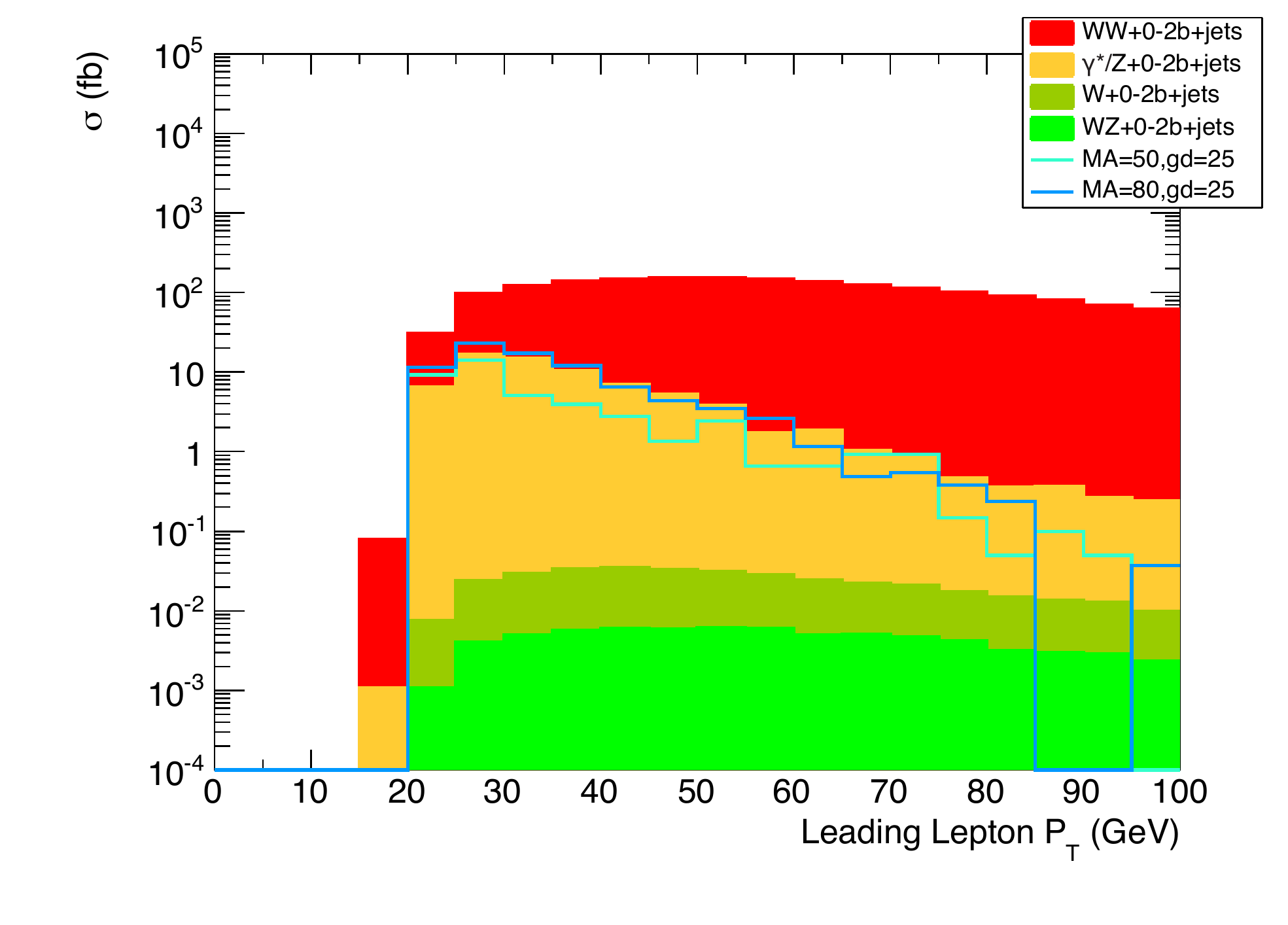}
\includegraphics[width=0.31\textwidth]{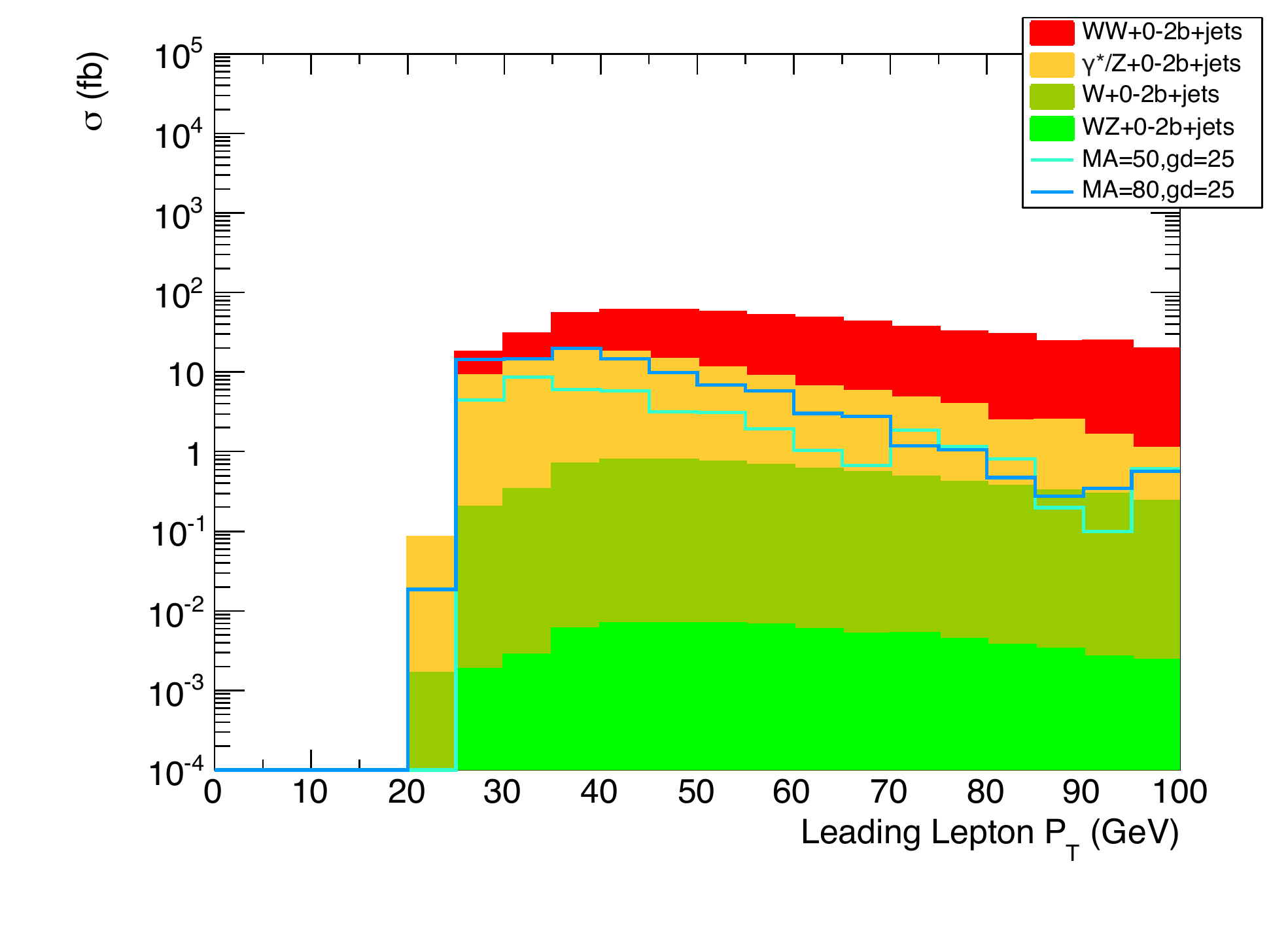}
\includegraphics[width=0.31\textwidth]{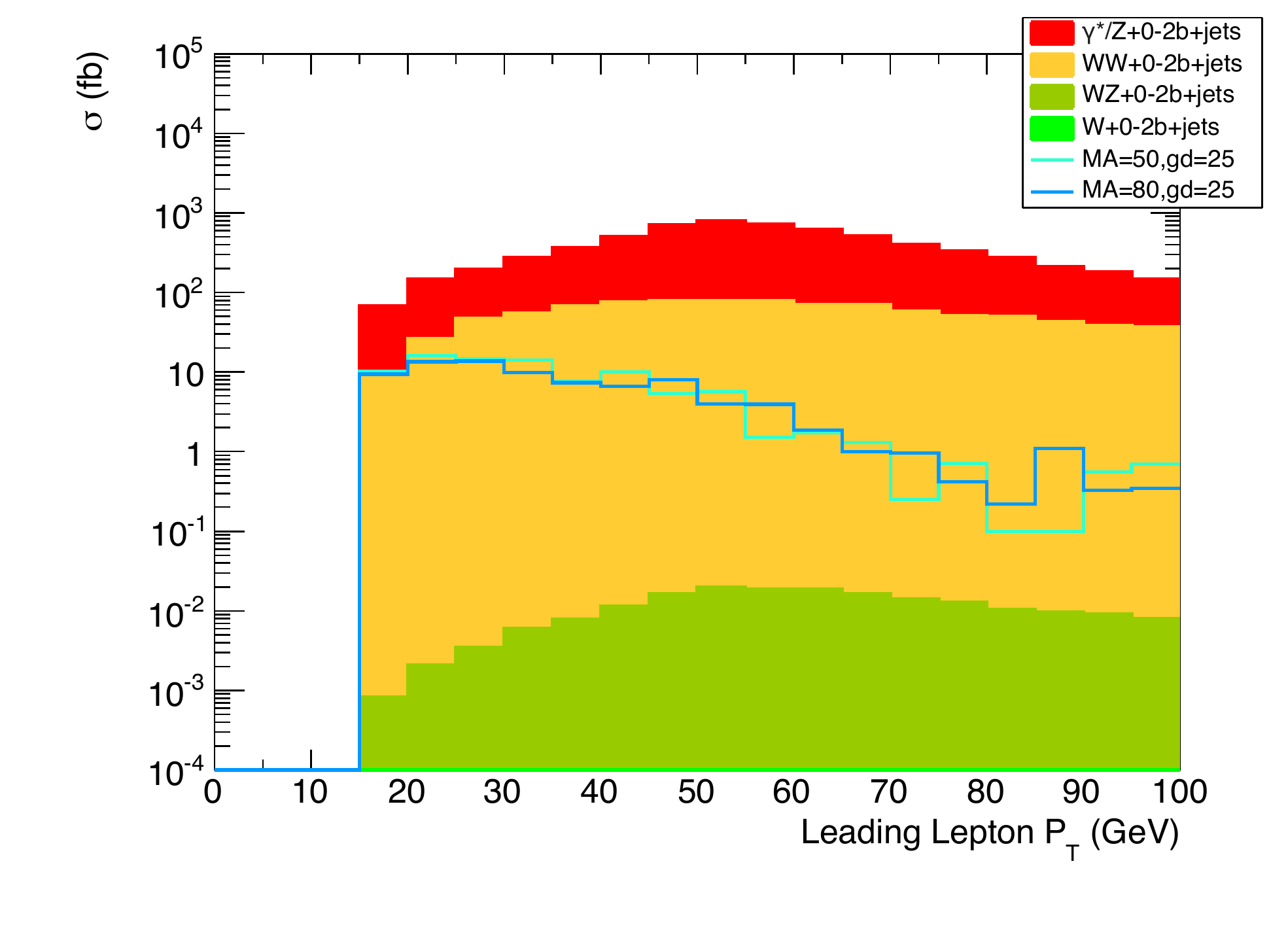}
   \end{center}
\caption{Leading lepton ($e$, $\mu$) $p_T$ distribution. For SR2, there is only one lepton. Trigger cuts and tagging are applied, but no other kinematic cuts are applied. From left to right, figures are for SR1 ($1e1\mu$), SR2 ($1\ell1\tau_h$) and SR3 ($2\mu$).} \label{fig:lptmax}
\end{figure}

\begin{figure}[!h]
   \begin{center}
\includegraphics[width=0.31\textwidth]{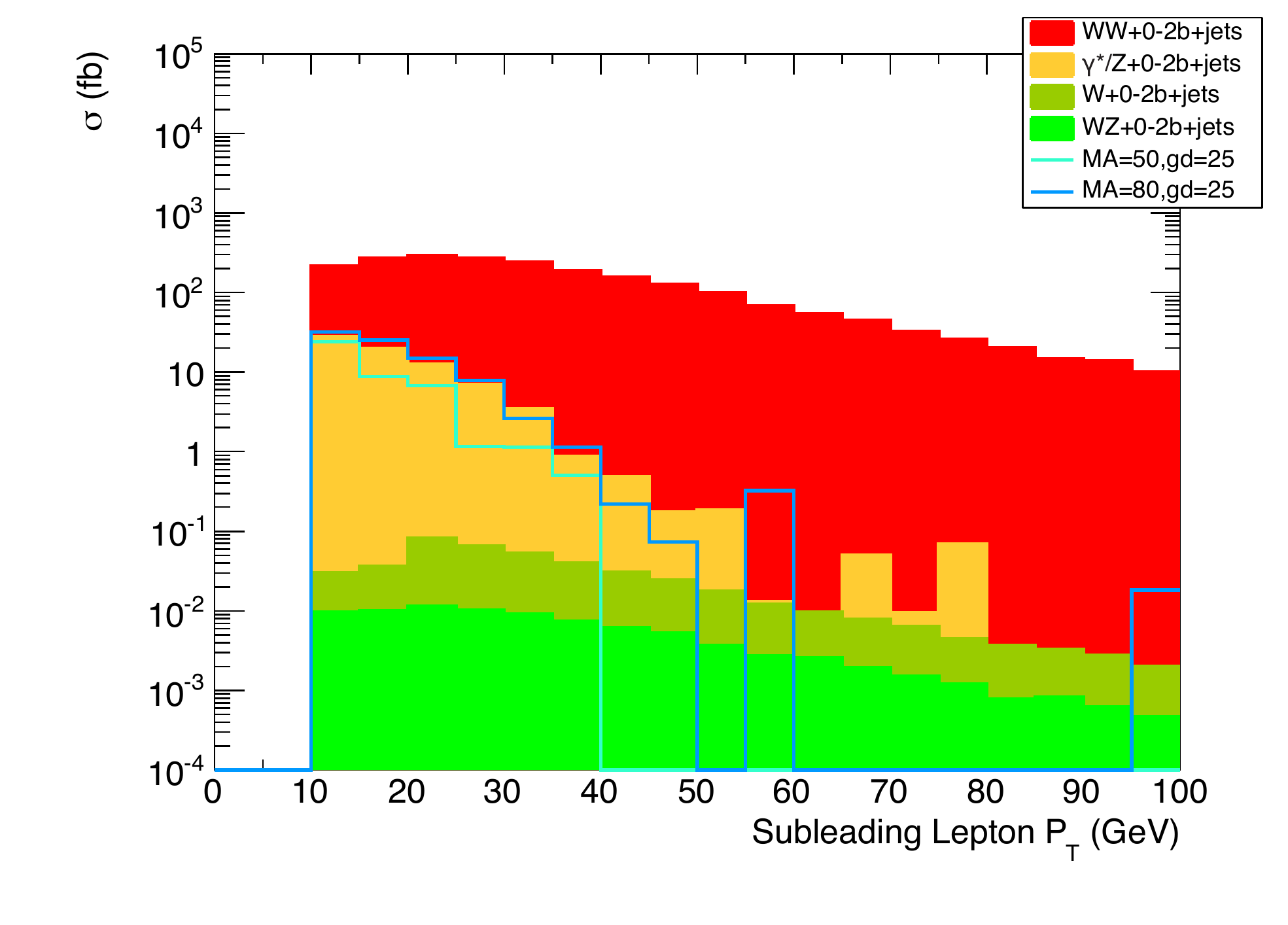}
\includegraphics[width=0.31\textwidth]{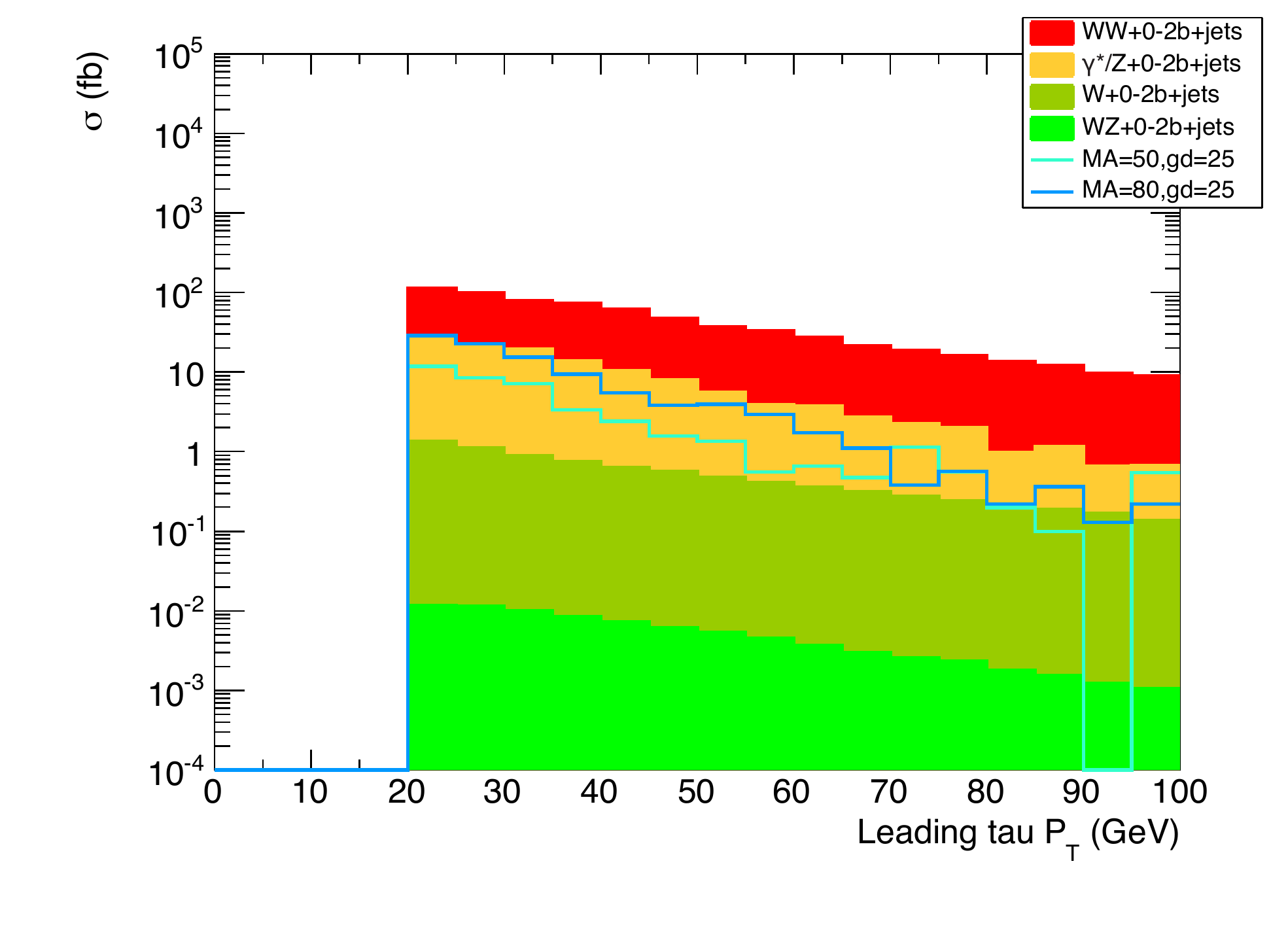}
\includegraphics[width=0.31\textwidth]{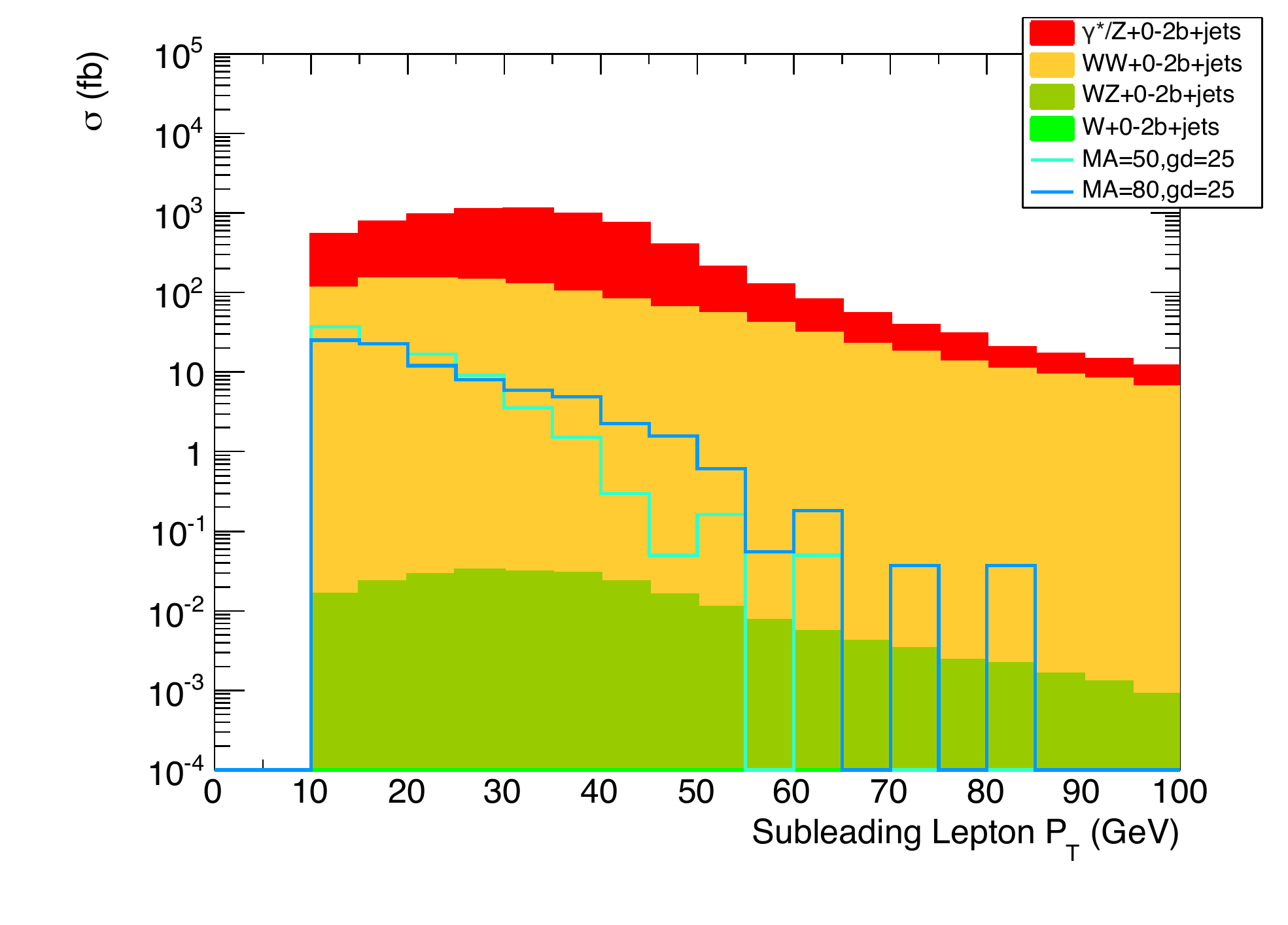}
   \end{center}
\caption{Sub-leading lepton $p_T$ distribution. For SR2, this is the $\tau$ $p_T$ distribution. Trigger cuts and tagging are applied, but no other kinematic cuts are applied. From left to right, figures are for SR1 ($1e1\mu$), SR2 ($1\ell1\tau_h$) and SR3 ($2\mu$).} \label{fig:lptnmax}
\end{figure}

\begin{figure}[!h]
   \begin{center}
\includegraphics[width=0.31\textwidth]{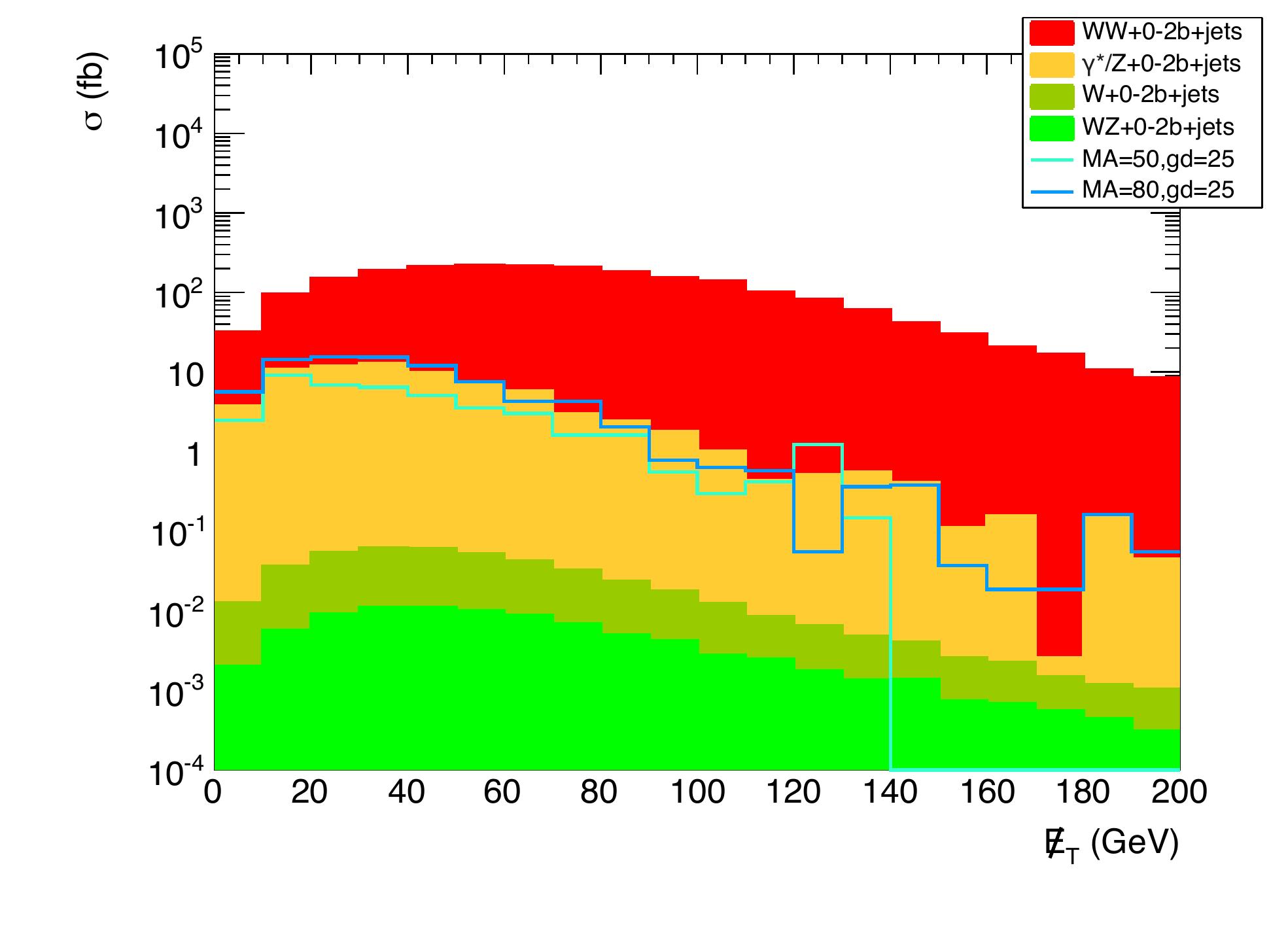}
\includegraphics[width=0.31\textwidth]{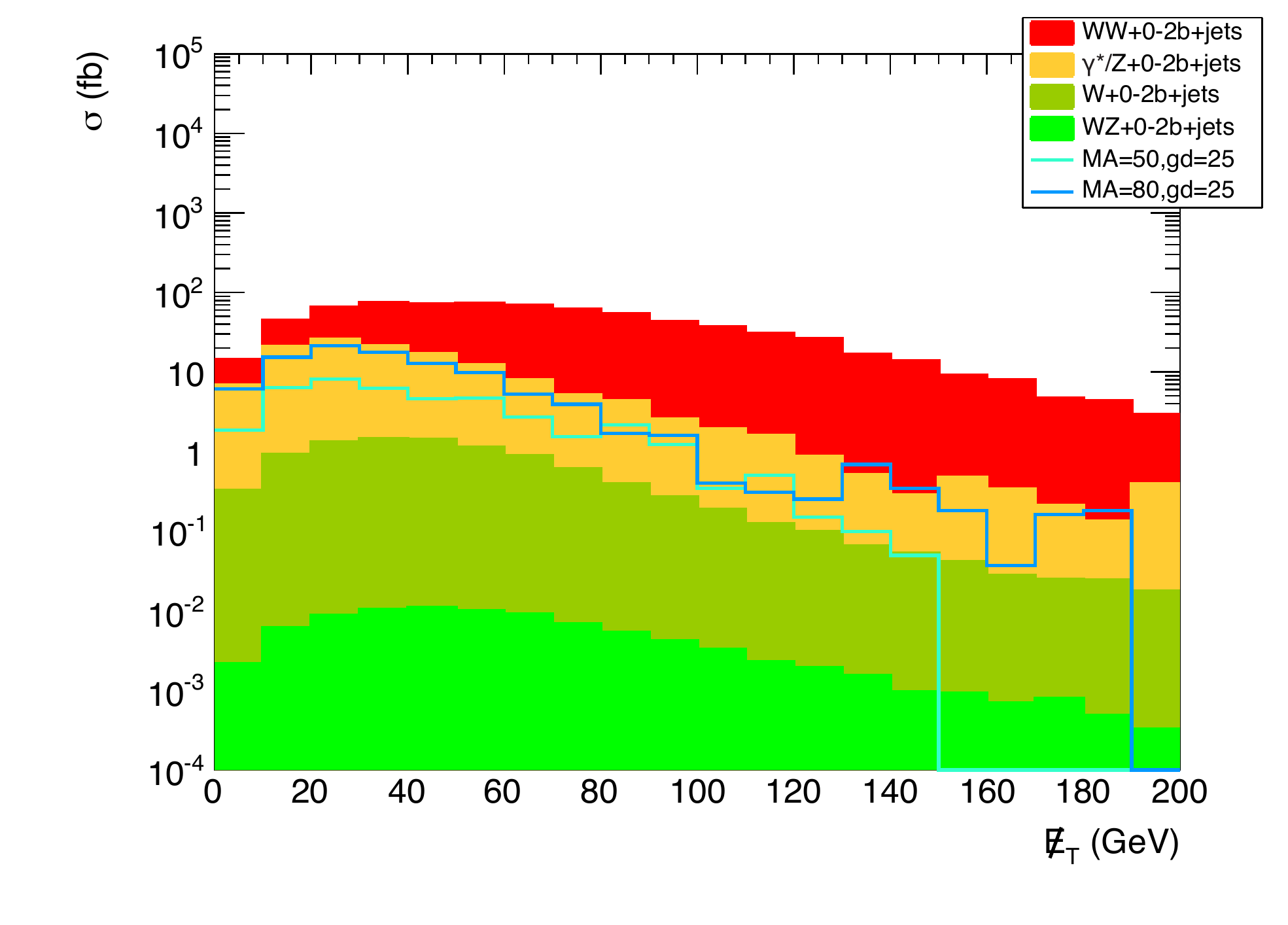}
\includegraphics[width=0.31\textwidth]{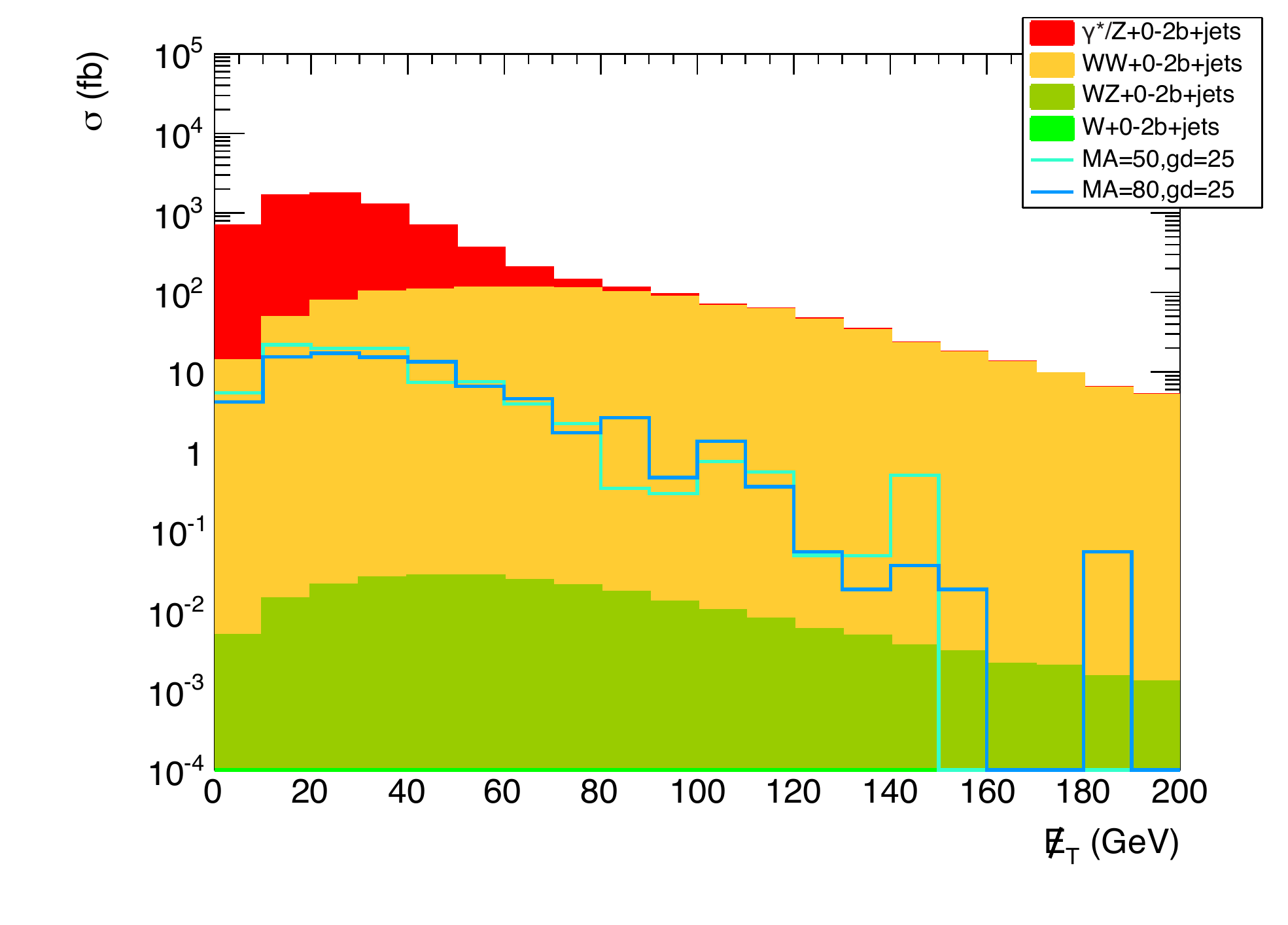}
   \end{center}
\caption{Leading $b$-jet $p_T$ distribution. Trigger cuts and tagging are applied, but no other kinematic cuts are applied. From left to right, figures are for SR1 ($1e1\mu$), SR2 ($1\ell1\tau_h$) and SR3 ($2\mu$).} \label{fig:met}
\end{figure}

\begin{figure}[!h]
   \begin{center}
\includegraphics[width=0.31\textwidth]{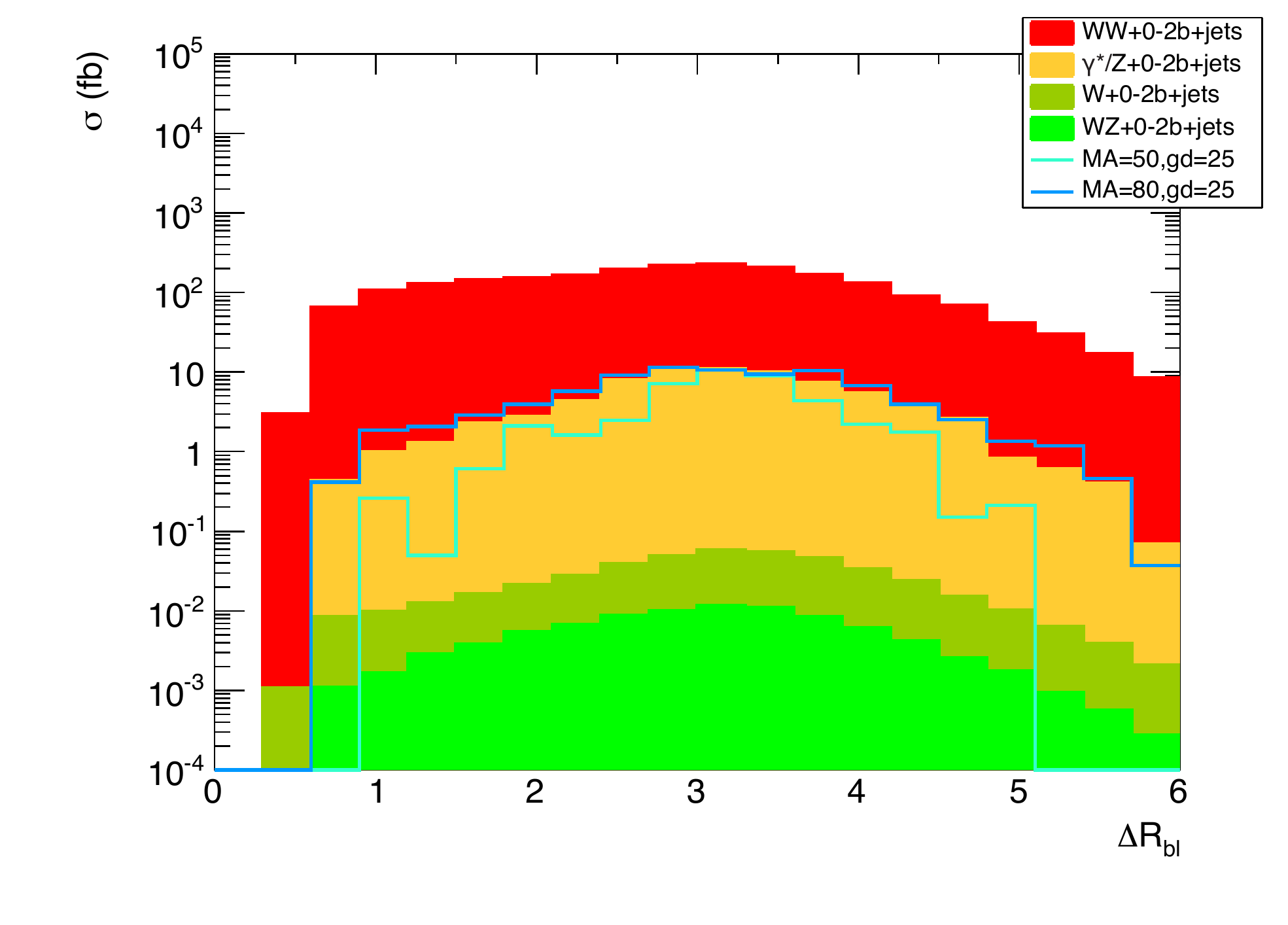}
\includegraphics[width=0.31\textwidth]{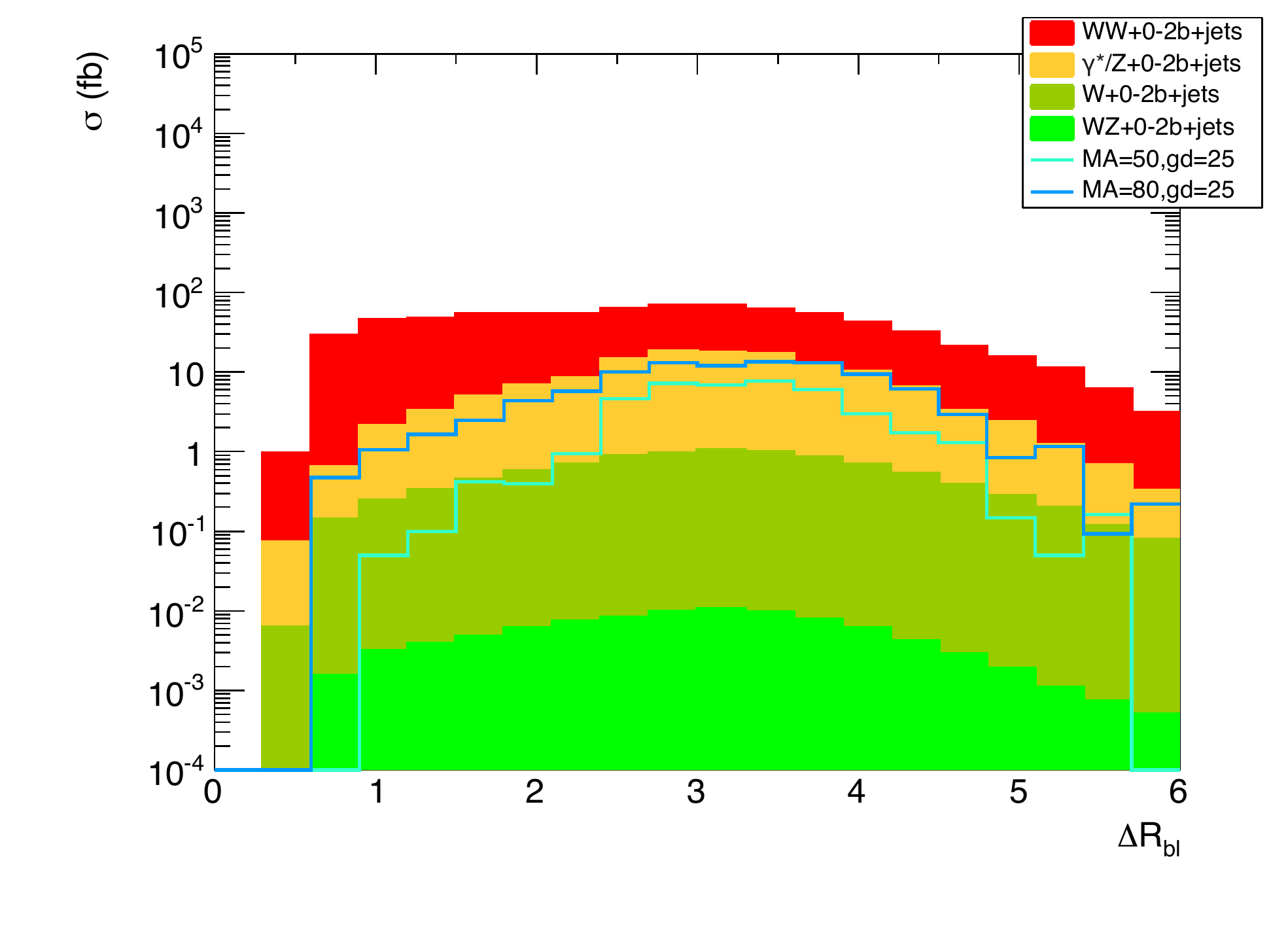}
\includegraphics[width=0.31\textwidth]{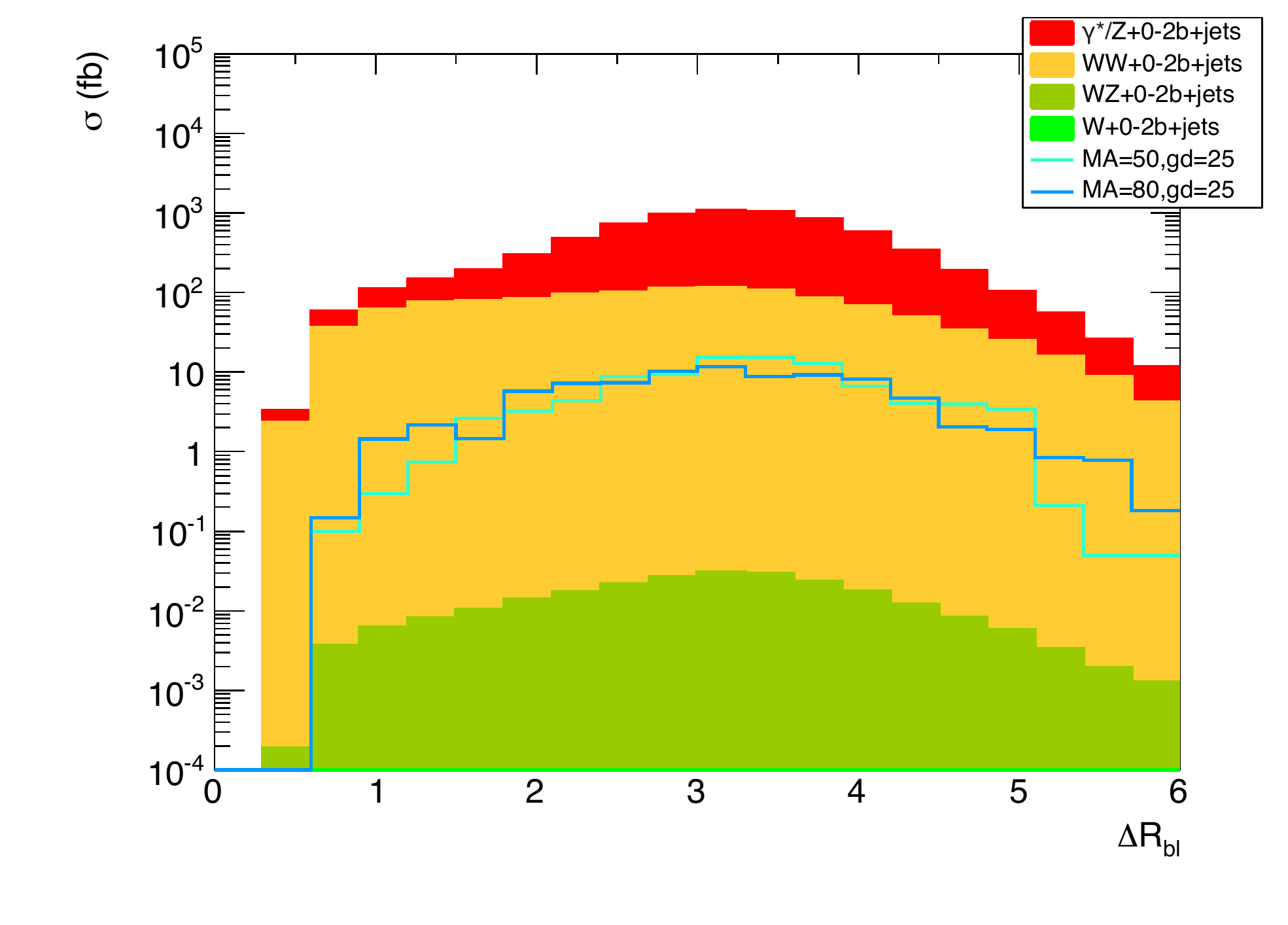}
   \end{center}
\caption{$\Delta R$ distribution between the leading $b$-jet and leading lepton. Trigger cuts and tagging are applied, but no other kinematic cuts are applied. From left to right, figures are for SR1 ($1e1\mu$), SR2 ($1\ell1\tau_h$) and SR3 ($2\mu$).} \label{fig:dRbl}
\end{figure}

\begin{figure}[!h]
   \begin{center}
\includegraphics[width=0.31\textwidth]{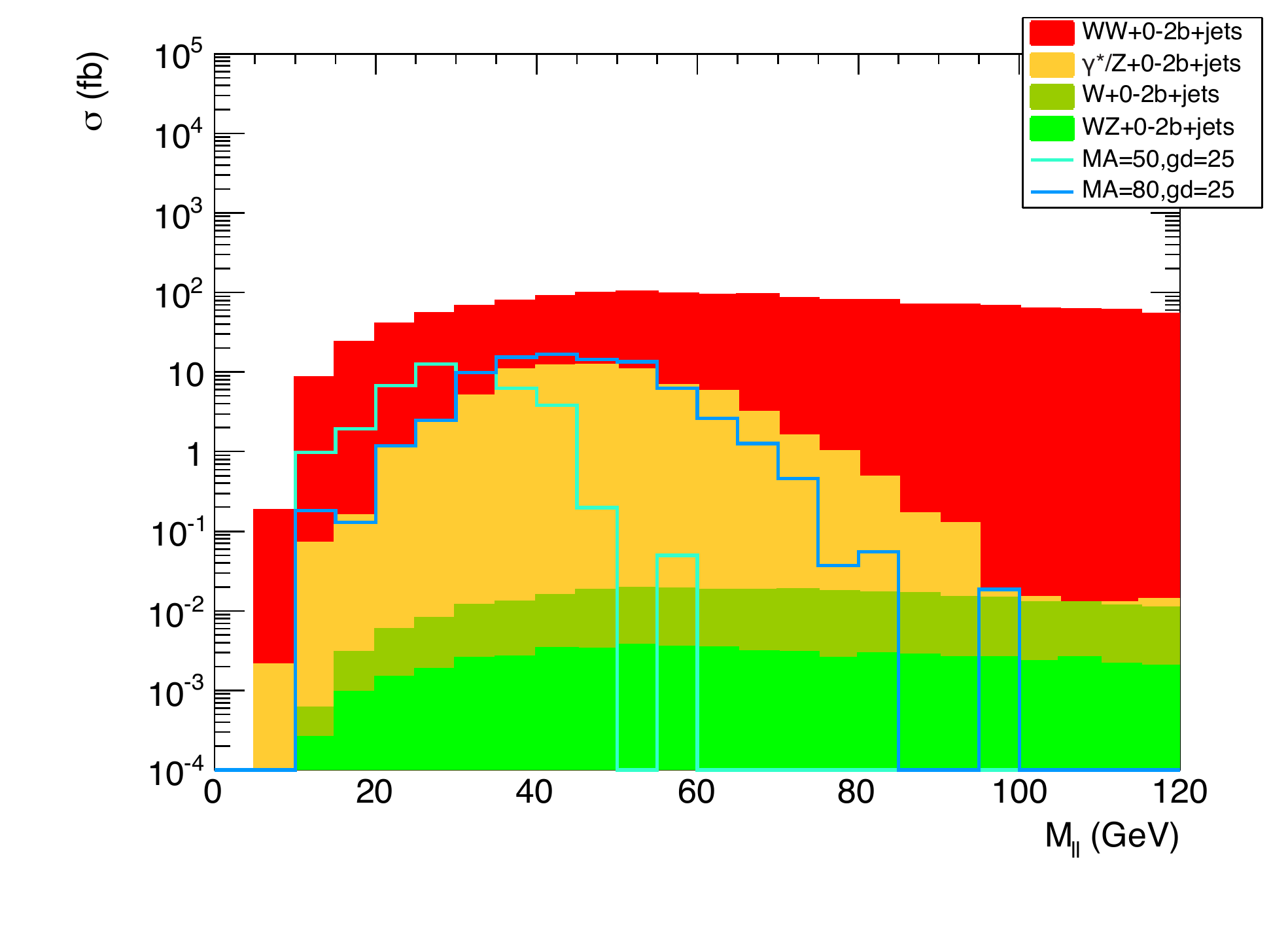}
\includegraphics[width=0.31\textwidth]{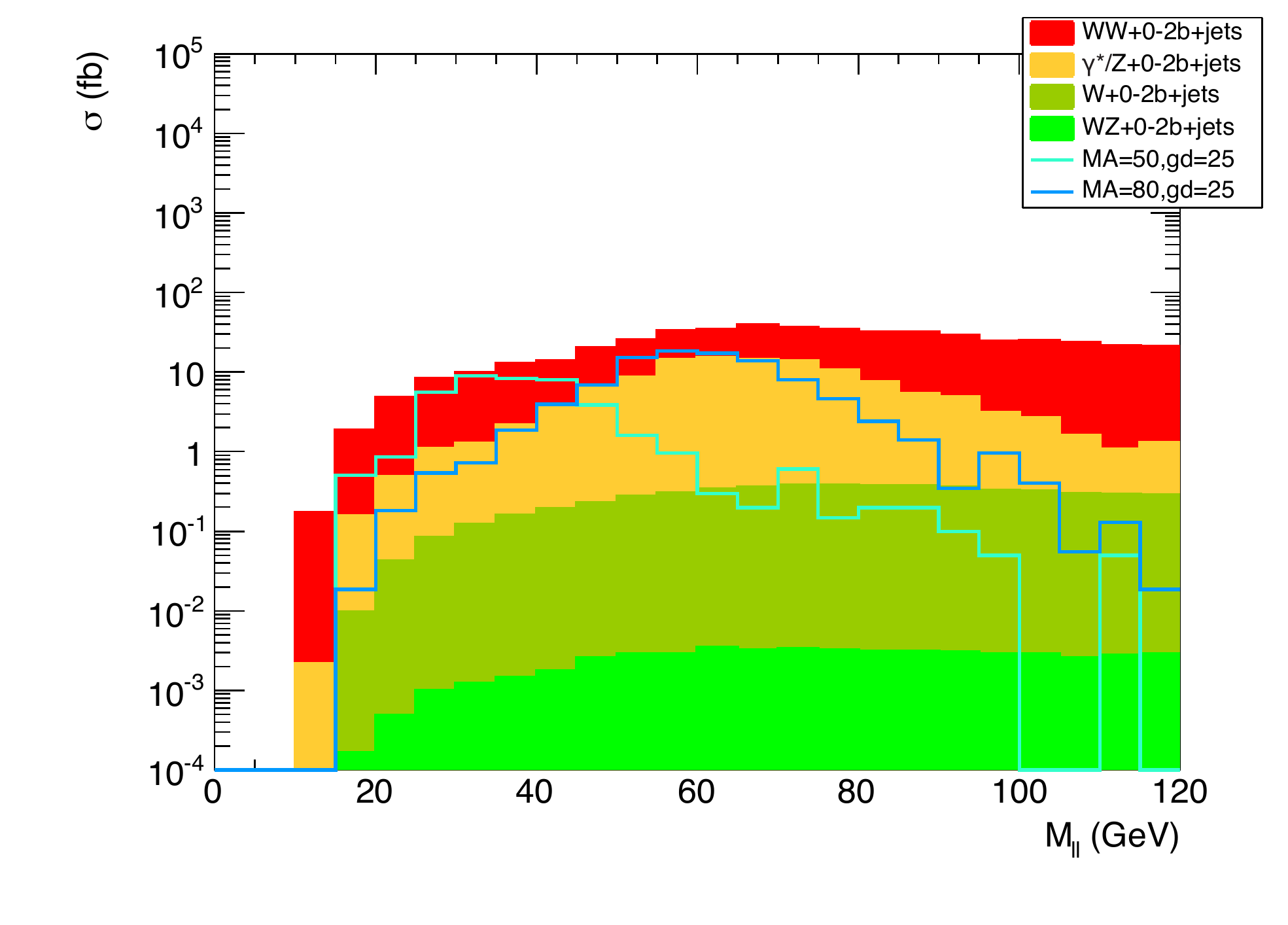}
\includegraphics[width=0.31\textwidth]{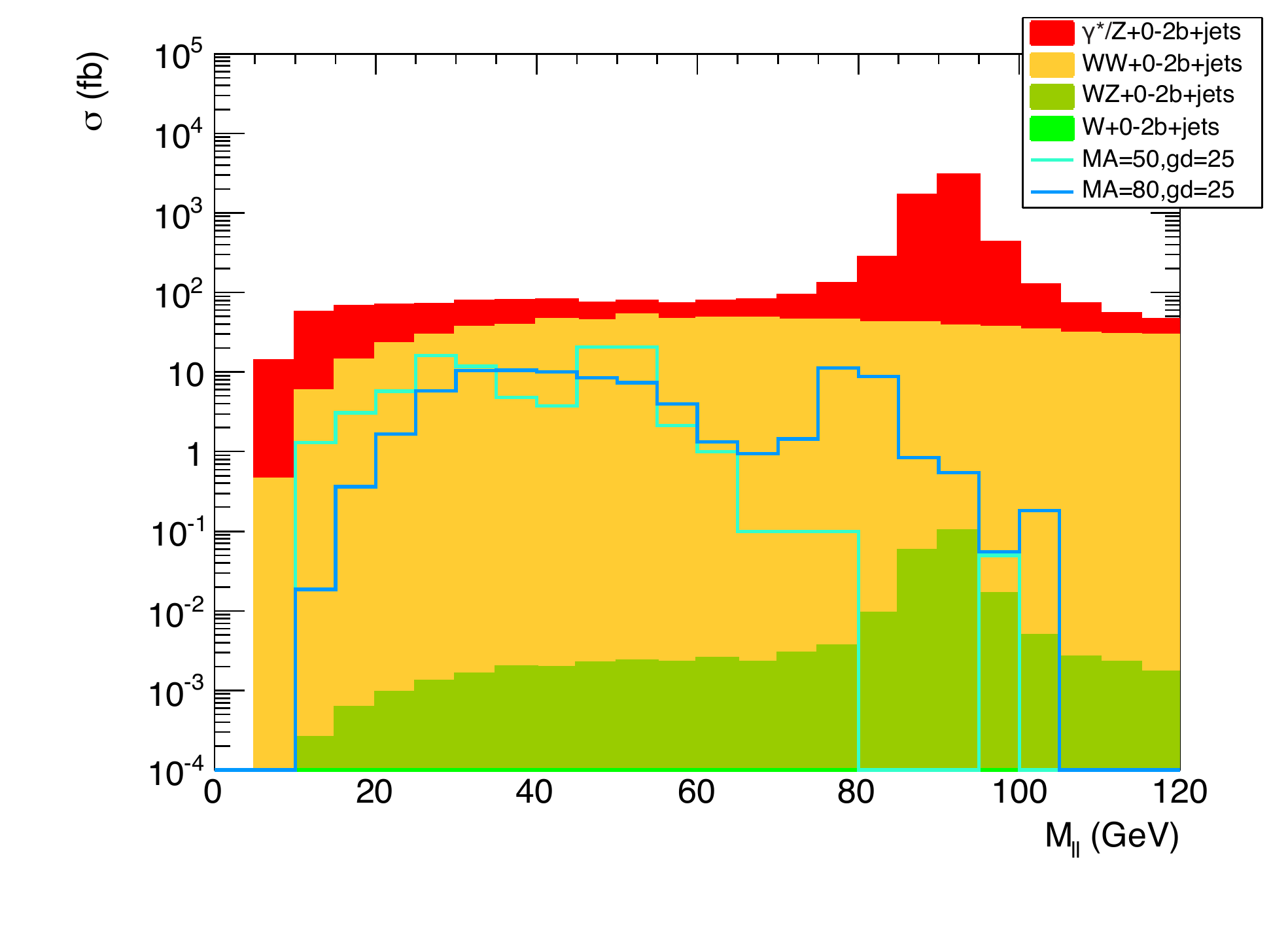}
   \end{center}
\caption{Dilepton invariant mass distribution. Trigger cuts and tagging are applied, but no other kinematic cuts are applied. For SR3 with $\mu^+\mu^-$ final states, the width of the $a$ is $\mathcal{O}(1)$~GeV, and thus the direct dimuon production peak is more pronounced than what is shown with 5~GeV bins. From left to right, figures are for SR1 ($1e1\mu$), SR2 ($1\ell1\tau_h$) and SR3 ($2\mu$).} \label{fig:llmass}
\end{figure}

\begin{figure}[!h]
   \begin{center}
\includegraphics[width=0.31\textwidth]{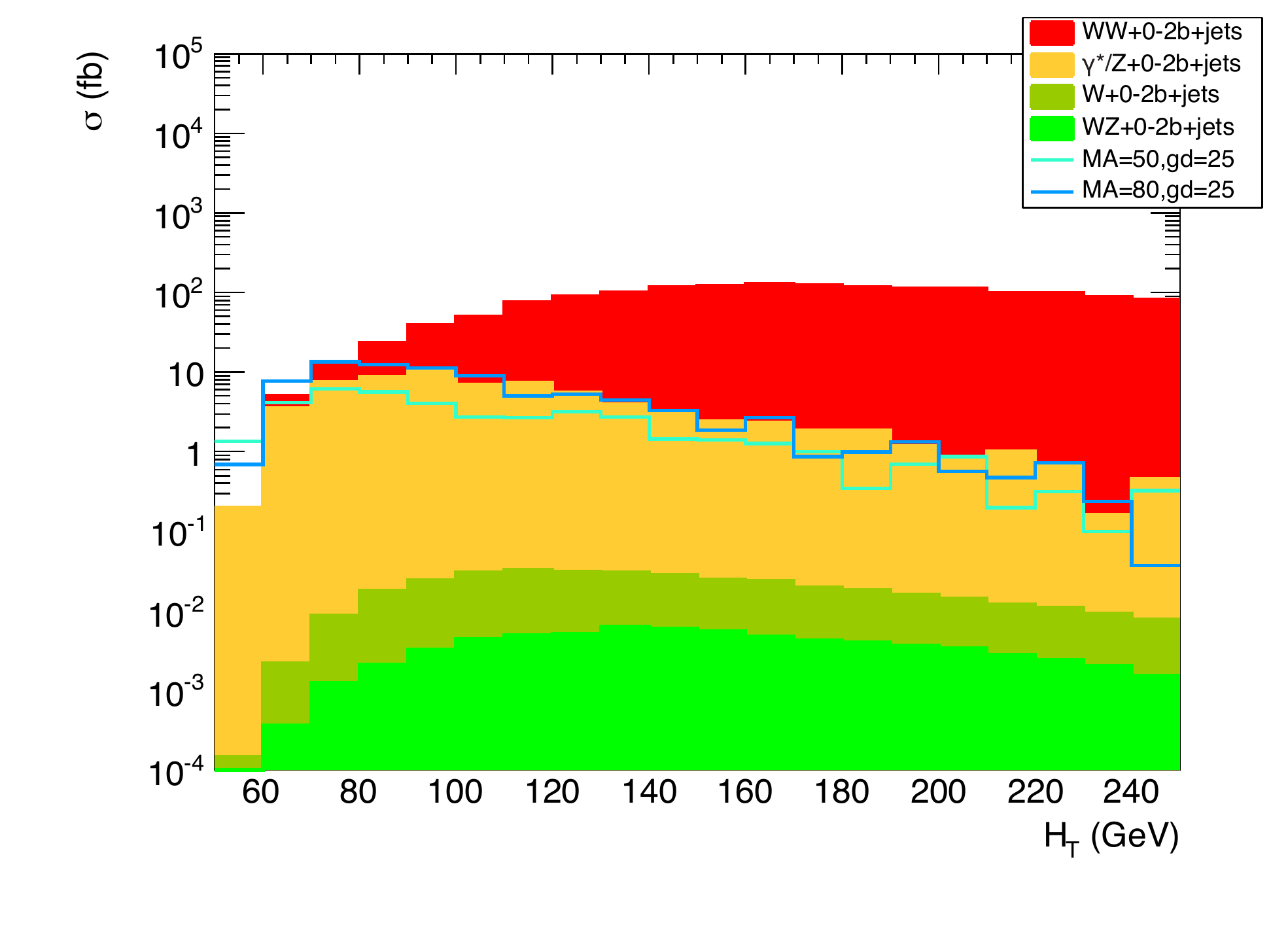}
\includegraphics[width=0.31\textwidth]{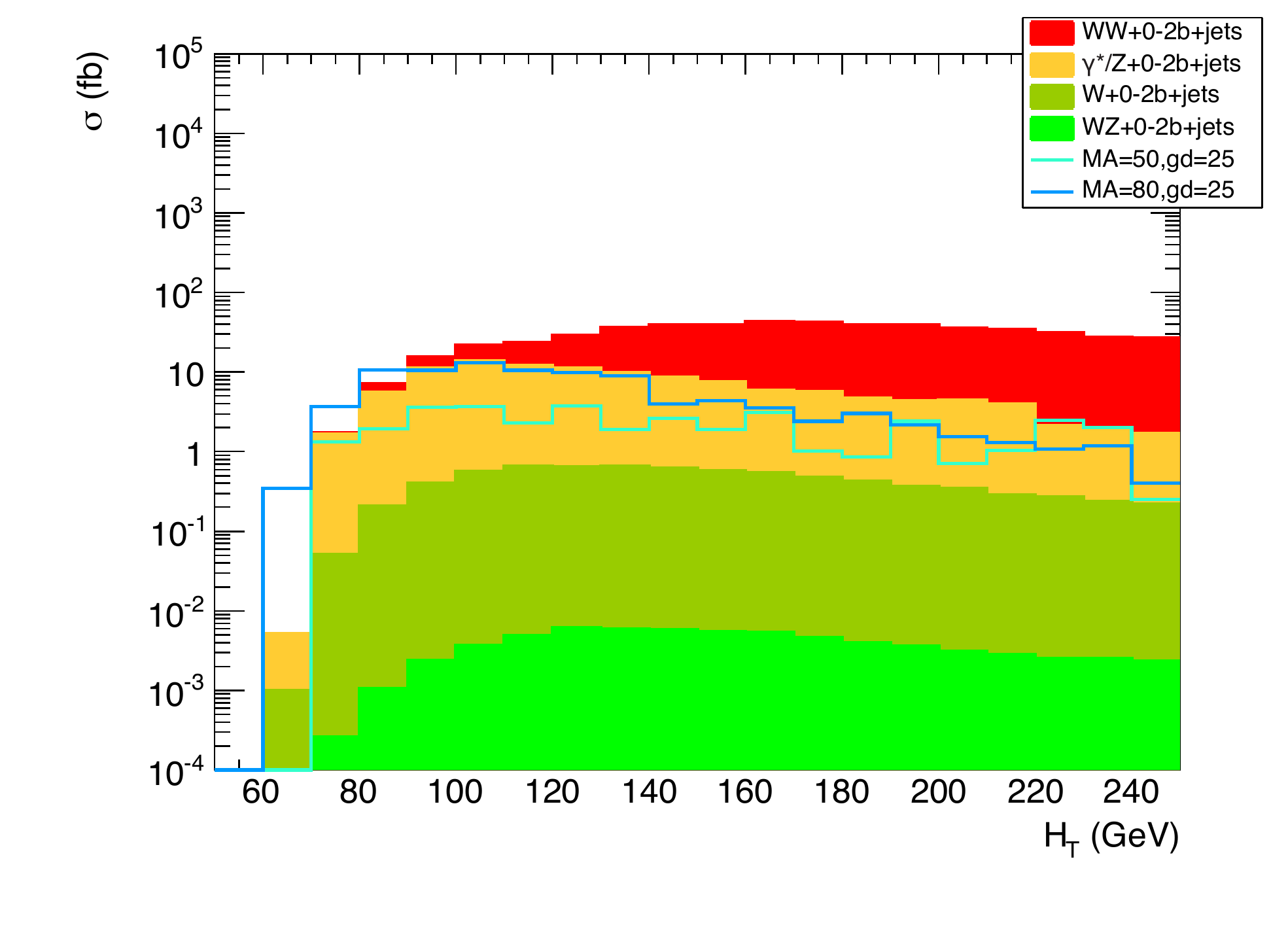}
\includegraphics[width=0.31\textwidth]{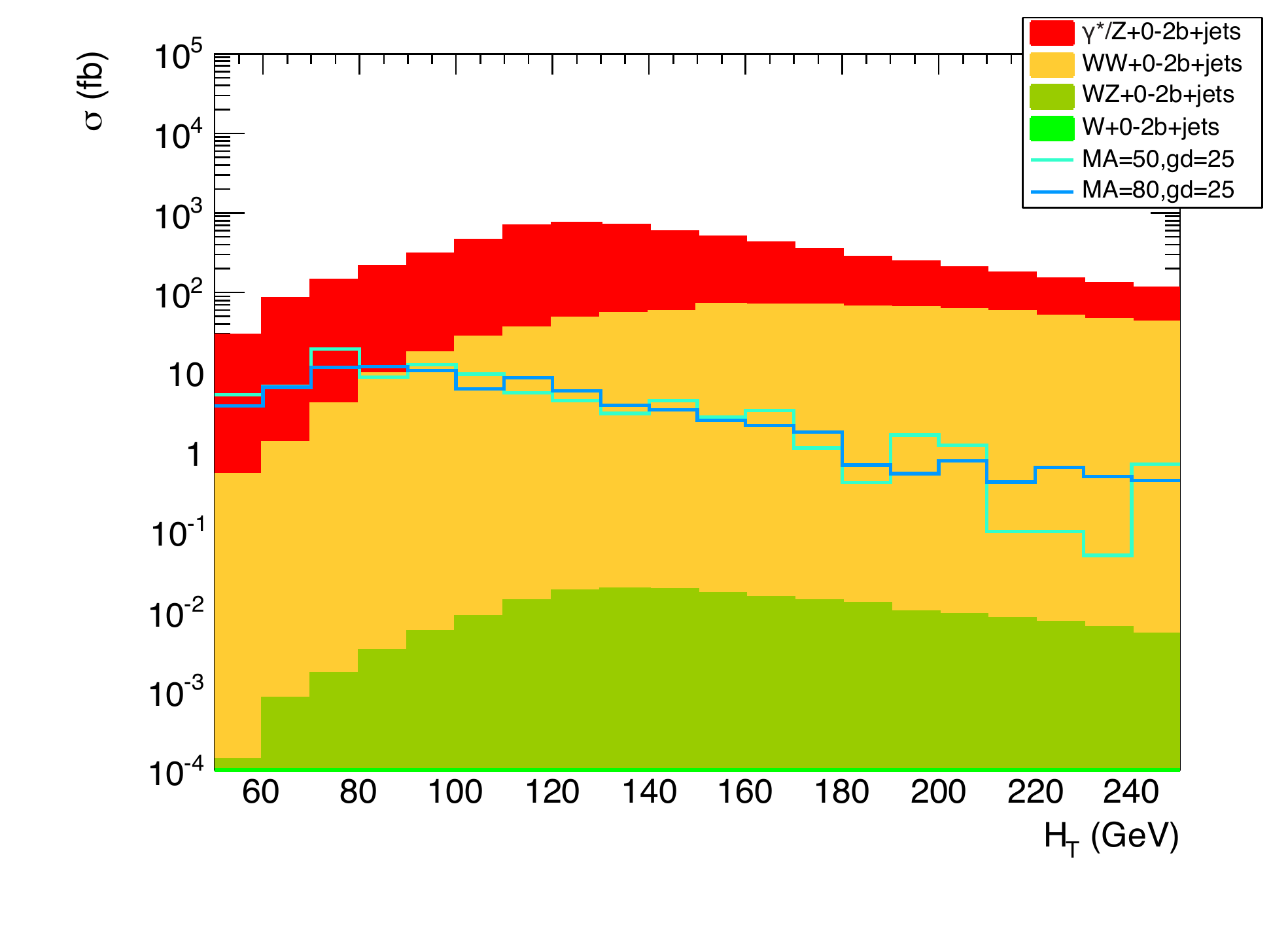}
   \end{center}
\caption{Scalar sum of visible transverse momenta distribution. Trigger cuts and tagging are applied, but no other kinematic cuts are applied. From left to right, figures are for SR1 ($1e1\mu$), SR2 ($1\ell1\tau_h$) and SR3 ($2\mu$).} \label{fig:ht}
\end{figure}

\begin{figure}[!h]
   \begin{center}
\includegraphics[width=0.31\textwidth]{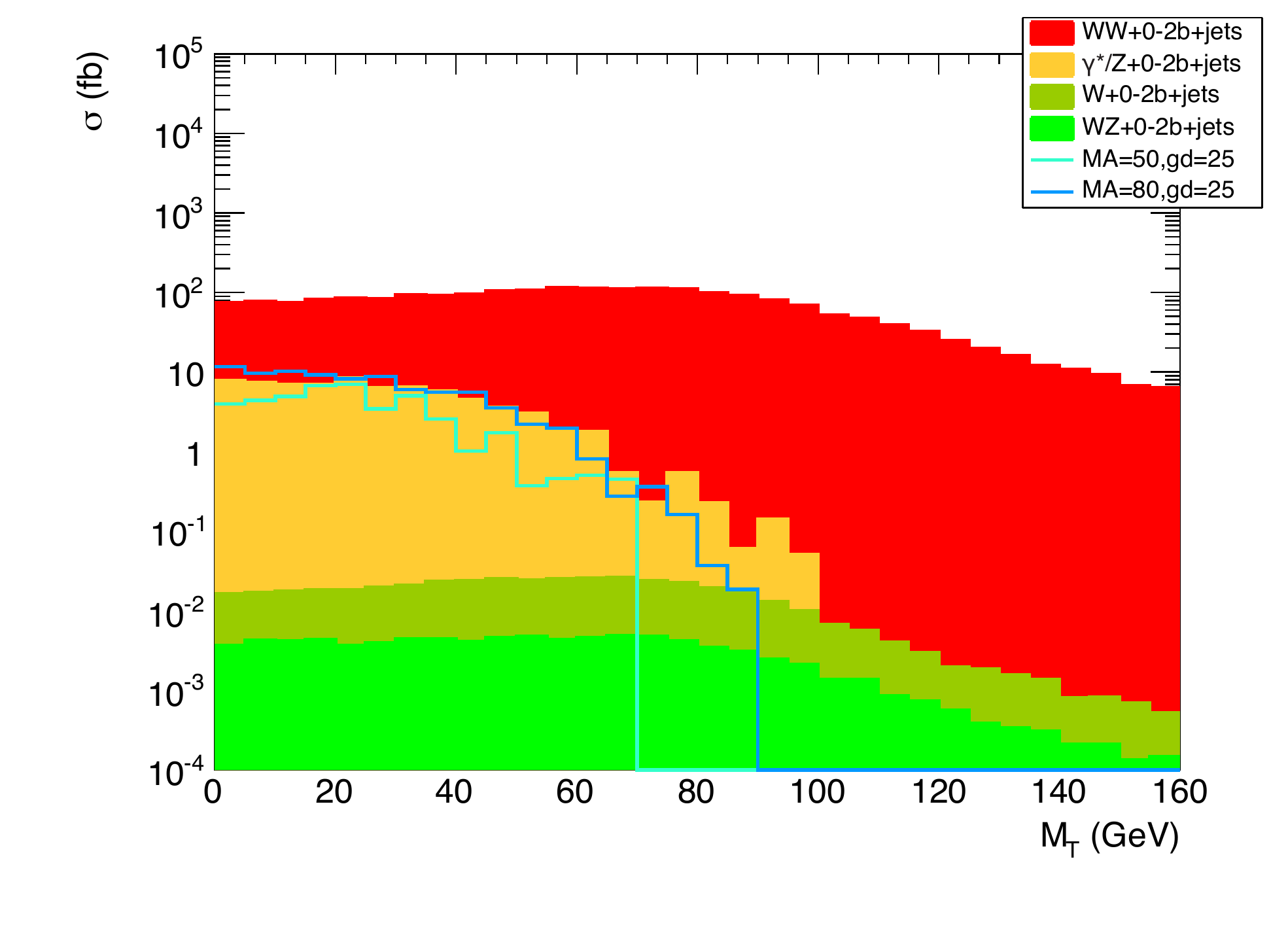}
\includegraphics[width=0.31\textwidth]{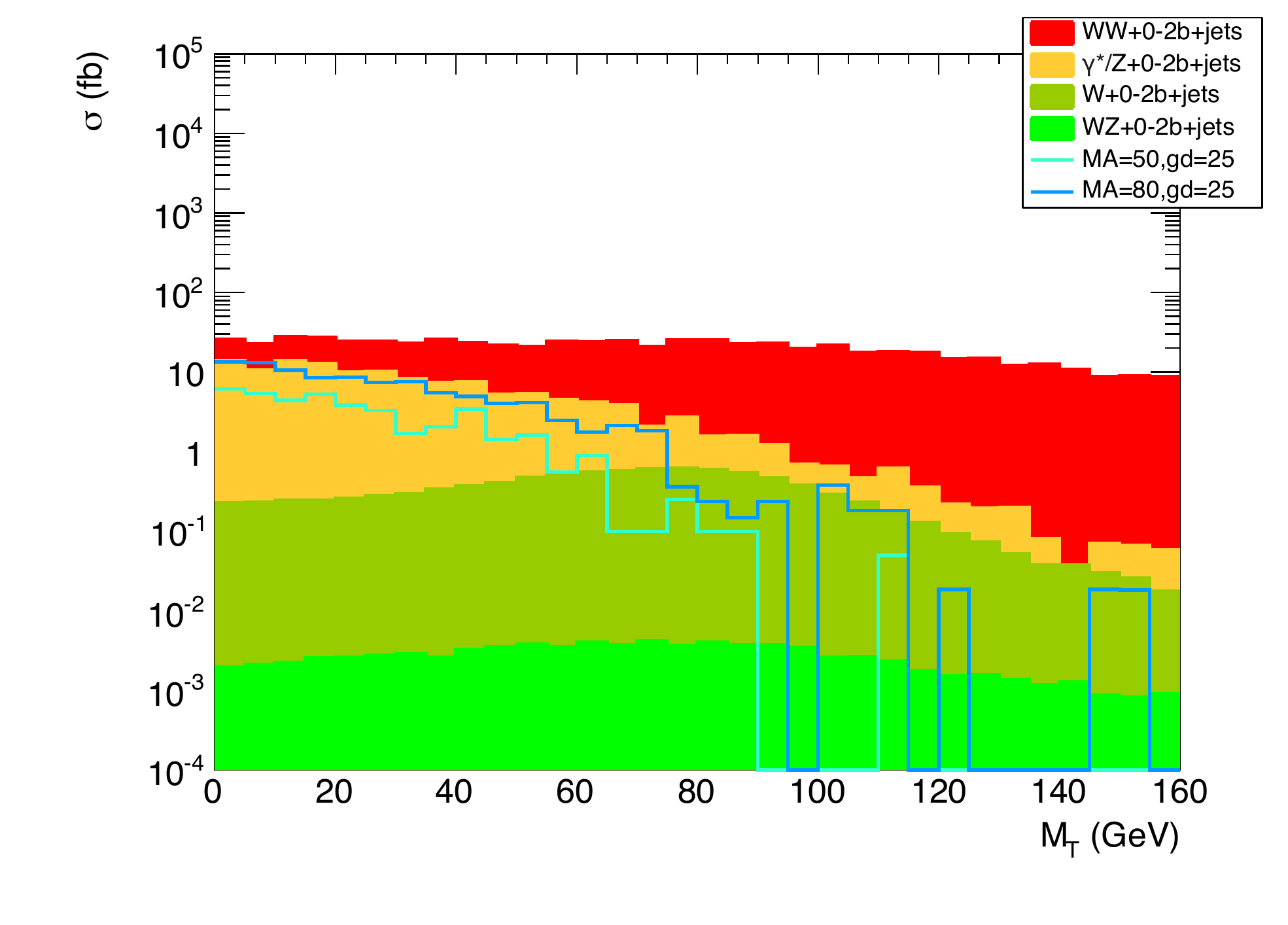}
\includegraphics[width=0.31\textwidth]{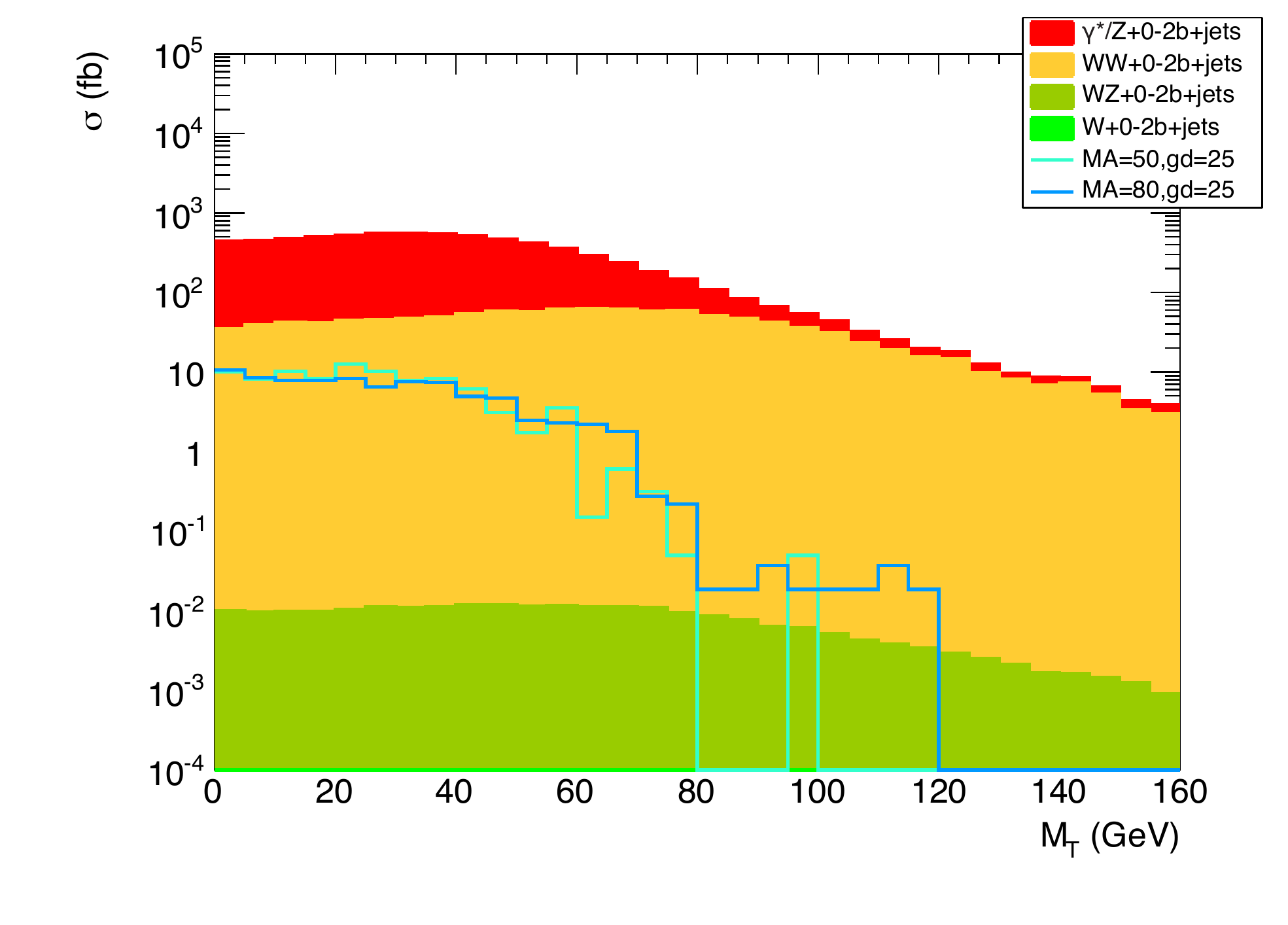}
   \end{center}
\caption{Transverse mass distribution. Trigger cuts and tagging are applied, but no other kinematic cuts are applied. From left to right, figures are for SR1 ($1e1\mu$), SR2 ($1\ell1\tau_h$) and SR3 ($2\mu$).} \label{fig:mtrans}
\end{figure}

\begin{figure}[!h]
   \begin{center}
\includegraphics[width=0.31\textwidth]{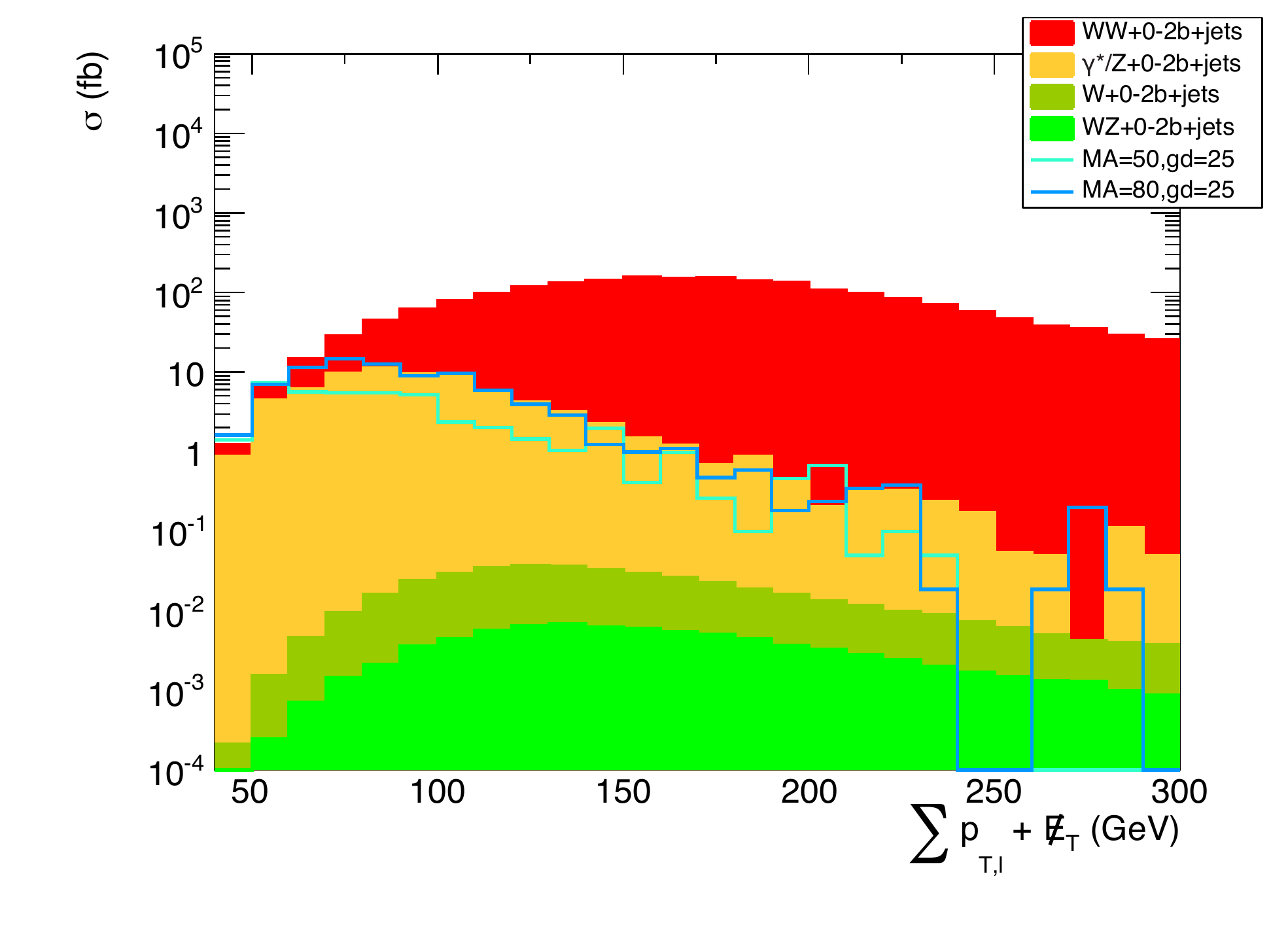}
\includegraphics[width=0.31\textwidth]{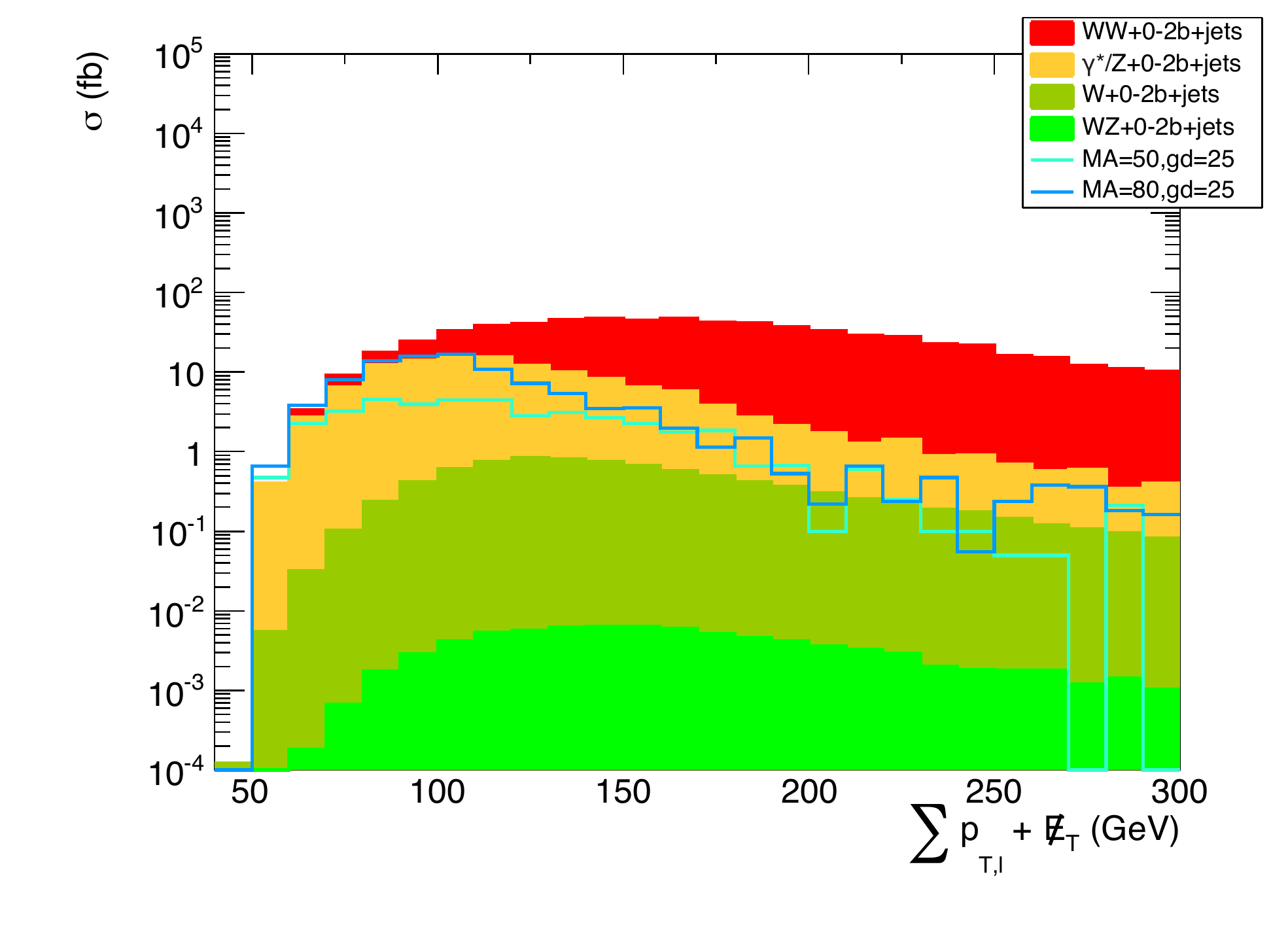}
\includegraphics[width=0.31\textwidth]{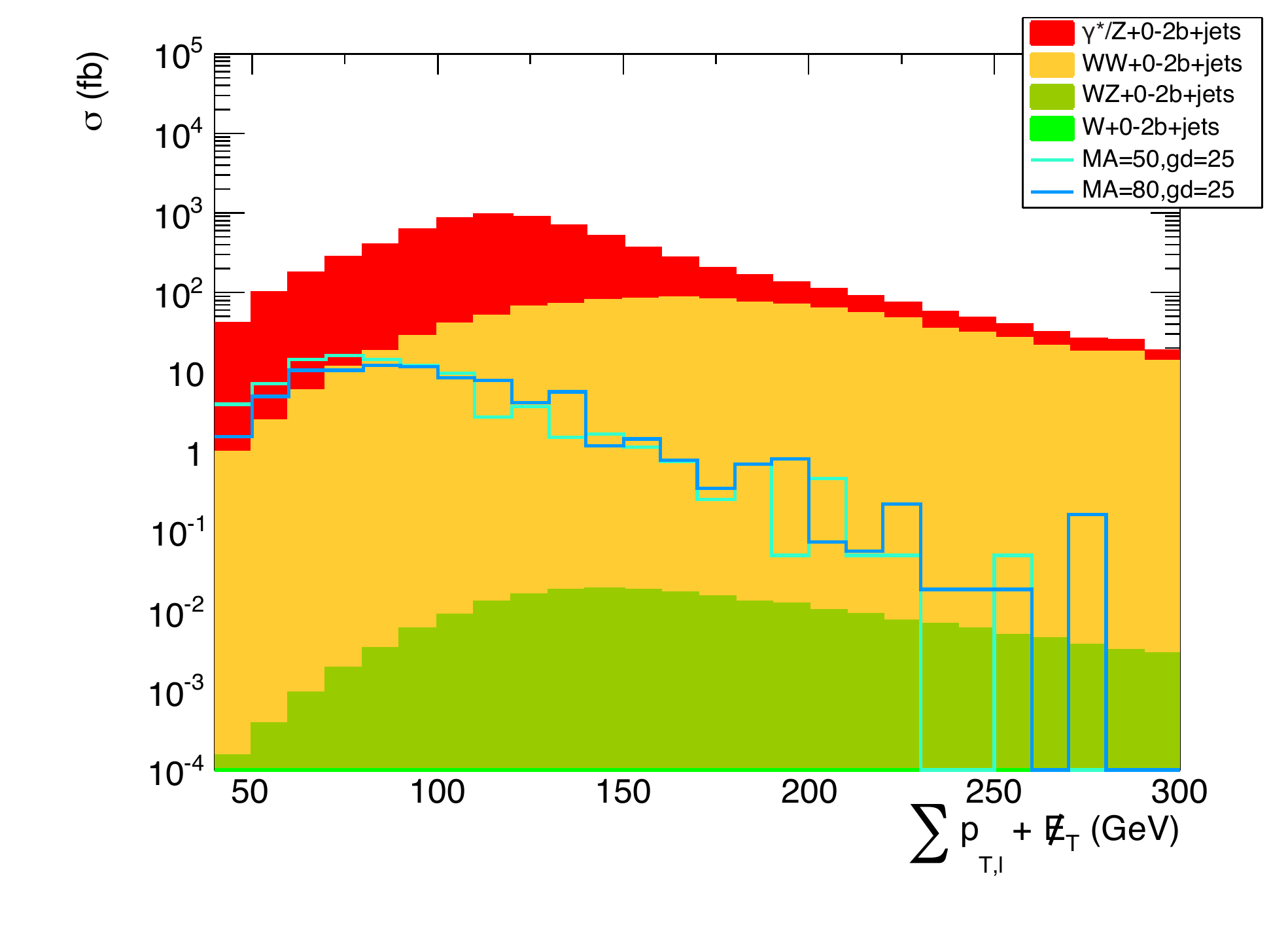}
   \end{center}
\caption{Scalar sum of lepton and missing transverse momenta distribution. Trigger cuts and tagging are applied, but no other kinematic cuts are applied. From left to right, figures are for SR1 ($1e1\mu$), SR2 ($1\ell1\tau_h$) and SR3 ($2\mu$).} \label{fig:ptlmet}
\end{figure}

\clearpage

\section{Cut Flow Matrices\label{sec:appb}}

When examining the potential of enhancing the visibility of the signal through cuts, we considered a variety of possible kinematic variable distributions, some of which are shown in Appendix \ref{sec:appa}. Of those examined, we chose to consider only those cuts in which the shape of the backgrounds was distinctly different than the shape of the signal for at least one of the signal regions so that cuts on the background had a larger fractional effect on the backgrounds than on the signal. 

The variables that most effectively improved the signal significance were $\met$, $p_{T}$ of the leading lepton ($p_{T_\ell}$), dilepton mass ($m_{\ell\ell}$), total scalar sum of visible momenta ($H_T$), transverse mass of the subleading lepton ($m_{T_{2nd\ell}}$), and the scalar sum of the lepton $p_T$ and $\met$ ($\not{\!\!H}_T^\ell$). Variables with $\met$ components were most effective at eliminating backgrounds containing decays of $W$ bosons, including $m_{T_{2nd\ell}}$ where backgrounds containing intermediate $W$'s have a longer tail on the distribution. Of note, we found that the transverse mass distribution based on the leading lepton $p_T$ had a longer tail for the signal, and so was not quite as effective. 

Since $\met$, for example, is a component of multiple cuts, we consider the correlation between the events passing each pair of cuts in cut flow matrices in Tables \ref{tab:matrixeuhard} through \ref{tab:matrixuusoft}. Diagonal entries are the acceptance rate for the single cut labeled in both the column and row headers, where red text indicates background acceptance rates and black text indicates signal acceptance rates. Each off-diagonal entry in these tables represents the acceptance rate ($A$) of the cut labeled by the row ($r$) header on the events remaining after performing the cut in the column ($c$) header, such that each entry is given by
\begin{equation}
A_{r,c}=A_r(A_c(\sigma))/A_c(\sigma).
\end{equation}
For example, the upper right most entry of Table \ref{tab:matrixeuhard} shows an 87.4\% acceptance rate for background events and 84.3\% acceptance rate for signal events from applying the $\met$ cut to the pool of events {\em that already passed the} $\not{\!\!H}_T^\ell$ cut. The lower left most entry shows that 17.2\% of background events and 87.7\% of signal events pass the $\not{\!\!H}_T^\ell$ cut after applying the $\met$ cut. Since the $\met$ cut removes a similar number of events for the signal and background once the $\not{\!\!H}_T^\ell$ cut has been applied, the $\met$ cut is superfluous once the $\not{\!\!H}_T^\ell$ cut has been applied, and thus should not be included in the final set of cuts. In fact, the $\met$ distribution for signals and backgrounds have similar shapes once the $\not{\!\!H}_T^\ell$ cut has been applied, and thus no $\met$ cut value will be effective.

\begin{table}[B]
\begin{center}

\begin{tabular}{ | l | c | c | c | c | c | c |}
\cline{2-7}
\multicolumn{1}{ c |}{} & \multicolumn{1}{ c | }{$\met$} & \multicolumn{1}{ c | }{$p_{T_\ell}$} & \multicolumn{1}{ c | }{$m_{\ell\ell}$} & \multicolumn{1}{ c | }{$H_T$} & \multicolumn{1}{ c | }{$m_{T_{2nd \ell}}$} & \multicolumn{1}{ c | }{$\not{\!\!H}_T^\ell$}	\\\hline
\multirow{2}{*}{$\met$}	& \cellcolor{black!8}\textcolor{red}{	0.130	} & \textcolor{red}{	0.158	} & \textcolor{red}{	0.115	} & \textcolor{red}{	0.335	} & \textcolor{red}{	0.272	} & \textcolor{red}{	0.874	}\\
	& \cellcolor{black!8}	0.431	 & 	0.523	 & 	0.246	 & 	0.619	 & 	0.452	 & 	0.843	\\\hline
\multirow{2}{*}{$p_{T_\ell}$}	& \textcolor{red}{	0.070	} & \cellcolor{black!8}\textcolor{red}{	0.058	} & \textcolor{red}{	0.203	} & \textcolor{red}{	0.512	} & \textcolor{red}{	0.088	} & \textcolor{red}{	0.451	}\\
	& 	0.596	 & \cellcolor{black!8}	0.490	 & 	0.483	 & 	0.742	 & 	0.477	 & 	0.689	\\\hline
\multirow{2}{*}{$m_{\ell\ell}$}	& \textcolor{red}{	0.076	} & \textcolor{red}{	0.305	} & \cellcolor{black!8}\textcolor{red}{	0.087	} & \textcolor{red}{	0.204	} & \textcolor{red}{	0.084	} & \textcolor{red}{	0.185	}\\
	& 	0.250	 & 	0.431	 & \cellcolor{black!8}	0.438	 & 	0.285	 & 	0.512	 & 	0.298	\\\hline
\multirow{2}{*}{$H_T$}	& \textcolor{red}{	0.050	} & \textcolor{red}{	0.172	} & \textcolor{red}{	0.046	} & \cellcolor{black!8}\textcolor{red}{	0.019	} & \textcolor{red}{	0.041	} & \textcolor{red}{	0.296	}\\
	& 	0.537	 & 	0.565	 & 	0.243	 & \cellcolor{black!8}	0.373	 & 	0.336	 & 	0.631	\\\hline
\multirow{2}{*}{$m_{T_{2nd \ell}}$}	& \textcolor{red}{	0.295	} & \textcolor{red}{	0.214	} & \textcolor{red}{	0.138	} & \textcolor{red}{	0.301	} & \cellcolor{black!8}\textcolor{red}{	0.141	} & \textcolor{red}{	0.478	}\\
	& 	0.582	 & 	0.539	 & 	0.648	 & 	0.499	 & \cellcolor{black!8}	0.554	 & 	0.538	\\\hline
\multirow{2}{*}{$\not{\!\!H}_T^\ell$}	& \textcolor{red}{	0.172	} & \textcolor{red}{	0.200	} & \textcolor{red}{	0.055	} & \textcolor{red}{	0.390	} & \textcolor{red}{	0.087	} & \cellcolor{black!8}\textcolor{red}{	0.026	}\\
	& 	0.877	 & 	0.629	 & 	0.648	 & 	0.756	 & 	0.435	 & \cellcolor{black!8}	0.448	\\\hline
\end{tabular}
\caption{Cut flow matrix for SR1: $1e1\mu + 1-2b + 0j$ signal with hard cuts. The cuts are: $\met < 30$~GeV, $p_T^{\ell_1} < 30$~GeV, $12 < m_{\ell\ell} < 35$~GeV, $H_T < 90$~GeV, $M_T^{\ell_2}<20$, $\not{\!\!H}_T^\ell < 80$. The red/top entry in each cell is the acceptance rate for all backgrounds combined, while the black/bottom entry shows the acceptance rate for the signal. The total cross sections for the $1e1\mu + 1-2b + 0j$ signal after applying the trigger cuts are $\sigma_{bkg} = 2187$ fb and $\sigma_{sig} = 60.4$ fb ($m_a=60$~GeV, $g_d=25$).
}
\label{tab:matrixeuhard}
\end{center}
\end{table}

\begin{table}
\begin{center}

\begin{tabular}{ | l | c | c | c | c | c | c |}
\cline{2-7}
\multicolumn{1}{ c |}{} & \multicolumn{1}{ c | }{$\met$} & \multicolumn{1}{ c | }{$p_{T_\ell}$} & \multicolumn{1}{ c | }{$m_{\ell\ell}$} & \multicolumn{1}{ c | }{$H_T$} & \multicolumn{1}{ c | }{$m_{T_{2nd \ell}}$} & \multicolumn{1}{ c | }{$\not{\!\!H}_T^\ell$}	\\\hline
\multirow{2}{*}{$\met$}	& \cellcolor{black!8}\textcolor{red}{	0.312	} & \textcolor{red}{	0.324	} & \textcolor{red}{	0.310	} & \textcolor{red}{	0.400	} & \textcolor{red}{	0.497	} & \textcolor{red}{	0.809	}\\
	& \cellcolor{black!8}	0.721	 & 	0.765	 & 	0.699	 & 	0.845	 & 	0.734	 & 	0.885	\\\hline
\multirow{2}{*}{$p_{T_\ell}$}	& \textcolor{red}{	0.188	} & \cellcolor{black!8}\textcolor{red}{	0.181	} & \textcolor{red}{	0.456	} & \textcolor{red}{	0.503	} & \textcolor{red}{	0.236	} & \textcolor{red}{	0.538	}\\
	& 	0.886	 & \cellcolor{black!8}	0.835	 & 	0.851	 & 	0.943	 & 	0.833	 & 	0.920	\\\hline
\multirow{2}{*}{$m_{\ell\ell}$}	& \textcolor{red}{	0.161	} & \textcolor{red}{	0.408	} & \cellcolor{black!8}\textcolor{red}{	0.162	} & \textcolor{red}{	0.320	} & \textcolor{red}{	0.187	} & \textcolor{red}{	0.336	}\\
	& 	0.850	 & 	0.893	 & \cellcolor{black!8}	0.877	 & 	0.871	 & 	0.898	 & 	0.863	\\\hline
\multirow{2}{*}{$H_T$}	& \textcolor{red}{	0.236	} & \textcolor{red}{	0.511	} & \textcolor{red}{	0.364	} & \cellcolor{black!8}\textcolor{red}{	0.184	} & \textcolor{red}{	0.249	} & \textcolor{red}{	0.535	}\\
	& 	0.909	 & 	0.876	 & 	0.771	 & \cellcolor{black!8}	0.776	 & 	0.764	 & 	0.923	\\\hline
\multirow{2}{*}{$m_{T_{2nd \ell}}$}	& \textcolor{red}{	0.487	} & \textcolor{red}{	0.398	} & \textcolor{red}{	0.353	} & \textcolor{red}{	0.413	} & \cellcolor{black!8}\textcolor{red}{	0.306	} & \textcolor{red}{	0.626	}\\
	& 	0.912	 & 	0.893	 & 	0.918	 & 	0.883	 & \cellcolor{black!8}	0.896	 & 	0.897	\\\hline
\multirow{2}{*}{$\not{\!\!H}_T^\ell$}	& \textcolor{red}{	0.404	} & \textcolor{red}{	0.463	} & \textcolor{red}{	0.323	} & \textcolor{red}{	0.453	} & \textcolor{red}{	0.319	} & \cellcolor{black!8}\textcolor{red}{	0.156	}\\
	& 	0.970	 & 	0.870	 & 	0.918	 & 	0.941	 & 	0.791	 & \cellcolor{black!8}	0.790	\\\hline
\end{tabular}
\caption{Cut flow matrix for SR1: $1e1\mu + 1-2b + 0j$ signal with soft cuts. The cuts are: $\met < 50$~GeV, $p_T^{\ell_1} < 40$~GeV, $12 < m_{\ell\ell} < 45$~GeV, $H_T < 140$~GeV, $M_T^{\ell_2}<40$, $\not{\!\!H}_T^\ell < 120$. The red/top entry in each cell is the acceptance rate for all backgrounds combined, while the black/bottom entry shows the acceptance rate for the signal. The total cross sections for the $1e1\mu + 1-2b + 0j$ signal after applying the trigger cuts are $\sigma_{bkg} = 2187$ fb and $\sigma_{sig} = 60.4$ fb ($m_a=60$~GeV, $g_d=25$).
}
\label{tab:matrixeusoft}
\end{center}
\end{table}

\begin{table}
\begin{center}

\begin{tabular}{ | l | c | c | c | c | c | c |}
\cline{2-7}
\multicolumn{1}{ c |}{} & \multicolumn{1}{ c | }{$\met$} & \multicolumn{1}{ c | }{$p_{T_\ell}$} & \multicolumn{1}{ c | }{$m_{\ell\ell}$} & \multicolumn{1}{ c | }{$H_T$} & \multicolumn{1}{ c | }{$m_{T_{2nd \ell}}$} & \multicolumn{1}{ c | }{$\not{\!\!H}_T^\ell$}	\\\hline
\multirow{2}{*}{$\met$}	& \cellcolor{black!8}\textcolor{red}{	0.170	} & \textcolor{red}{	0.221	} & \textcolor{red}{	0.151	} & \textcolor{red}{	0.389	} & \textcolor{red}{	0.328	} & \textcolor{red}{	0.814	}\\
	& \cellcolor{black!8}	0.396	 & 	0.441	 & 	0.267	 & 	0.488	 & 	0.320	 & 	0.715	\\\hline
\multirow{2}{*}{$p_{T_\ell}$}	& \textcolor{red}{	0.339	} & \cellcolor{black!8}\textcolor{red}{	0.262	} & \textcolor{red}{	0.517	} & \textcolor{red}{	0.644	} & \textcolor{red}{	0.403	} & \textcolor{red}{	0.723	}\\
	& 	0.671	 & \cellcolor{black!8}	0.602	 & 	0.570	 & 	0.840	 & 	0.602	 & 	0.857	\\\hline
\multirow{2}{*}{$m_{\ell\ell}$}	& \textcolor{red}{	0.063	} & \textcolor{red}{	0.140	} & \cellcolor{black!8}\textcolor{red}{	0.071	} & \textcolor{red}{	0.086	} & \textcolor{red}{	0.087	} & \textcolor{red}{	0.104	}\\
	& 	0.288	 & 	0.405	 & \cellcolor{black!8}	0.428	 & 	0.306	 & 	0.472	 & 	0.340	\\\hline
\multirow{2}{*}{$H_T$}	& \textcolor{red}{	0.328	} & \textcolor{red}{	0.353	} & \textcolor{red}{	0.173	} & \cellcolor{black!8}\textcolor{red}{	0.143	} & \textcolor{red}{	0.243	} & \textcolor{red}{	0.716	}\\
	& 	0.596	 & 	0.673	 & 	0.345	 & \cellcolor{black!8}	0.483	 & 	0.408	 & 	0.808	\\\hline
\multirow{2}{*}{$m_{T_{2nd \ell}}$}	& \textcolor{red}{	0.344	} & \textcolor{red}{	0.274	} & \textcolor{red}{	0.218	} & \textcolor{red}{	0.303	} & \cellcolor{black!8}\textcolor{red}{	0.178	} & \textcolor{red}{	0.451	}\\
	& 	0.401	 & 	0.495	 & 	0.546	 & 	0.419	 & \cellcolor{black!8}	0.496	 & 	0.395	\\\hline
\multirow{2}{*}{$\not{\!\!H}_T^\ell$}	& \textcolor{red}{	0.373	} & \textcolor{red}{	0.215	} & \textcolor{red}{	0.114	} & \textcolor{red}{	0.389	} & \textcolor{red}{	0.197	} & \cellcolor{black!8}\textcolor{red}{	0.078	}\\
	& 	0.759	 & 	0.597	 & 	0.546	 & 	0.702	 & 	0.334	 & \cellcolor{black!8}	0.420	\\\hline
\end{tabular}
\caption{Cut flow matrix for SR2: $1\ell1\tau + 1-2b + 0j$ signal with hard cuts. The cuts are: $\met < 30$~GeV, $p_T^{\ell_1} < 40$~GeV, $12 < m_{\ell\ell} < 45$~GeV, $H_T < 130$~GeV, $M_T^{\ell_2}<25$, $\not{\!\!H}_T^\ell < 100$. The red/top entry in each cell is the acceptance rate for all backgrounds combined, while the black/bottom entry shows the acceptance rate for the signal. The total cross sections for the $1\ell1\tau + 1-2b + 0j$ signal after applying the trigger cuts are $\sigma_{bkg} = 742$ fb and $\sigma_{sig} = 84.3$ fb ($m_a=60$~GeV, $g_d=25$).
}
\label{tab:matrixlthard}
\end{center}
\end{table}

\begin{table}
\begin{center}

\begin{tabular}{ | l | c | c | c | c | c | c |}
\cline{2-7}
\multicolumn{1}{ c |}{} & \multicolumn{1}{ c | }{$\met$} & \multicolumn{1}{ c | }{$p_{T_\ell}$} & \multicolumn{1}{ c | }{$m_{\ell\ell}$} & \multicolumn{1}{ c | }{$H_T$} & \multicolumn{1}{ c | }{$m_{T_{2nd \ell}}$} & \multicolumn{1}{ c | }{$\not{\!\!H}_T^\ell$}	\\\hline
\multirow{2}{*}{$\met$}	& \cellcolor{black!8}\textcolor{red}{	0.422	} & \textcolor{red}{	0.456	} & \textcolor{red}{	0.462	} & \textcolor{red}{	0.530	} & \textcolor{red}{	0.642	} & \textcolor{red}{	0.860	}\\
	& \cellcolor{black!8}	0.778	 & 	0.821	 & 	0.762	 & 	0.874	 & 	0.776	 & 	0.912	\\\hline
\multirow{2}{*}{$p_{T_\ell}$}	& \textcolor{red}{	0.564	} & \cellcolor{black!8}\textcolor{red}{	0.521	} & \textcolor{red}{	0.844	} & \textcolor{red}{	0.717	} & \textcolor{red}{	0.693	} & \textcolor{red}{	0.817	}\\
	& 	0.899	 & \cellcolor{black!8}	0.851	 & 	0.850	 & 	0.944	 & 	0.871	 & 	0.955	\\\hline
\multirow{2}{*}{$m_{\ell\ell}$}	& \textcolor{red}{	0.198	} & \textcolor{red}{	0.293	} & \cellcolor{black!8}\textcolor{red}{	0.181	} & \textcolor{red}{	0.257	} & \textcolor{red}{	0.255	} & \textcolor{red}{	0.334	}\\
	& 	0.864	 & 	0.880	 & \cellcolor{black!8}	0.882	 & 	0.872	 & 	0.899	 & 	0.874	\\\hline
\multirow{2}{*}{$H_T$}	& \textcolor{red}{	0.587	} & \textcolor{red}{	0.642	} & \textcolor{red}{	0.663	} & \cellcolor{black!8}\textcolor{red}{	0.467	} & \textcolor{red}{	0.575	} & \textcolor{red}{	0.816	}\\
	& 	0.878	 & 	0.867	 & 	0.772	 & \cellcolor{black!8}	0.781	 & 	0.795	 & 	0.957	\\\hline
\multirow{2}{*}{$m_{T_{2nd \ell}}$}	& \textcolor{red}{	0.467	} & \textcolor{red}{	0.408	} & \textcolor{red}{	0.433	} & \textcolor{red}{	0.378	} & \cellcolor{black!8}\textcolor{red}{	0.307	} & \textcolor{red}{	0.572	}\\
	& 	0.766	 & 	0.786	 & 	0.782	 & 	0.781	 & \cellcolor{black!8}	0.767	 & 	0.777	\\\hline
\multirow{2}{*}{$\not{\!\!H}_T^\ell$}	& \textcolor{red}{	0.596	} & \textcolor{red}{	0.458	} & \textcolor{red}{	0.539	} & \textcolor{red}{	0.511	} & \textcolor{red}{	0.545	} & \cellcolor{black!8}\textcolor{red}{	0.292	}\\
	& 	0.889	 & 	0.851	 & 	0.782	 & 	0.930	 & 	0.768	 & \cellcolor{black!8}	0.759	\\\hline
\end{tabular}
\caption{Cut flow matrix for SR2: $1\ell1\tau + 1-2b + 0j$ signal with soft cuts. The cuts are: $\met < 55$~GeV, $p_T^{\ell_1} < 55$~GeV, $12 < m_{\ell\ell} < 60$~GeV, $H_T < 190$~GeV, $M_T^{\ell_2}<45$, $\not{\!\!H}_T^\ell < 140$. The red/top entry in each cell is the acceptance rate for all backgrounds combined, while the black/bottom entry shows the acceptance rate for the signal. The total cross sections for the $1\ell1\tau + 1-2b + 0j$ signal after applying the trigger cuts are $\sigma_{bkg} = 742$ fb and $\sigma_{sig} = 84.3$ fb ($m_a=60$~GeV, $g_d=25$).
}
\label{tab:matrixltsoft} 
\end{center}
\end{table}

\begin{table}
\begin{center}

\begin{tabular}{ | l | c | c | c | c | c |}
\cline{2-6}
\multicolumn{1}{ c |}{} & \multicolumn{1}{ c | }{$\met$} & \multicolumn{1}{ c | }{$p_{T_\ell}$} & \multicolumn{1}{ c | }{$H_T$} & \multicolumn{1}{ c | }{$m_{T_{2nd \ell}}$} & \multicolumn{1}{ c | }{$\not{\!\!H}_T^\ell$}	\\\hline
\multirow{2}{*}{$\met$}	& \cellcolor{black!8}\textcolor{red}{	0.882	} & \textcolor{red}{	0.883	} & \textcolor{red}{	0.962	} & \textcolor{red}{	0.950	} & \textcolor{red}{	0.992	}\\
	& \cellcolor{black!8}	0.888	 & 	0.902	 & 	0.950	 & 	0.885	 & 	0.962	\\\hline
\multirow{2}{*}{$p_{T_\ell}$}	& \textcolor{red}{	0.317	} & \cellcolor{black!8}\textcolor{red}{	0.317	} & \textcolor{red}{	0.770	} & \textcolor{red}{	0.349	} & \textcolor{red}{	0.546	}\\
	& 	0.878	 & \cellcolor{black!8}	0.864	 & 	0.969	 & 	0.859	 & 	0.907	\\\hline
\multirow{2}{*}{$H_T$}	& \textcolor{red}{	0.288	} & \textcolor{red}{	0.642	} & \cellcolor{black!8}\textcolor{red}{	0.264	} & \textcolor{red}{	0.318	} & \textcolor{red}{	0.496	}\\
	& 	0.815	 & 	0.855	 & \cellcolor{black!8}	0.762	 & 	0.767	 & 	0.853	\\\hline
\multirow{2}{*}{$m_{T_{2nd \ell}}$}	& \textcolor{red}{	0.688	} & \textcolor{red}{	0.703	} & \textcolor{red}{	0.767	} & \cellcolor{black!8}\textcolor{red}{	0.639	} & \textcolor{red}{	0.818	}\\
	& 	0.858	 & 	0.856	 & 	0.866	 & \cellcolor{black!8}	0.860	 & 	0.887	\\\hline
\multirow{2}{*}{$\not{\!\!H}_T^\ell$}	& \textcolor{red}{	0.532	} & \textcolor{red}{	0.815	} & \textcolor{red}{	0.888	} & \textcolor{red}{	0.606	} & \cellcolor{black!8}\textcolor{red}{	0.473	}\\
	& 	0.947	 & 	0.918	 & 	0.979	 & 	0.901	 & \cellcolor{black!8}	0.874	\\\hline
\end{tabular}
\caption{Cut flow matrix for SR3: $2\mu + 1-2b + 0j$ signal. The cuts are: $\met < 60$~GeV, $p_T^{\ell_1} < 50$~GeV, $H_T < 120$~GeV, $M_T^{\ell_2}<45$, $\not{\!\!H}_T^\ell < 120$. The red/top entry in each cell is the acceptance rate for all backgrounds combined, while the black/bottom entry shows the acceptance rate for the signal. The total cross sections for the $2\mu + 1-2b + 0j$ signal after applying the trigger cuts are $\sigma_{bkg} = 7249$ fb and $\sigma_{sig} = 108$ fb ($m_a=60$~GeV, $g_d=25$).
}
\label{tab:matrixuusoft} 
\end{center}
\end{table}

\clearpage

\section{Variation of Exclusions}\label{sec:appc}

As discussed in section \ref{sec:production}, calculations of signal events using the 5FS is quite strongly dependent on the factorization and renormalization scales used. In \texttt{MadGraph5}, we employed a dynamic scale scheme that we then varied by an overall scaling factor between 0.5 and 2.0 (see Eq.~\ref{eq:scale}). This factor had the largest effect for low mass pseudoscalar calculations, with a factor of 0.5 reducing the total cross section by approximately 22\% for $m_a=20$~GeV, while only reducing the total cross section by a factor of 4\% at $m_a=80$~GeV. Alternatively, the authors of \cite{Dawson:2005vi} use a fixed renormalization and factorization scale scheme based on the sum of the masses of the pseudoscalar and the on-shell $b$ quark masses. Variations of this scale by a factor between 0.5 and 2.0 results in a cross section reduced by as much as 50\%.

\begin{figure}[!h]
   \begin{center}
\includegraphics[width=0.45\textwidth]{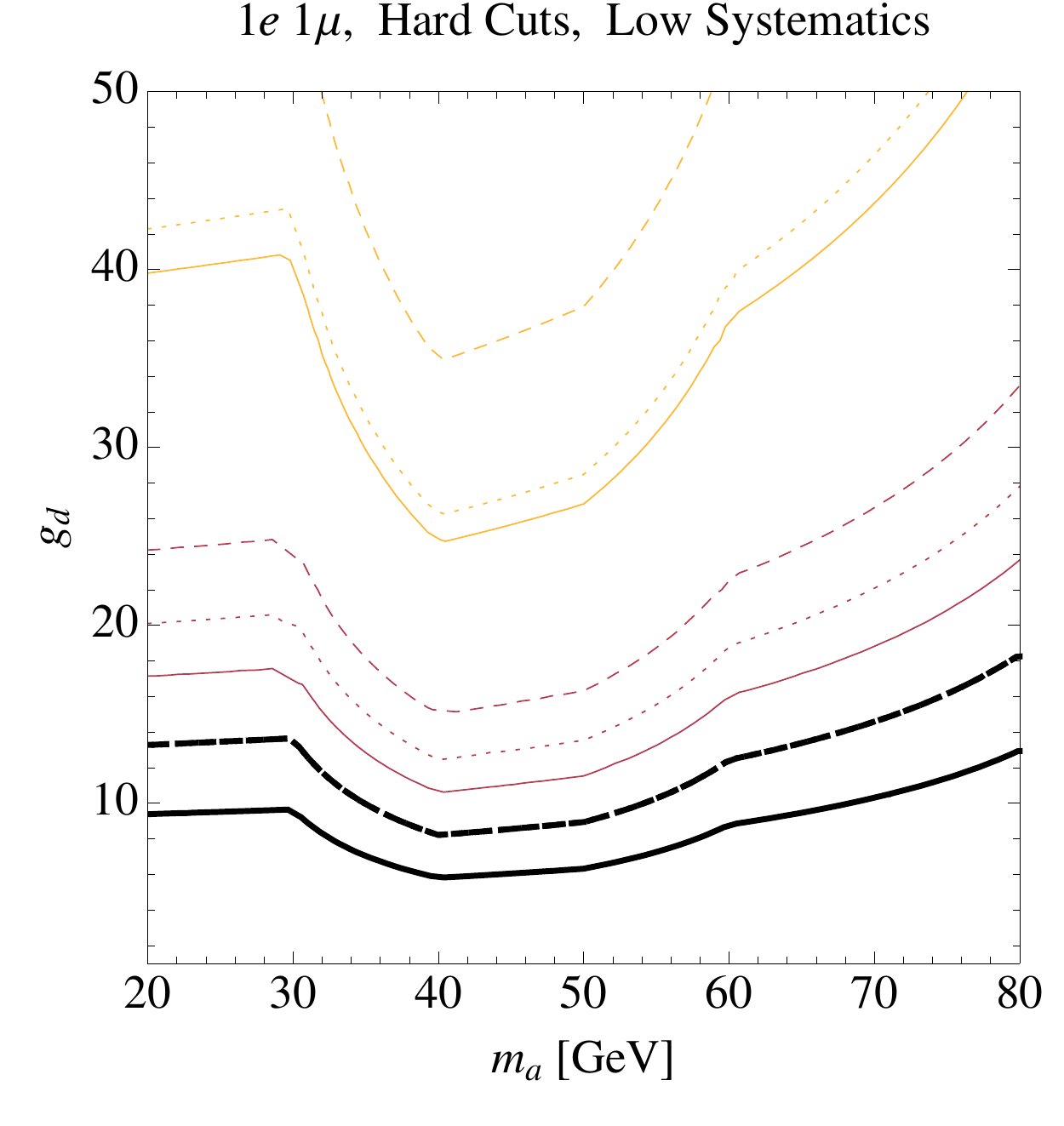}
\includegraphics[width=0.45\textwidth]{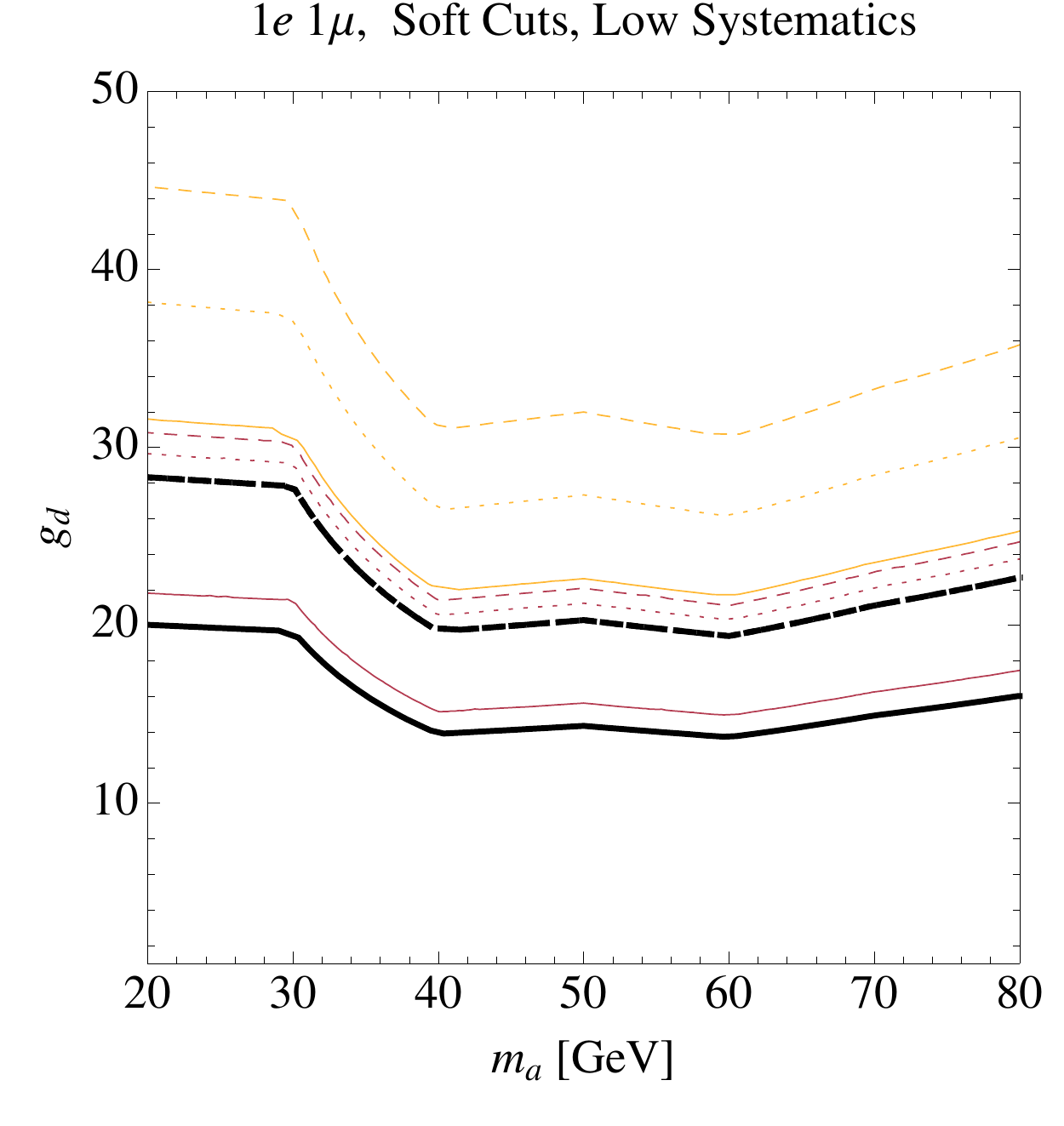}\\
\includegraphics[width=0.45\textwidth]{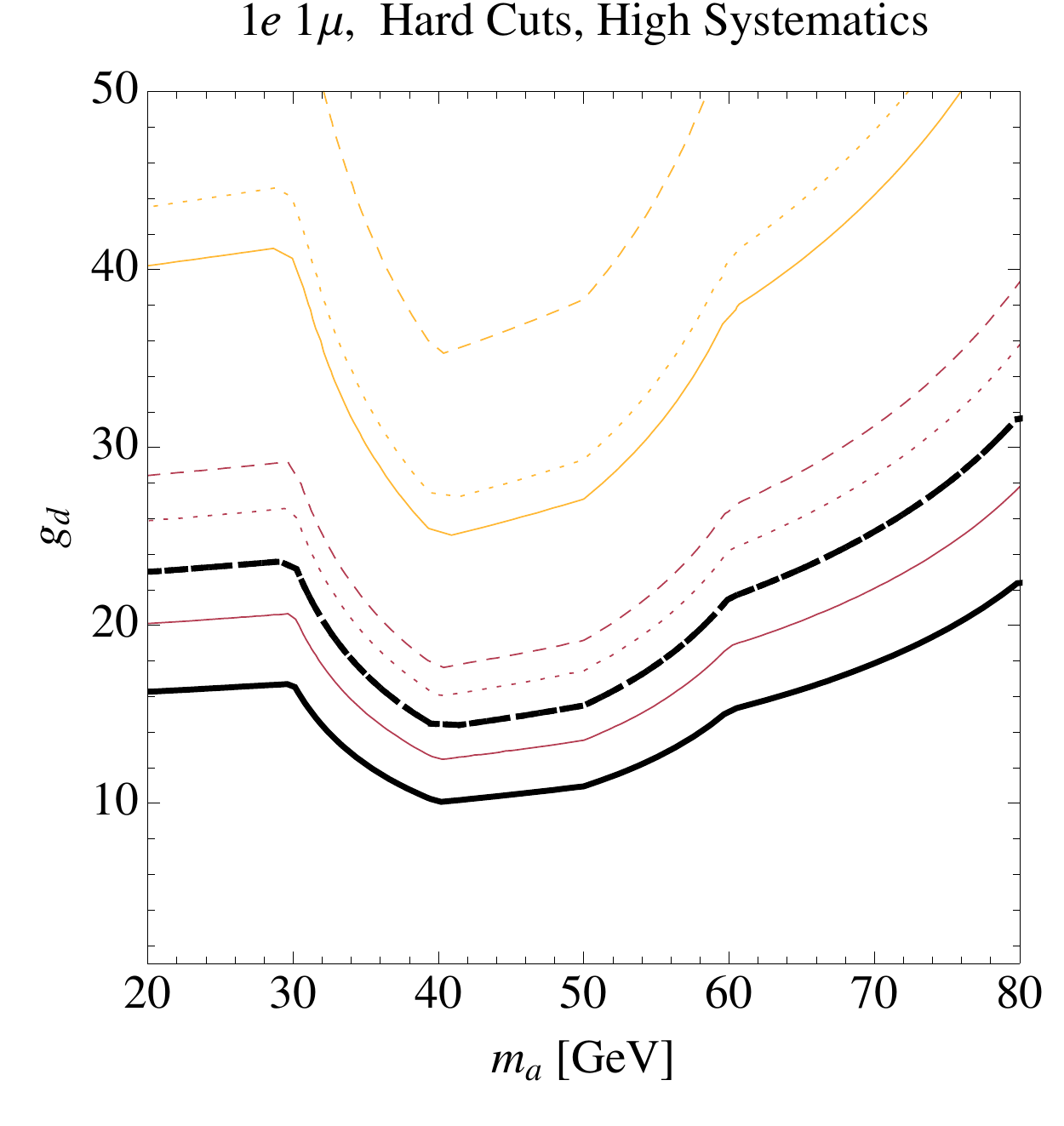}
\includegraphics[width=0.45\textwidth]{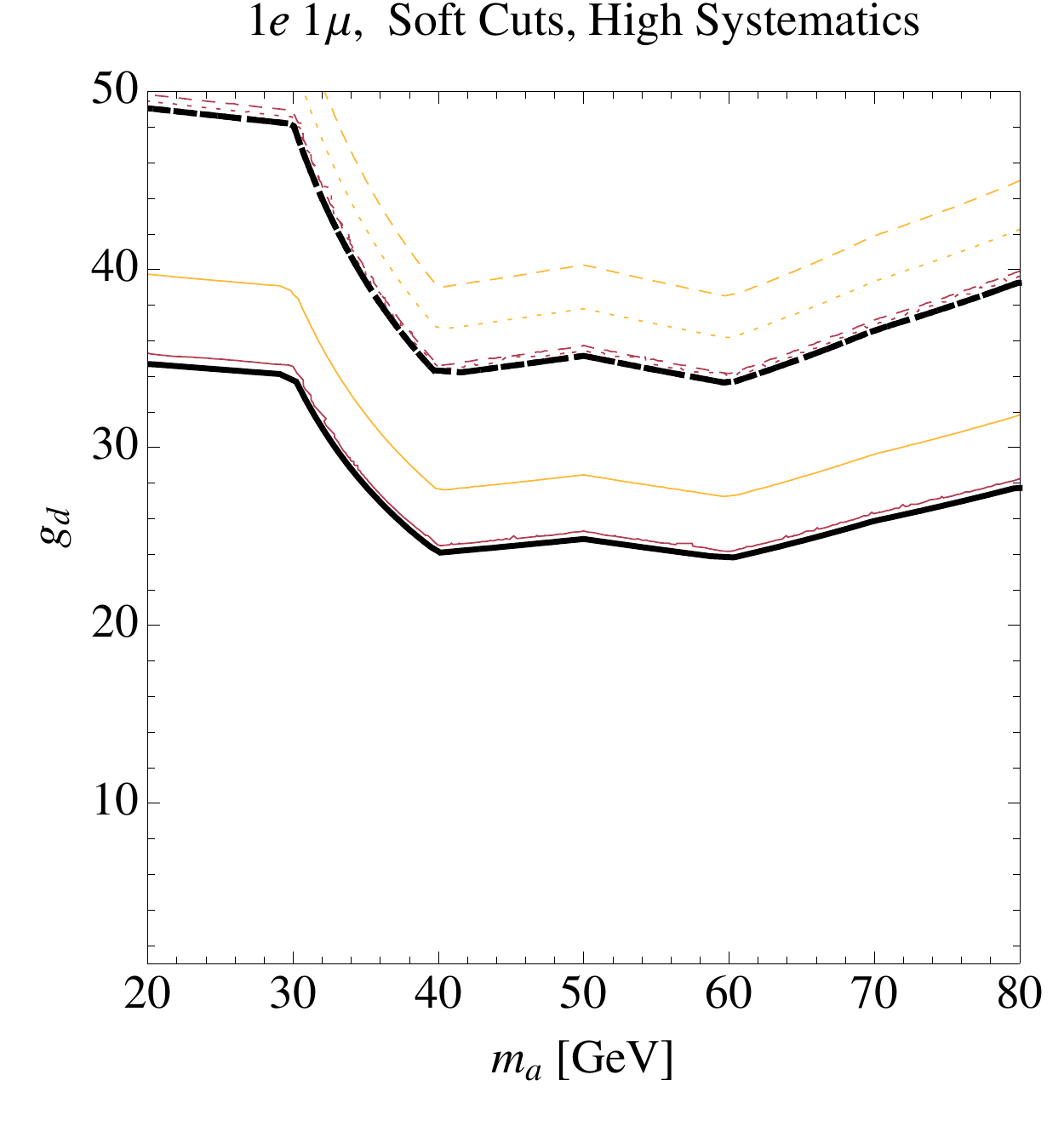}
   \end{center}
\caption{Discovery potential contours for the SR1 ($1e1\mu$) signal region with hard (left) and soft (right) cuts for $\epsilon_{sys}=0.1 (0.3)$ (top (bottom)) and conservative factors applied to the signal (dotted) and backgrounds (dashed) (solid lines show the original bounds without any factor applied to the signal or background). Contours correspond to constant values of $\log(L\times \mathrm{fb})$ needed to achieve $k=3$. The black lines represent the boundary of the systematics dominated region, the red lines represent the discovery potential at $L$=10/fb, while the yellow lines represent the discovery potential for $L$=1/fb.} \label{fig:1e1udisc_k}
\end{figure}

\begin{figure}[!h]
   \begin{center}
\includegraphics[width=0.45\textwidth]{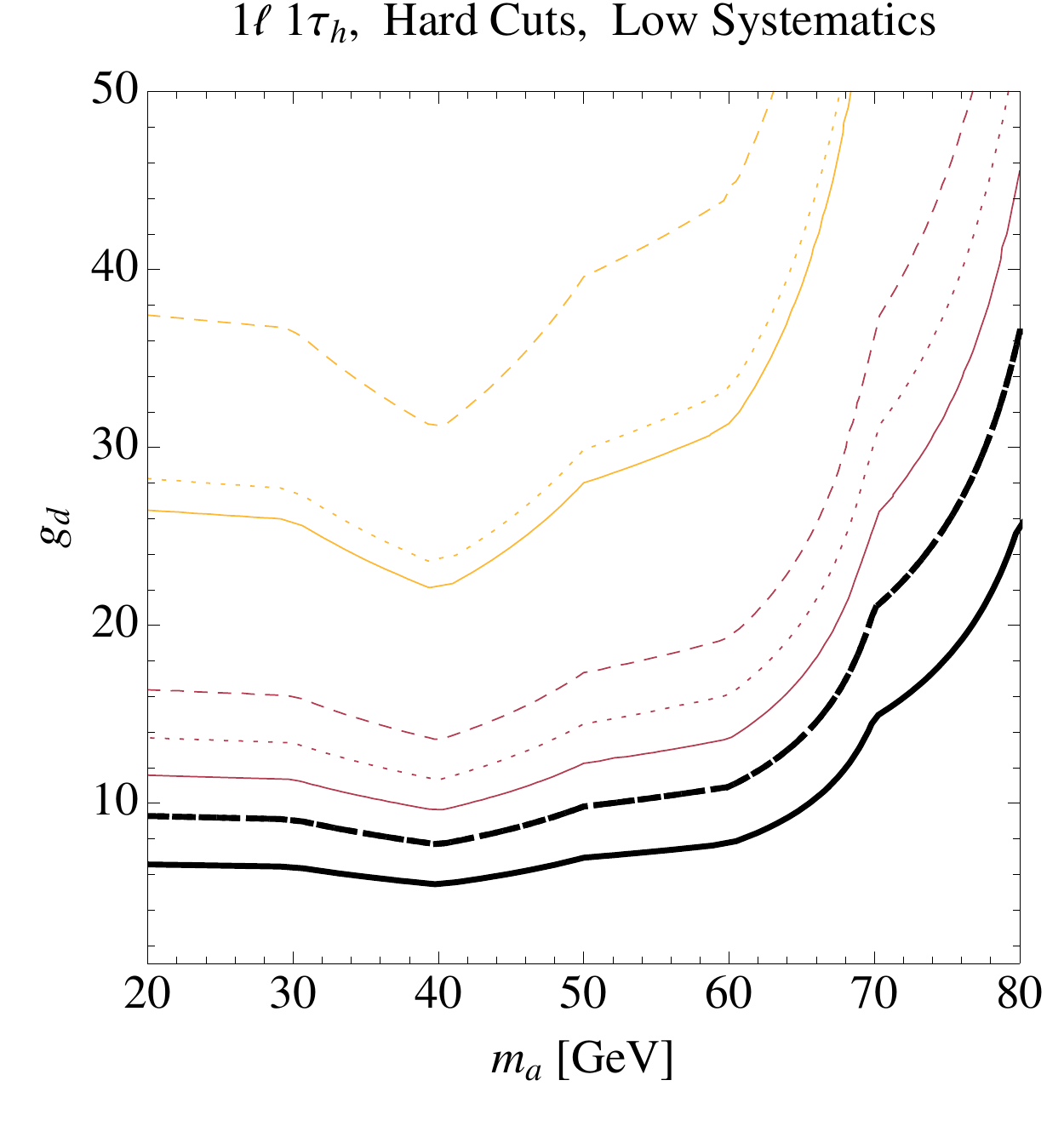}
\includegraphics[width=0.45\textwidth]{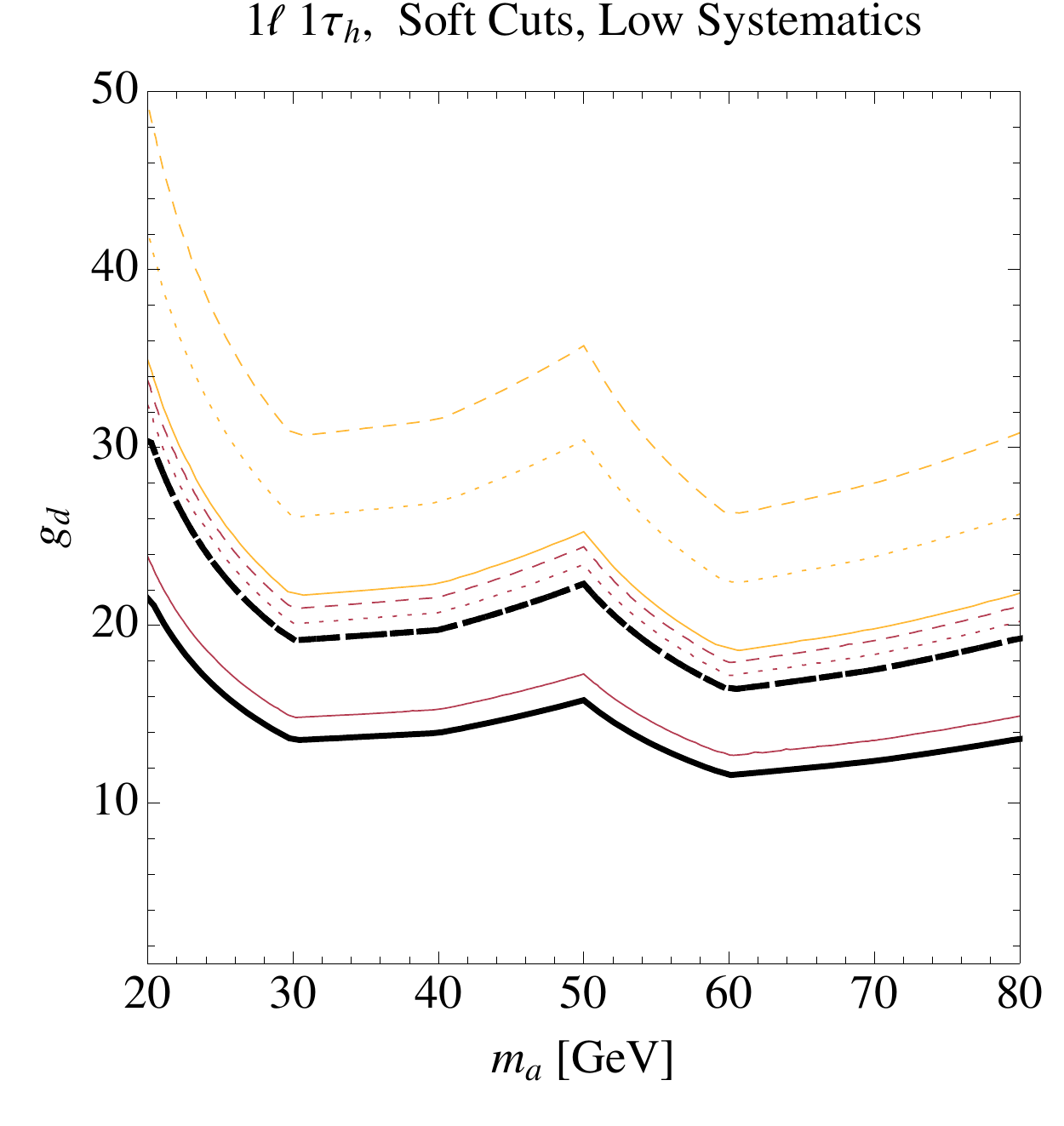}\\
\includegraphics[width=0.45\textwidth]{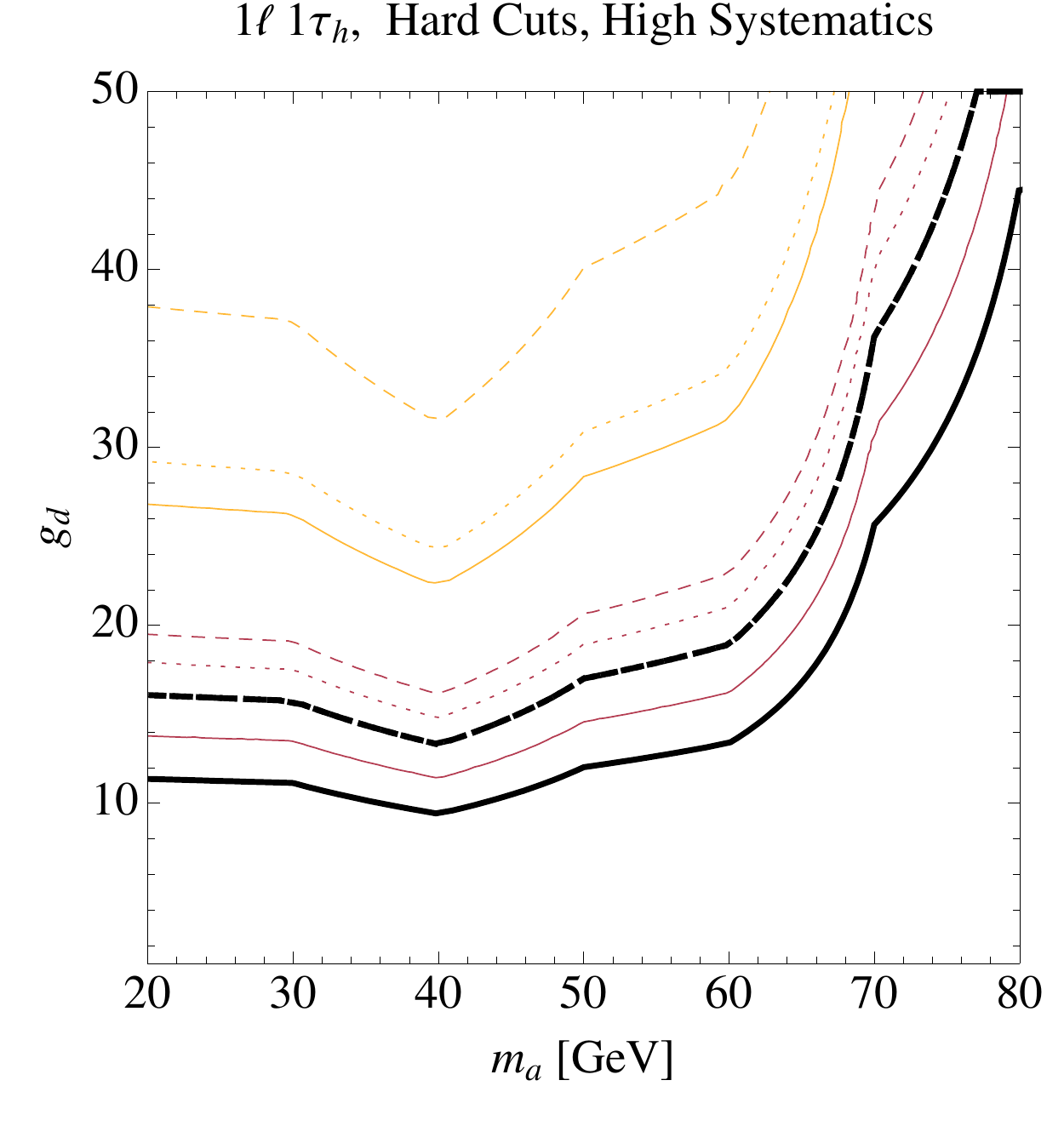}
\includegraphics[width=0.45\textwidth]{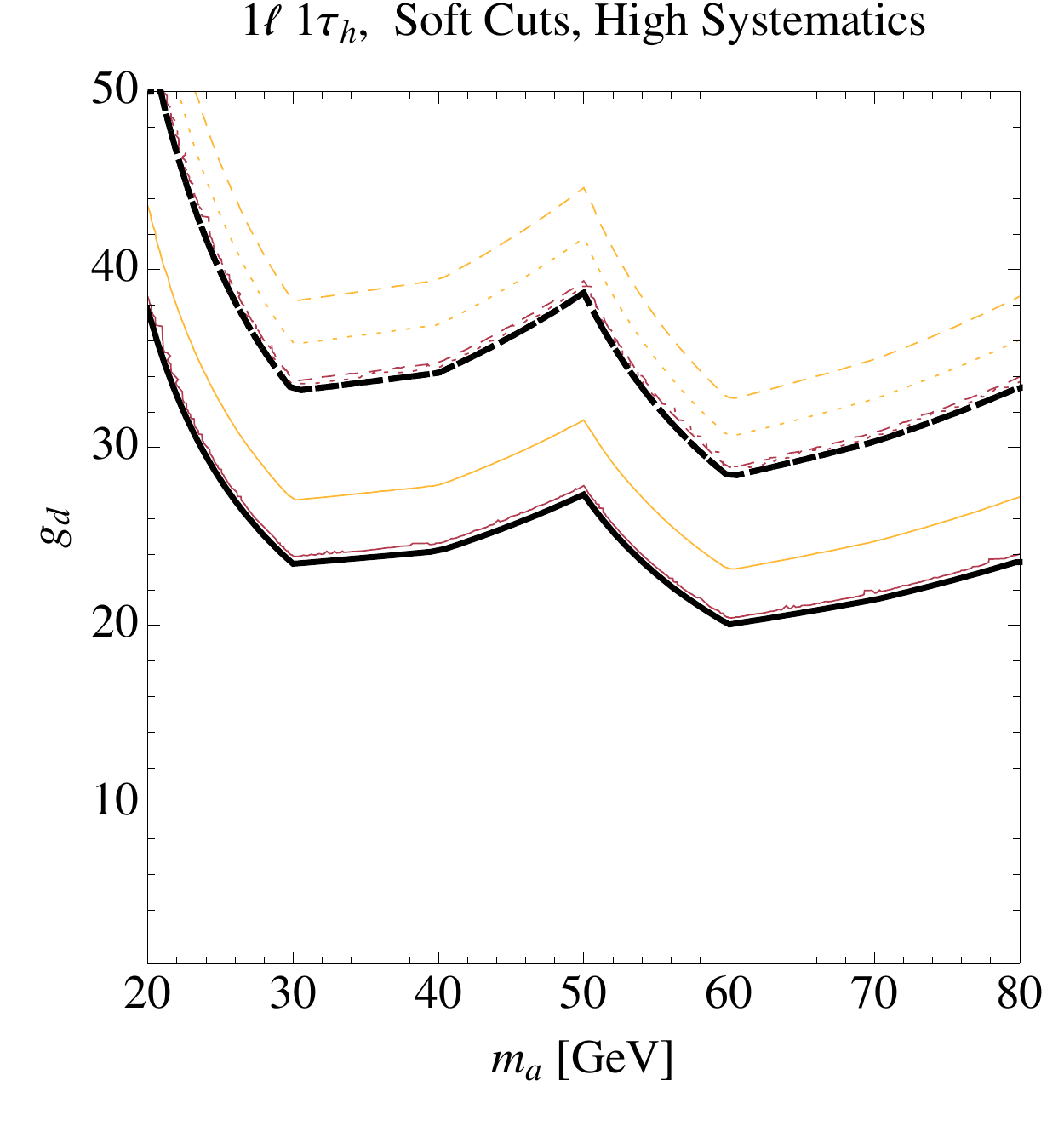}
   \end{center}
\caption{Discovery potential contours for the SR2 ($1\ell1\tau$) signal region with hard (left) and soft (right) cuts for $\epsilon_{sys}=0.1 (0.3)$ (top (bottom)) and conservative factors applied to the signal (dotted) and backgrounds (dashed) (solid lines show the original bounds without any factor applied to the signal or background). Contours correspond to constant values of $\log(L\times \mathrm{fb})$ needed to achieve $k=3$. The black lines represent the boundary of the systematics dominated region, the red lines represent the discovery potential at $L=$10/fb, while the yellow lines represent the discovery potential for $L=$1/fb.} \label{fig:1l1tdisc_k}
\end{figure}

\begin{figure}[!h]
   \begin{center}
\includegraphics[width=0.45\textwidth]{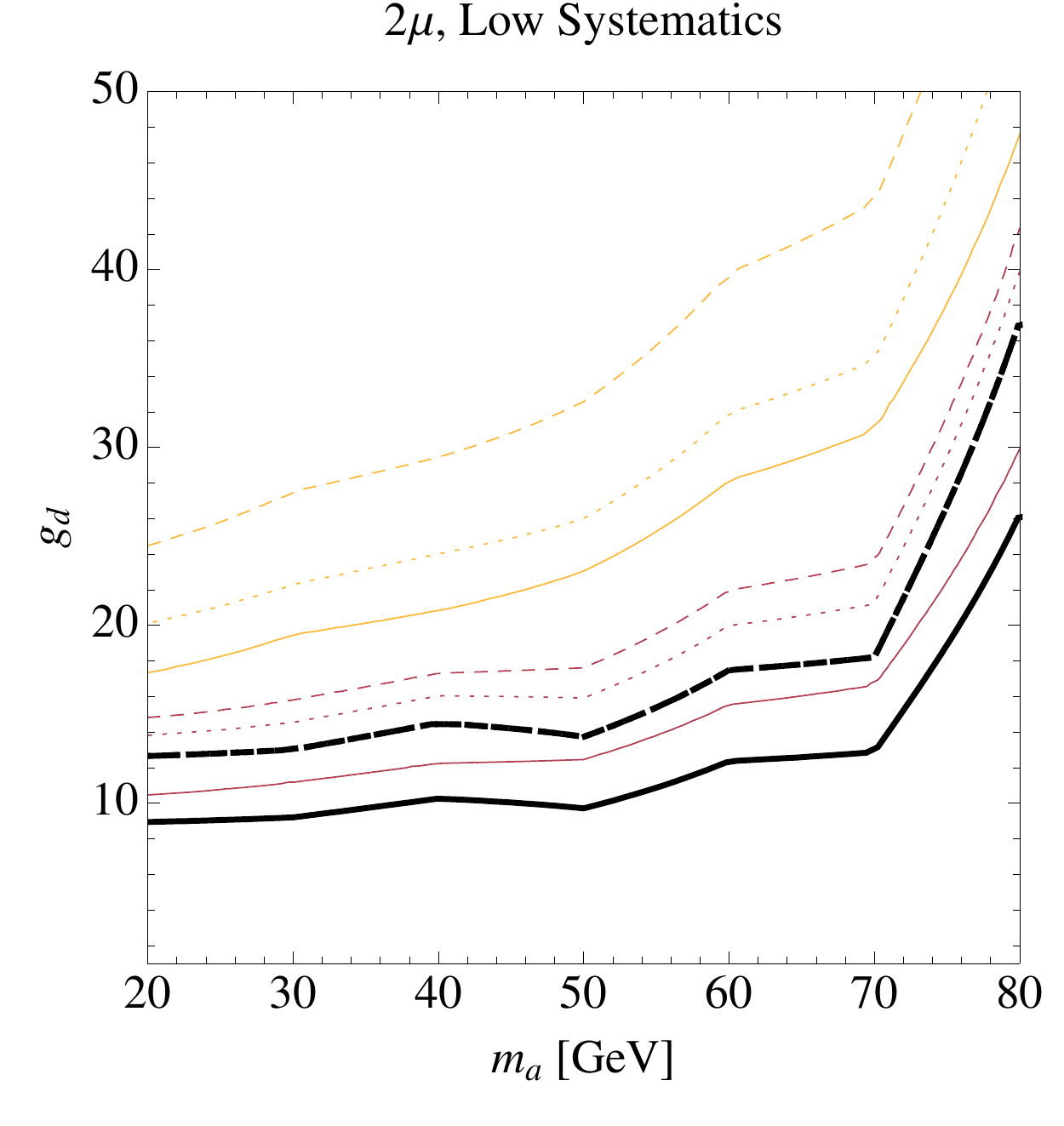}
\includegraphics[width=0.45\textwidth]{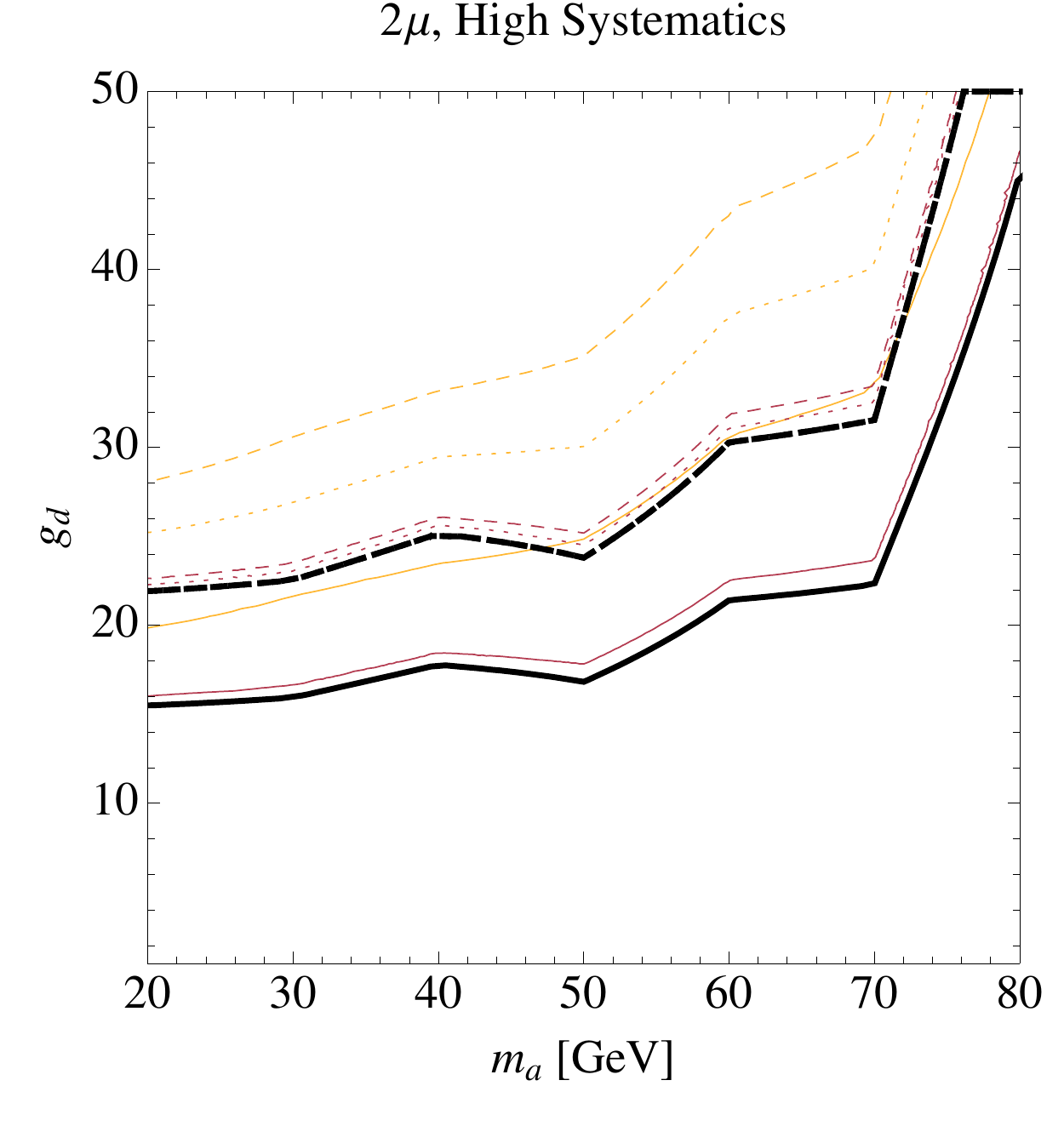}
   \end{center}
\caption{Discovery potential contours for the SR3 ($2\mu$) signal region for $\epsilon_{sys}=0.1 (0.3)$ (left (right)) and conservative factors applied to the signal (dotted) and backgrounds (dashed) (solid lines show the original bounds without any factor applied to the signal or background). Contours correspond to constant values of $\log(L\times \mathrm{fb})$ needed to achieve $k=3$. The black lines represent the boundary of the systematics dominated region, the red lines represent the discovery potential at $L=$10/fb, while the yellow lines represent the discovery potential for $L=$1/fb.} \label{fig:1u1udisc_k}
\end{figure}

In addition, our calculations of the backgrounds were performed at leading order. Higher order effects, as well as possible unaccounted-for experimental issues, may result in larger backgrounds than we predict. In order to address concerns regarding these two issues, we explore much more conservative contours determined by performing the same calculations but with a factor of 2.0 larger backgrounds, and separately with a factor of 0.5 smaller signal. Figures \ref{fig:1e1udisc_k}, \ref{fig:1l1tdisc_k} and \ref{fig:1u1udisc_k} give these results. Of note, many regions of parameter space are still explorable at the LHC with 100/fb of integrated luminosity even in the more pessimistic scenarios.

\end{document}